\documentclass[twocolumn, tighten]{aastex62}

\newcommand{\um}{$\mu$m}

\submitjournal{ApJ}

\shorttitle{Parallaxes with \textit{WISE}}
\shortauthors{Theissen}

\begin{document}

\title{Parallaxes of Cool Objects with \textit{WISE}: Filling in for \textit{Gaia}}

\email{ctheissen@ucsd.edu}

\author{Christopher A. Theissen}
\affiliation{Center for Astrophysics and Space Sciences, University of California, San Diego, 9500 Gilman Dr., Mail Code 0424, La Jolla, CA 92093, USA}

\begin{abstract}

This paper uses the multi-epoch astrometry from the \textit{Wide-field Infrared Survey Explorer} (\textit{WISE}) to demonstrate a method to measure proper motions and trigonometric parallaxes with precisions of $\sim$4 mas yr$^{-1}$ and $\sim$7 mas, respectively, for low-mass stars and brown dwarfs. This method relies on \textit{WISE} single exposures (Level 1b frames) and a Markov Chain Monte Carlo method. The limitations of \textit{Gaia} in observing very low-mass stars and brown dwarfs are discussed, and it is shown that \textit{WISE} will be able to measure astrometry past the 95\% completeness limit and magnitude limit of \textit{Gaia} (L, T, and Y dwarfs fainter than $G\approx19$ and $G=21$, respectively). This method is applied to \textit{WISE} data of 20 nearby ($\lesssim17$~pc) dwarfs with spectral types between M6--Y2 and previously measured trigonometric parallaxes. Also provided are \textit{WISE} astrometric measurements for 23 additional low-mass dwarfs with spectral types between M6--T7 and estimated photometric distances $<17$~pc. Only nine of these objects contain parallaxes within \textit{Gaia} Data Release 2.

\end{abstract}

\keywords{techniques: miscellaneous --- astrometry --- parallaxes --- proper motions --- brown dwarfs --- stars: low-mass}

\section{Introduction}
\label{intro}

	With the recent release of proper motions and trigonometric parallax measurements for over a billion sources from the \textit{Gaia} satellite \citep{arenou:2018:,gaia-collaboration:2018:,lindegren:2018:,luri:2018:}, it is important to understand what objects are not included in the recent catalog, and what objects will be missing from the final catalog. \citet{theissen:2017:92} investigated the \textit{Gaia} shortfall and found that \textit{Gaia} will be limited in its ability to observe ultracool dwarfs (spectral-types later than mid-L) at distances $\gtrsim10$~pc due to its relatively blue bandpass.
	
	A number of projects are aimed at measuring trigonometric parallaxes of these ultracool dwarfs \citep[e.g.,][]{dupuy:2012:19, beichman:2013:101, faherty:2012:56, kirkpatrick:2014:122, dupuy:2016:23, skinner:2016:36, smart:2017:3764, weinberger:2016:24}. These projects typically rely on either: 1) numerous epochs of ground-based and space-based observations, using facilities such as the \textit{Spitzer Space Telescope} \citep{werner:2004:1}; or 2) survey data spanning multiple epochs, such as the Digitized Sky Survey (DSS), the \textit{Wide-field Infrared Survey Explorer} \citep[\textit{WISE};][]{wright:2010:1868}, or the Two Micron All-Sky Survey \citep[2MASS;][]{skrutskie:2006:1163}. In this paper, I present a method to measure proper motions and trigonometric parallaxes of nearby ($\lesssim$17~pc), ultracool objects using publicly available \textit{WISE} data.
	
	In Section~\ref{gaia}, the properties and limitations of \textit{Gaia} and \textit{WISE} are discussed. The method for measuring proper motions and parallaxes is described in Section~\ref{method}. Comparisons between this method and previous literature measurements for 20 nearby, low-mass dwarfs are provided in Section~\ref{literature}. In Section~\ref{new}, I also provide new measurements for 23 nearby, low-mass dwarfs, nine of which are contained within \textit{Gaia} Data Release 2. Lastly, I discuss the utility of this method for the immediate future in Section~\ref{discussion}.

\section{\text{WISE} Multi-epoch Data}
\label{wisedata}

	The all-sky observations made by \textit{WISE} are ideal for astrometric studies because they span multiple epochs, most separated by $\sim$ 6 months, with the survey strategy of observing fields close to 90$^\circ$ Solar elongation; this places observed objects close to their maximum parallax factors \citep{kirkpatrick:2014:122}. The original \textit{WISE} mission surveyed the entire sky in four bands: 3.4, 4.6, 12, and 22~\um\ (hereafter $W1$, $W2$, $W3$, and $W4$). The original mission lasted from December 2009 to August 2010, after which the cryogen was depleted, and \textit{WISE} observed in $W1$, $W2$, and $W3$ until September (3-band survey; $\sim$30\% of the sky\footnote{\url{http://wise2.ipac.caltech.edu/docs/release/3band/}}).
	
	 \textit{WISE} continued to observe in $W1$ and $W2$ as part of the Near-Earth Object Wide-field Infrared Survey Explorer \citep[NEOWISE;][]{mainzer:2011:53} mission until February 2011, when it was put into hibernation upon completion of its mission. In December 2013, \textit{WISE} was reactivated to ``continue rapidly surveying and obtaining measurements of minor planet physical properties" \citep{mainzer:2014:30}. This mission, dubbed the NEOWISE-Reactivation \citep[NEOWISE-R;][]{mainzer:2014:30} mission, continued surveying the entire sky in $W1$ and $W2$. The NEOWISE-R mission is currently ongoing. The combined \textit{WISE} dataset contains $\geqslant10$ epochs for every source, with a cadence of $\sim$6 months and a time baseline of $\sim$7 years, with approximately 2.5 years of deactivation in-between. Each single epoch has 12--13 7.7 second exposures in both $W1$ and $W2$\footnote{\url{http://wise2.ipac.caltech.edu/docs/release/allsky/}}, and possibly additional exposures dependent on depth-of-coverage for a given line-of-sight.
	
\begin{figure}[htbp!]
\centering
\includegraphics[]{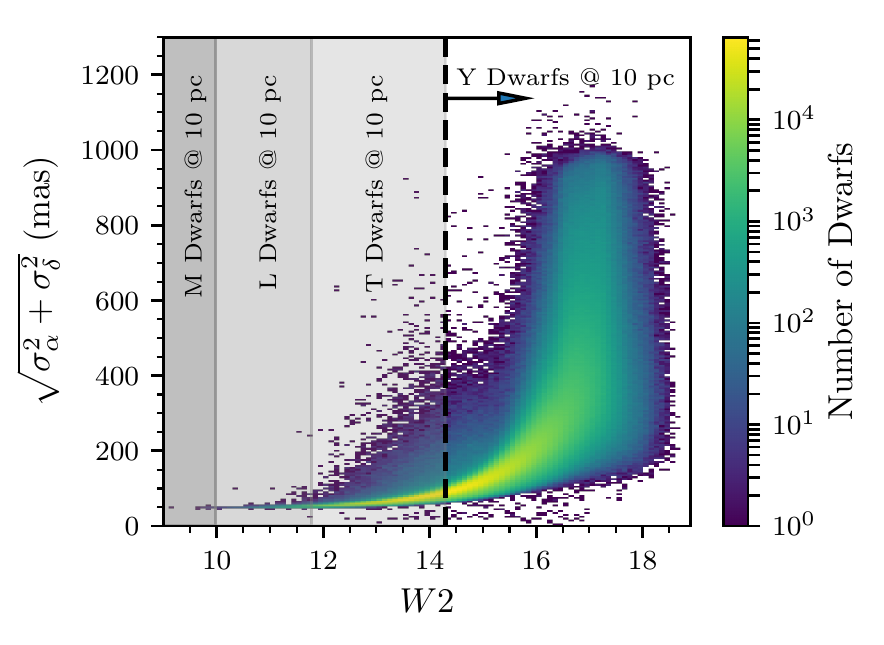}
\caption{
2-D histogram of total \textit{WISE} positional uncertainty (single frames) versus $W2$ magnitude for the MoVeRS catalog. Bin areas are 0.1~mag $\times$ 5~mas. Also shown are the $M_{W2}$ ranges (gray shaded areas) for M, L, T, and Y dwarfs at a distance of 10 pc, using typical values from \citet{filippazzo:2015:158} and \citet{tinney:2014:39}. The Y dwarf region is not well constrained due to the scarcity of Y dwarfs with measured parallaxes, and is denoted with an arrow to illustrate the approximate region Y dwarfs are expected to inhabit. The astrometric precision hits a floor of $\sim$50~mas for relatively bright sources. Y dwarfs typically have a single-band measurement ($W2$), with low signal-to-noise, which will push them to higher positional uncertainties ($>400$~mas).
\label{fig:dwarfs}}
\end{figure}

	Figure~\ref{fig:dwarfs} shows the total \textit{WISE} positional uncertainty for a single frame (i.e., $\sqrt{\sigma_\alpha^2+\sigma_\delta^2}$) as a function of $W2$ magnitude using the Motion Verified Red Stars catalog catalog \citep[MoVeRS;][]{theissen:2016:41}. The $W2$-absolute magnitude ranges (gray shaded regions) for M, L, T, and Y dwarfs at a distance of 10~pc are shown using the spectral-type--absolute magnitude relationships from \citet{filippazzo:2015:158} and \citet{tinney:2014:39}. The positional uncertainty can be reduced within the relative \textit{WISE} frame-of-reference by $\sim\sqrt{N}$, where $N$ is the number of frames an object was observed in. For the typical value of $N=13$, this can reduce positional uncertainties to $\approx 14$ mas for M, L, and early T dwarfs. Fainter sources such as late T dwarfs will have relative positional uncertainties of $\approx 50$ mas, and extremely faint Y dwarfs with single-band detections will typically have relative uncertainties of $\approx 150$ mas.

	Many studies have computed parallaxes using \textit{WISE} data---combined with either higher-positional precision observations (e.g., \textit{Spitzer}, Keck), and/or data providing a longer time baseline (e.g., DSS, 2MASS)---for nearby objects \citep[e.g.,][]{beichman:2013:101, luhman:2013:l1, kirkpatrick:2014:122, scholz:2014:a113}. However, the numerous epochs of current \textit{WISE} data now allow relatively precise ($\lesssim10$~mas) parallax measurements to be made without the need for further data.

\section{\textit{Gaia} Limitations for Observing Ultracool Dwarfs}
\label{gaia}

	\textit{Gaia} is currently conducting the largest astrometric mission to date, with an expected yield of over 1.3 billion sources with measured proper motions and precise trigonometric parallaxes \citep[$\approx$0.1~mas for sources with $G\approx17$ and $\approx$0.7~mas for sources with $G\approx20$;][]{luri:2018:}. \citet{theissen:2017:92} quantified the shortfall of ultracool objects within \textit{Gaia} Data Release 1 \citep[DR1;][]{gaia-collaboration:2016:a1, gaia-collaboration:2016:a2}, using matches between the Late-Type Extension to the Motion Verified Red Stars catalog (LaTE-MoVeRS) and the full \textit{Gaia} DR1 (catalog of positions and magnitudes, but not necessarily parallaxes). They found that \textit{Gaia} is limited in its ability to observe spectral types later than $\sim$L5 farther than $\sim$10~pc.
	
\begin{figure*}[htbp!]
\centering
\includegraphics[]{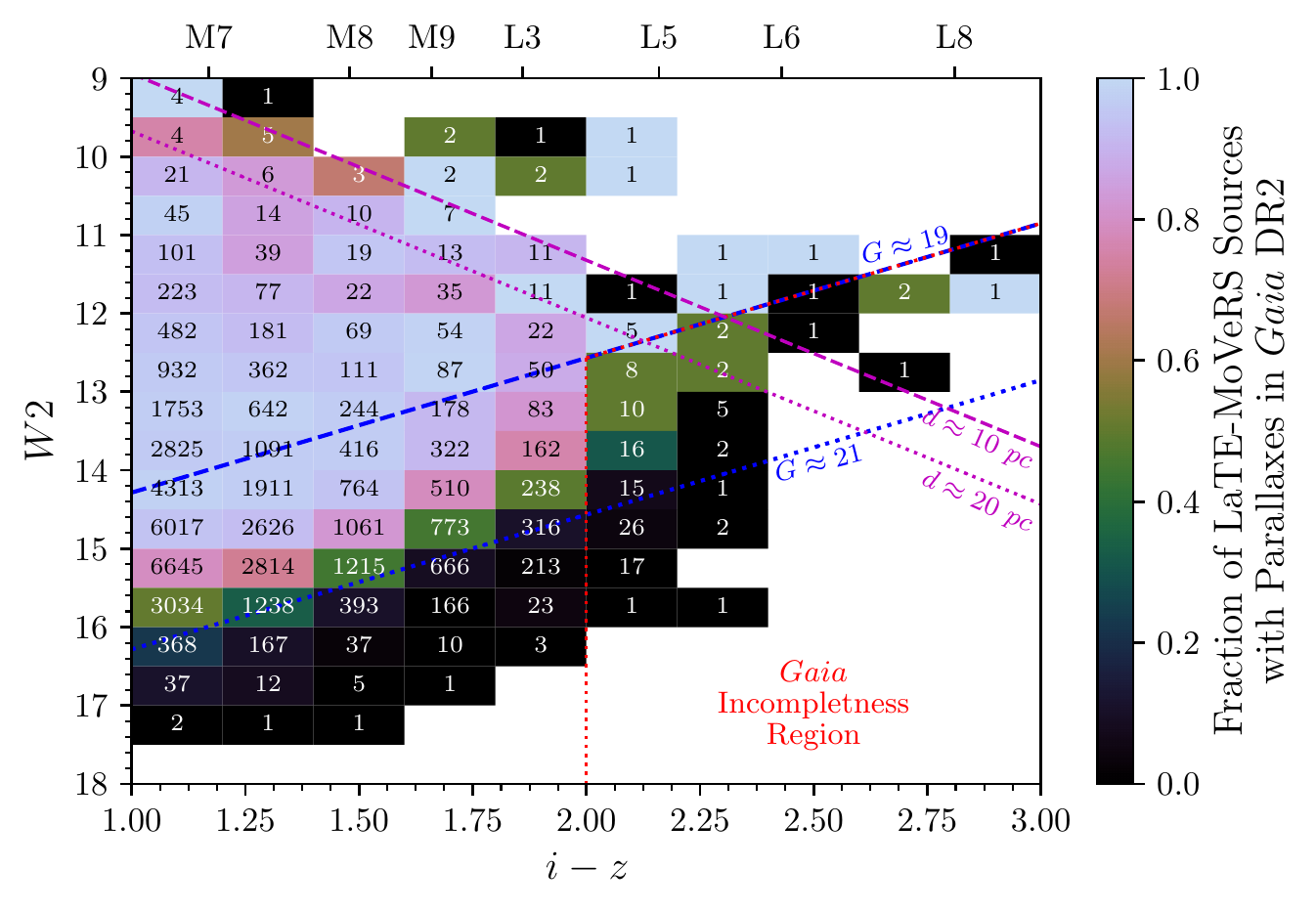}
\caption{
Fraction of matches between the LaTE-MoVeRS sample and \textit{Gaia} DR2 as a function of $W2$ and $i-z$, requiring \textit{Gaia} matches have $\pi/\sigma_\pi \geqslant3$. The number on each bin indicates the total number of LaTE-MoVeRS sources within that bin, and the color of the bin corresponds to the fraction of stars with matches in \textit{Gaia} DR2 (black colored text indicates a fraction $\geqslant0.7$). 
Approximate \textit{Gaia} iso-magnitude lines are shown with the blue dashed line ($G \approx 19$; $\sim$95\% 5-parameter astrometric solution completeness limit; see \citealt{arenou:2018:} Figure A.1) and blue dotted line ($G \approx 21$; limiting magnitude from \citealt{arenou:2018:}).
The fraction of \textit{Gaia} DR2 matches drops below $\sim$50\% for spectral types later than $\sim$L4 with $G\geqslant19$ (area enclosed with red-dotted lines). 
Approximate spectral types from \citet{schmidt:2015:158} are listed on the top. Magenta lines indicate distances using the photometric relationship from Schmidt et al. (2018, in preparation).
\label{fig:w2}}
\end{figure*}

	The \textit{Gaia} shortfall was reevaluated here using the LaTE-MoVeRS sample, by comparing the fraction of LaTE-MoVeRS sources with a counterpart found within 6\arcsec\ in \textit{Gaia} Data Release 2 \citep[DR2;][]{gaia-collaboration:2018:,luri:2018:}. Figure~\ref{fig:w2} shows the fraction of matched sources with significant parallaxes ($\pi / \sigma_\pi \geqslant 3$) as a function of $W2$ magnitude and Sloan Digital Sky Survey \citep[SDSS;][]{york:2000:1579} $i-z$ color. To put this in a \textit{Gaia} context, the matched sample between \textit{Gaia} DR2 and LaTE-MoVeRS was used to compute first order linear transformations between $W2$ and $G$ magnitudes as a function of $i-z$ color, shown in Figure~\ref{fig:w2} (blue lines), and given by the equation,
\begin{equation}
G=19: W2 = 16.00 - 1.72\, (i-z).
\end{equation}	

	The fraction of matched sources between LaTE-MoVeRS and \textit{Gaia} DR2 typically drops below 50\% for objects with $i-z > 2$ and $G\gtrsim19$ ($W2\gtrsim11$--12.5), which corresponds to the \textit{Gaia} DR2 95\% completeness limit for sources with 5-parameter astrometric solutions (see \citealt{arenou:2018:} Figure A.1). The \textit{Gaia} limiting magnitude for sources with computed 5-parameter astrometric solutions ($G = 21$) is also shown in Figure~\ref{fig:w2}. The linear fit to $G=21$ is a good approximation for the observed dropout of fainter sources without \textit{Gaia} parallaxes. Approximate distances for late-type dwarfs as a function of $W2$ and $i-z$ are also shown in Figure~\ref{fig:w2} (magenta lines) using the photometric distance relationships from Schmidt et al. (2018, in preparation). Future \textit{Gaia} data releases \emph{may} have measurements for ultracool dwarfs within 10~pc and with $G<21$, but much of the \textit{Gaia} incompleteness region will be lacking sources due to the \textit{Gaia} 95\% completeness limit and magnitude limit.

\section{Parallaxes using \text{WISE} Multi-epoch Data}
\label{method}

\subsection{Single Epoch Position Measurements}
\label{singleepoch}

\begin{figure}[htbp!]
\centering
\includegraphics[width=\linewidth]{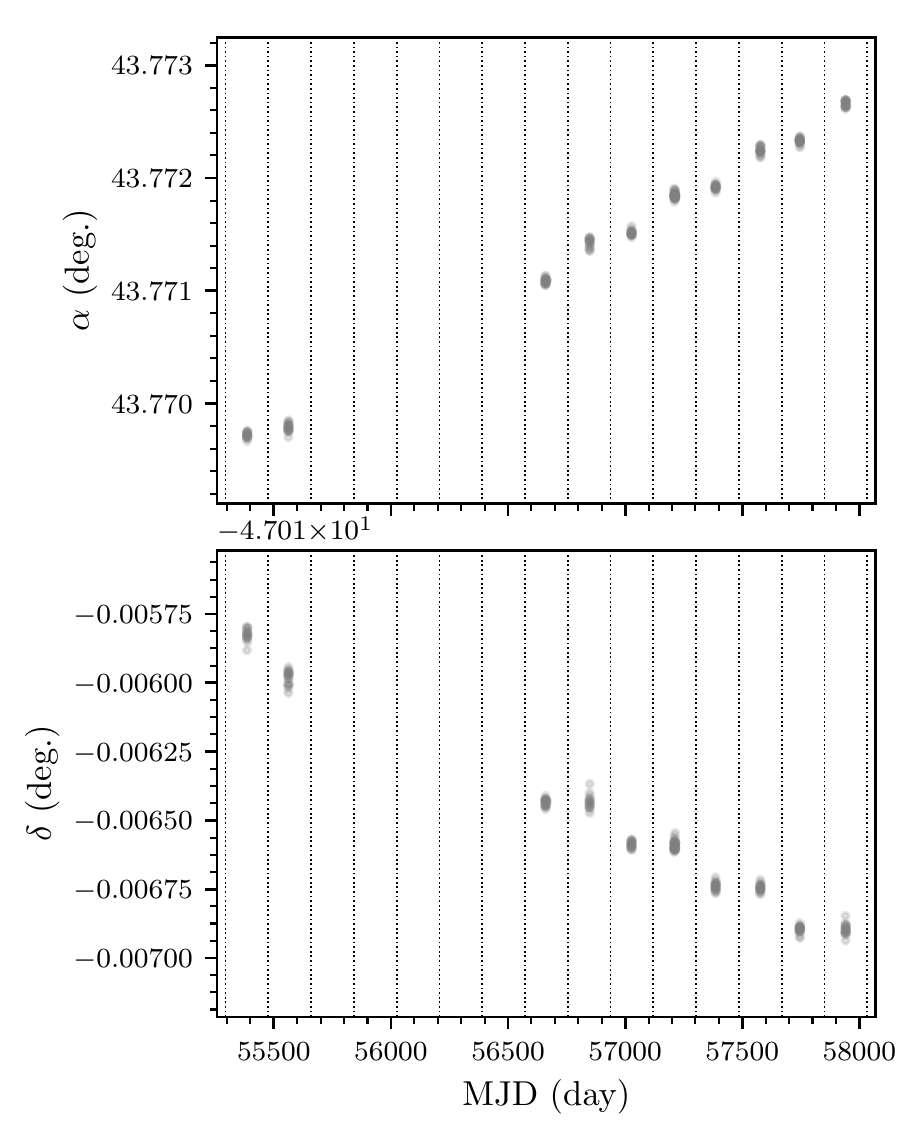}
\includegraphics[width=\linewidth]{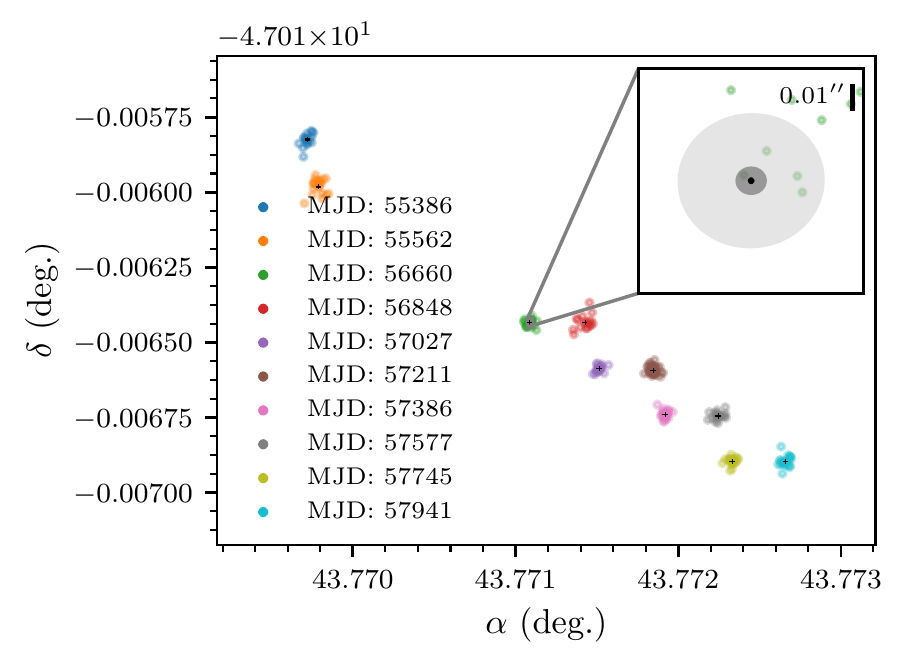}
\caption{Grouping method for \textit{WISE} parallaxes for 2MASS J02550357$-$4700509.
\textit{Top and Middle}: Groupings by date among the All-Sky, 3-band, NEOWISE Post-Cryo, and NEOWISE-R releases. Each translucent gray point represents a single observation.
\textit{Bottom}: Grouping by $\alpha$ and $\delta$ for each observational epoch. Each translucent colored point is a single L1b measurement, with the color representing the observational epoch. The inset plot is a $0.1\arcsec \times 0.1\arcsec$ box centered around the uncertainty weighted average position for the MJD = 56660 epoch (black point). The median error ellipse among the individual observations within the MJD $= 56660$ epoch is shown in light gray. The weighted positional uncertainty is shown with the dark gray ellipse.
\label{fig:method}}
\end{figure}

	An illustration of the parallax method described here is shown in Figure~\ref{fig:method} for 2MASS J02550357$-$4700509, a nearby ($\sim$5 pc) L8 dwarf \citep{martin:1999:2466, patten:2006:502, kirkpatrick:2008:1295, faherty:2012:56}. First, all Level 1b (L1b) source catalogs (i.e., All-Sky, 3-band, NEOWISE Post-Cryo, and NEOWISE-R) are queried for objects within 30\arcsec\ of the expected position of 2MASS J02550357$-$4700509. The L1b source catalogs contain sources extracted from each single exposure\footnote{\url{http://wise2.ipac.caltech.edu/docs/release/allsky/expsup/sec4\_1.html}}. Sources were grouped by epoch, demarcated by 6 month periods starting 91 days after the mean modified Julian date (MJD) of the first epoch (shown as dotted lines in the top and middle panels of Figure~\ref{fig:method}). The \textit{WISE} pipeline already accounts for registering dithers by obtaining a global astrometric solution to each frame using 2MASS positions\footnote{\url{http://wise2.ipac.caltech.edu/docs/release/prelim/expsup/sec4\_3d.html}}.
	
	A registration method similar to \citet{dupuy:2012:19} was investigated for residual frame-to-frame offsets that were not removed during the dither registration method already implemented within the \textit{WISE} pipeline. First, a list of potential reference objects was created from the AllWISE catalog \citep{cutri:2013:}, selecting all sources within a 10\arcmin\ radius of the target object \citep[\textit{WISE} field-of-view $= 47\arcmin \times 47\arcmin$;][]{wright:2010:1868}. Reference objects were then selected by requiring the following:
\begin{enumerate}
\item Each object has no saturated pixels in $W1$ and $W2$ (\textsc{w1sat} $=0$ \& \textsc{w2sat} $=0$);
\item Each object is free from contamination and confusion flags (\textsc{cc\_flag} $=0000$);
\item Each object does not have an extended object flag (\textsc{ext\_flag} $=0$); and
\item Each object is within the magnitude limit $8 \leqslant W1 \leqslant 12$.
\end{enumerate}
These criteria ensured unsaturated sources with pristine astrometry within \textit{WISE}. A minimum of 40 reference sources within each frame was required, and the search radius was incrementally increased by 1\arcmin\ until enough reference sources were detected.

	Using the final list of reference objects, sources were then extracted from each L1b frame, and positional offsets ($\Delta\alpha$ and $\Delta\delta$) were computed for each object between the first frame and each subsequent frame within the given epoch. The residual $\alpha$ and $\delta$ between the first frame and each subsequent frame ($\Delta\alpha$ and $\Delta\delta$) were then binned in a 2-dimensional histogram, starting with an arbitrarily large bin size and iteratively decreasing the bin size until $\leqslant 70\%$ of the sources were contained within a single bin (typical bin size $\approx 100$ mas $\times$ 100 mas). The median $\Delta\alpha$ and $\Delta\delta$ offsets were then taken from the peak bin, which should contain the actual offsets between two frames. These shifts were typically very small ($\lesssim 10$ mas), and did not have a significant effect on the final computed position and uncertainty. Therefore, this correction was not applied for the remainder of this study.

	To obtain the true relative position within the \textit{WISE} frame of reference, the uncertainty weighted average position for each epoch was computed using the \textit{WISE} reported position values ($\alpha$, $\delta$, $\sigma_\alpha$, and $\sigma_\delta$), using a 3$\sigma$ clip to remove outliers. This method gives a statistical measure of the true position within the relative \textit{WISE} frame of reference, and is based on the fact that each frame is independently calibrated using the exact same method (i.e., the \textit{WISE} processing pipeline). Each epoch typically spans a period of $\sim$1--10 days. The true observational epoch time is selected to be the average MJD for each epoch.
	
	Uncertainties ($\sigma_\alpha$ and $\sigma_\delta$) were computed using the weighted positional uncertainty per epoch, as illustrated in the inset figure of the bottom panel of Figure~\ref{fig:method} and given by, 
\begin{equation}
\langle\sigma_{\alpha, \delta}\rangle = \frac{1}{\sqrt{\sum_{N} 1 / \sigma_i^2}},
\end{equation}
where $N$ is the number of frames within the given epoch. This method reduces the relative astrometric uncertainty by a factor of $\sim \sqrt{N}$ where $N$ is the number of frames a source was observed in during a given epoch.

\subsection{Registering Astrometry between Epochs}
\label{registration2}

	To obtain a relative astrometric solution, each epoch must be registered using sources common to all epochs that exhibit little to no proper motion. In principle, this step is already done through the \textit{WISE} processing pipeline, similar to registration within a given epoch (i.e., dither registration). However, there is an observed dipole residual in the astrometry between forward and backward pointing frames taken with \textit{WISE} \citep{meisner:2017:}. This astrometric shift between 6 month periods can be as large as 600 mas along a given line-of-sight, inducing a potential parallax signature in objects with no measurable parallax \citep{meisner:2017:}. \citet{meisner:2017:} posit this astrometric shift is due to an asymmetry in the \textit{WISE} point-spread function (PSF) models with respect to scan direction. This astrometric shift can be accounted for by registering epochs using a common set of reference objects across all epochs for a given line-of-sight. 
	
	To correct for this systematic effect, again, a procedure similar to the one outlined in \citet{dupuy:2012:19} was used. This method follows the same procedure discussed in Section~\ref{singleepoch}. First, a list of potential reference objects was created from the AllWISE catalog using the same selection criteria as Section~\ref{singleepoch}. Using the final list of reference objects, sources were then extracted from each L1b frame, similar to the target object, and uncertainty weighted positions were computed for each object within each epoch. Only sources found within $>$50\% of the epochs were kept. Then, the positional shifts between the first epoch and each subsequent epoch were computed. Sources with large proper motions ($\gtrsim 30$ mas yr$^{-1}$) were masked where proper motion information was available or where a proper motion was measured.
	
	The residual $\alpha$ and $\delta$ between two epochs ($\Delta\alpha$ and $\Delta\delta$) were then binned in a 2-dimensional histogram, starting with an arbitrarily large bin size and iteratively decreasing the bin size until $\leqslant 70\%$ of the sources were contained within a single bin (typical bin size $\approx 100$ mas $\times$ 100 mas). The median $\Delta\alpha$ and $\Delta\delta$ offsets were then taken from the peak bin, which should contain the true offsets between epochs. Offsets from the first epoch were then applied to each subsequent epoch of the target object's uncertainty weighted position.

\subsubsection{Testing the Efficacy of the Epoch Registration}

	To test the efficacy of the above registration procedure in reducing positional scatter, the above registration procedure was applied to the SDSS Data Release 14 Spectroscopic Quasar Catalog\footnote{\url{http://www.sdss.org/dr14/data\_access/value-added-catalogs/?vac_id=the-sloan-digital-sky-survey-quasar-catalog-fourteenth-data-release}}. This catalog contains 526,356 quasars confirmed based on their optical spectroscopy. These sources should exhibit no proper motion, and offer an unbiased sample---distributed approximately uniformly across one-third of the sky---on which to test the registration procedure from Section~\ref{registration2}.
	
\begin{figure*}[htbp!]
\centering
\includegraphics{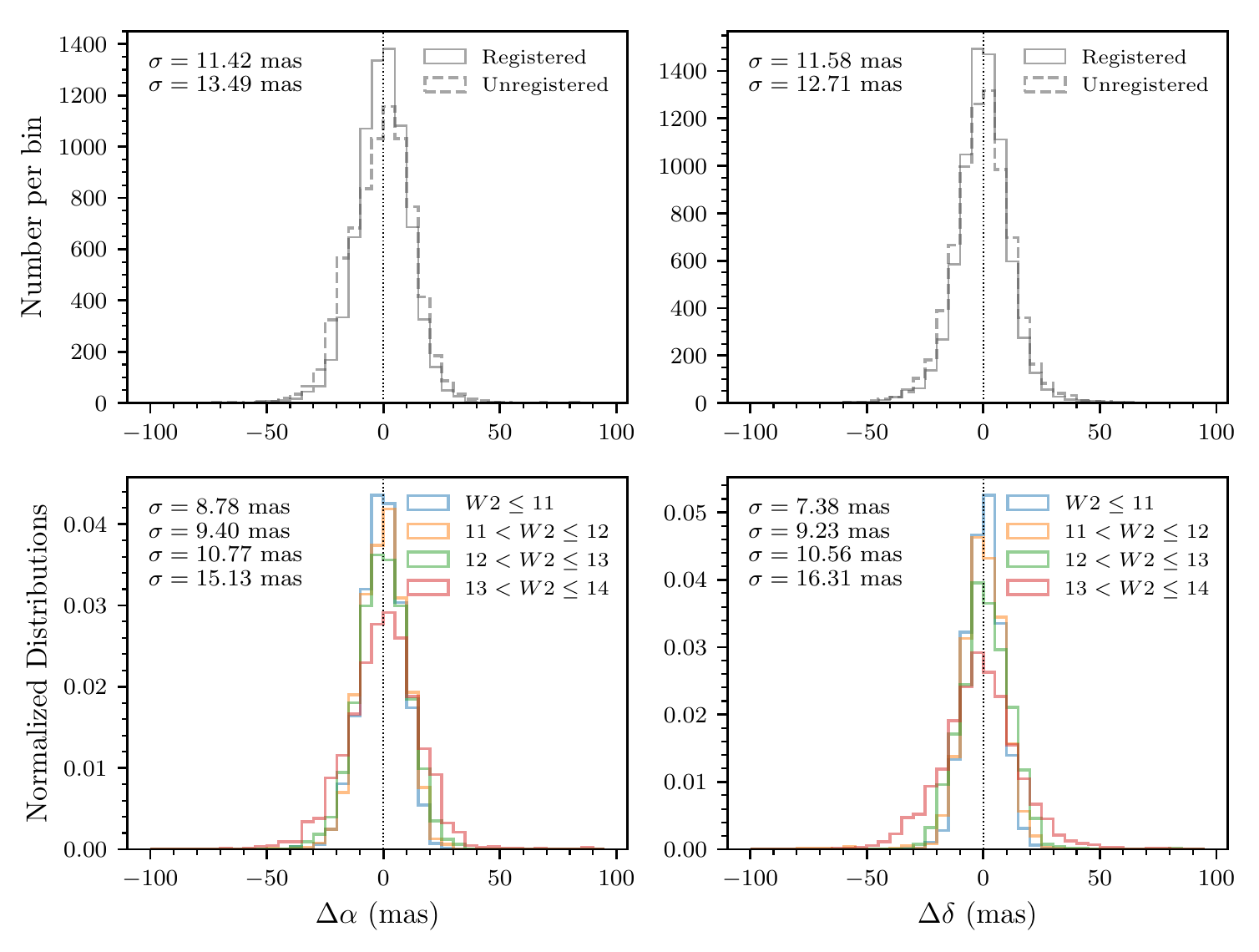}
\caption{Positional differences between epochs for 1000 quasars from the SDSS DR14 spectroscopic QSO catalog.
\textit{Top}: The $\alpha$ (left) and $\delta$ (right) distributions of the uncertainty weighted positions between each epoch relative to the uncertainty weighted position of the first epoch. Using the registration procedure from Section~\ref{registration2} reduces the scatter relative to the position of the first epoch (standard deviations shown on the top left of each subplot).
\textit{Bottom}: The $\alpha$ (left) and $\delta$ (right) normalized distributions of registered sources separated by $W2$ magnitude. The relative scatter about the position of the first epoch is reduced for brighter sources.
\label{fig:qsos}}
\end{figure*}

	Only QSOs with $W2 \leqslant 14$ (50,834 QSOs; 36 with $W2 < 10$) were selected to coincide with the magnitude range of objects in this study. One thousand QSOs were randomly sampled within the aforementioned magnitude range, with QSOs selected in roughly equal numbers between the limits of $W2 \leqslant 11$, $11 < W2 \leqslant 12$, $12 < W2 \leqslant 13$, and $13 < W2 \leqslant 14$. The uncertainty weighted position for each epoch was computed, epochs were then registered, and then shifts between the first epoch and each subsequent epoch were computed. Figure~\ref{fig:qsos} (top) shows the distribution of shifts for both registered and unregistered epochs in both $\alpha$ and $\delta$, weighted by the number of epochs available. The scatter relative to the position of the first epoch is reduced using the registration method from Section~\ref{registration2}. Figure~\ref{fig:qsos} (bottom) shows registered sources separated into $W2$ magnitude bins, showing increasing positional uncertainty with decreasing source brightness. A similar trend was not observed with $\alpha$ or $\delta$ position, suggesting astrometric precision is primarily a function of source magnitude rather than on-sky position. Typical precision using this registration method is $\sim$12 mas, or 1/200 of a \textit{WISE} L1b pixel. These values will be revisited in terms of positional precision in Section~\ref{literature}.

\subsection{Astrometric Solutions}

	Using the registered positions and uncertainties, the astrometric solution was computed using the following equations, 
\begin{equation}
\label{palpha}
(\alpha_i - \alpha_0) \cos \delta_0 = \Delta\alpha + \mu_\alpha (t_i-t_0) + \pi (P_{\alpha,i} - P_{\alpha,0}),
\end{equation}
\begin{equation}
\label{pdelta}
\delta_i - \delta_0 = \Delta\delta + \mu_\delta (t_i-t_0) + \pi (P_{\delta,i} - P_{\delta,0}),
\end{equation}
where the subscript 0 denotes the first epoch, and the subscript $i$ denotes each subsequent epoch. $\Delta\alpha$ and $\Delta\delta$ represent the mean positions in $\alpha$ and $\delta$, respectively. $P_{\alpha, \delta}$ represents the parallax factors \citep{van-dekamp:1967:} given by \citep{green:1985:},
\begin{equation}
P_{\alpha} = X \sin \alpha - Y \cos \alpha,
\end{equation}
\begin{equation}
P_{\delta} = X \cos \alpha \sin \delta + Y \sin \alpha \sin \delta - Z \cos \delta,
\end{equation}
where $X,Y$, and $Z$ are the components of the barycentric position vector of the Earth obtained from the JPL DE430 solar system ephemeris. These equations were solved using a Markov Chain Monte Carlo (MCMC) routine built on the \textit{emcee} code \citep{foreman-mackey:2013:306}, assuming normally distributed parameters and uniform priors. The parameter space was sampled using 200 walkers, with each walker taking 200 steps. Convergence typically occurred after 50--75 steps, and the first 50\% of chains were discarded as burn in. The posterior distributions were observed to be normally distributed, and reported values throughout this study represent the median values of the posterior distributions and the largest deviation between the median values and the 16th and 84th percentile values (corresponding to the 68\%---or 1$\sigma$---values).

	The above steps provide astrometric solutions \emph{relative} to the \textit{WISE} reference frame. To convert parallax values from relative to absolute, the finite motions of the calibration sources must be known to shift to an absolute reference frame. The absolute astrometric shifts of the calibration sources were estimated using the Besan\c{c}on model \citep{robin:2003:523, czekaj:2014:a102}, using the \textit{Spitzer} Infrared Array Camera \citep[IRAC;][]{fazio:2004:10} 3.4 \um\ band as a proxy for $W1$. Applying the magnitude cuts and search radii from Section~\ref{registration2} to select calibration stars, the average parallax of the calibration sources was found to be $\lesssim$1 mas, with $\gtrsim$90\% of calibration sources having parallaxes $<$1 mas. These shifts are typically much smaller than the average parallax error, and are therefore negligible.

\section{Comparison to Literature Astrometric Measurements}
\label{literature}

	The MCMC routine described in the previous section was applied to 20 known, nearby, low-mass objects with generally well-determined parallaxes ($<$15\% uncertainty). Sources were chosen to cover a range of spectral types, distances, and $W2$ magnitudes. Table~\ref{tbl:parallax} reports the values from the MCMC posterior distributions. The computed positional uncertainty listed in Table~\ref{tbl:parallax} is typically larger than the relative positional scatter of QSOs with similar magnitudes (see Figure~\ref{fig:qsos}). Adding the positional QSO scatter (from Figure~\ref{fig:qsos}) in quadrature with the computed positional uncertainties does not change results significantly, indicating that further corrections or larger positional uncertainties are not warranted.
	
\startlongtable
\begin{deluxetable*}{lccccccc}
\tabletypesize{\footnotesize}
\tablecolumns{8}
\tablecaption{\textit{WISE} Astrometry Measurements Compared to Literature Astrometry
\label{tbl:parallax}}
\tablehead{
\colhead{Parallax} & \colhead{$\alpha_0$} & \colhead{$\delta_0$} & \colhead{$\langle\sigma_{\alpha,\delta}\rangle$} & \colhead{$\mu_\alpha \cos \delta$} & \colhead{$\mu_\delta$} & \colhead{$\pi$} & \colhead{Baseline} \\
\colhead{Source} & \colhead{(deg)} & \colhead{(deg)} & \colhead{(mas)} & \colhead{(mas yr$^{-1}$)} & \colhead{(mas yr$^{-1}$)} & \colhead{(mas)} & \colhead{(yr)} 
}
\startdata
\multicolumn{8}{c}{WISE J104915.57$-$531906.1 (Luhman 16AB); L7.5+T0.5 \citep{burgasser:2013:129}; $W2=7.284\pm0.019$}\\ \hline
\textit{WISE}				& $162.31553$ 	& $-53.31851$ 		& $\sim$8 	& $-2761\pm3$		& $358\pm3$		& $500\pm6$			& $\sim$7\\
\citet{luhman:2013:l1}		& ...			 	& ...			 	& 60--400 		& $-2759\pm6$		& $354\pm6$		& $496\pm37$			& $\sim$33\\ 
\citet{bedin:2017:1140}		& ...			 	& ...			 	& $\sim$0.3 	& $-2762.2\pm2.3$	& $354.5\pm2.8$	& $501.118\pm0.093$	& $\sim$3\\ \hline 
\multicolumn{8}{c}{2MASS J10481463$-$3956062 (DENIS J10480$-$3956); M9 \citep{faherty:2009:1}; $W2=7.798\pm0.020$}\\ \hline 
\textit{WISE}				& $162.05621$		& $-39.93808$ 		& $\sim$8 	& $-1179\pm3$			& $-986\pm3$			& $241\pm7$		& $\sim$7\\ 
\citet{lurie:2014:91}			& ... 				& ... 				& 2.1--3.5 		& $-1165\pm1$			& $-995\pm1$			& $248.08\pm0.61$	& $\sim$12\\ 
\citet{weinberger:2016:24}		& ... 				& ... 				& $\sim$0.43 	& $-1159.36\pm0.24$	& $-986.08\pm0.31$		& $246.36\pm0.60$	& $\sim$7.9\\ 
\textit{Gaia} DR2			& ...			 	& ...			 	& $\sim$0.01 	& $-1179.18\pm0.15$	& $-988.10\pm0.18$		& $247.22	\pm0.90$	& $\sim$1.75\\ \hline 
\multicolumn{8}{c}{2MASS J00113182$+$5908400; M6.5 \citep{lepine:2009:4109}; $W2=8.651\pm0.019$}\\ \hline 
\textit{WISE}				& $2.87761$ 		& $59.14114$ 		& $\sim$8 	& $-905\pm3$				& $-1166\pm3$				& $115\pm7$		& $\sim$7\\ 
\citet{lepine:2009:4109}		& ... 				& ... 				& $\sim$6 	& $-901$\tablenotemark{a}	& $-1167$\tablenotemark{a}	& $108.3\pm1.4$	& $\sim$3 \\ 
\citet{dittmann:2014:156}		& ... 				& ... 				& $\sim$5 	& $-908.4$\tablenotemark{a}	& $-1162.0$\tablenotemark{a}	& $113.6\pm3.7$	& $\sim$4 \\ 
\textit{Gaia} DR2			& ...			 	& ...			 	& $\sim$0.01 	& $-905.70\pm0.09$			& $-1166.81\pm0.08$		& $107.42	\pm0.56$	& $\sim$1.75\\ \hline
\multicolumn{8}{c}{2MASS J02461477$-$0459182 (LHS 0017); M6 \citep{reid:1995:1838}; $W2=9.699\pm0.020$}\\ \hline 
\textit{WISE}				& $41.56683$		& $-4.99427$ 		& $\sim$10 	& $1687\pm3$			& $-1875\pm3$			& $54\pm7$			& $\sim$7\\ 
\citet{weinberger:2016:24}		& ... 				& ... 				& $\sim$0.43 	& $1676.06\pm0.27$		& $-1856.16\pm0.16$	& $60.10\pm0.91$		& $\sim$6.2\\ 
\textit{Gaia} DR2			& ...			 	& ...			 	& $\sim$0.01 	& $1691.12\pm0.15$		& $-1881.06\pm0.14$	& $59.68\pm0.48$		& $\sim$1.75\\ \hline 
\multicolumn{8}{c}{2MASS J23062928$-$0502285 (TRAPPIST-1); M8 \citep{cruz:2003:2421}; $W2=9.807\pm0.020$}\\ \hline
\textit{WISE}				& $346.62505$ 	& $-5.04274$		& $\sim$10 	& $924\pm4$		& $-467\pm3$		& $81\pm8$			& $\sim$7\\ 
\citet{costa:2006:1234}		& ... 				& ... 				& 3--5 		& $922.1\pm1.8$	& $-471.9\pm1.8$	& $82.58\pm2.58$		& $\sim$3.3\\ 
\textit{Gaia} DR2			& ...			 	& ...			 	& $\sim$0.01 	& $930.88\pm0.25$	& $-479.40\pm0.17$	& $80.45\pm0.73$		& $\sim$1.75\\ \hline 
\multicolumn{8}{c}{2MASS J02550357$-$4700509; L8 \citep{kirkpatrick:2008:1295}; $W2=10.207\pm0.021$}\\ \hline 
\textit{WISE}				& $43.76972$ 		& $-47.01582$ 		& $\sim$9 	& $997\pm3$		& $-539\pm3$		& $206\pm7$			& $\sim$7\\ 
\citet{costa:2006:1234}		& ... 			 	& ... 				& $\sim$13	& $999.6\pm2.7$	& $-565.6\pm3.7$	& $201.37\pm3.89$		& $\sim$2--4\\ 
\citet{weinberger:2016:24}		& ... 			 	& ... 				& $\sim$0.43	& $999.09\pm0.45$	& $-547.61\pm0.13$	& $205.83\pm0.53$		& $\sim$5.1\\ 
\textit{Gaia} DR2			& ...			 	& ...			 	& $\sim$0.01 	& $1011.24\pm0.39$	& $-554.77\pm0.47$	& $205.32\pm0.54$		& $\sim$1.75\\ \hline 
\multicolumn{8}{c}{2MASS J08354193$-$0819227; L5 \citep{faherty:2009:1}; $W2=10.407\pm0.022$}\\ \hline 
\textit{WISE}				& $128.92562$ 	& $-8.32227$ 		& $\sim$10 	& $-527\pm4$		& $306\pm3$		& $150\pm8$			& $\sim$7\\ 
\citet{andrei:2011:54}		& ... 			 	& ... 				& $\sim$6		& $-519.8\pm7.7$	& $285.4\pm10.5$	& $117.3\pm11.2$		& $\sim$1.96\\ 
\citet{weinberger:2016:24}		& ... 			 	& ... 				& $\sim$0.43	& $-527.88\pm0.11$	& $298.20\pm0.14$	& $137.49\pm0.39$		& $\sim$6.2\\ 
\textit{Gaia} DR2			& ...			 	& ...			 	& $\sim$0.01 	& $-535.66\pm0.44$	& $302.74\pm0.41$	& $138.60\pm1.23$		& $\sim$1.75\\ \hline
\multicolumn{8}{c}{2MASS J08173001$-$6155158; T6 \citep{artigau:2010:l38}; $W2=11.265\pm020$}\\ \hline
\textit{WISE}				& $124.37394$		& $-61.91781$ 		& $\sim$11 		& $-159\pm3$	& $1107\pm4$		& $208\pm7$			& $\sim$7\\ 
\citet{artigau:2010:l38}		& ... 				& ... 				& $\sim$20--300 	& $-158\pm54$	& $1095\pm41$	& $203\pm13$			& $\sim$14 \\ 
\textit{Gaia} DR2			& ...			 	& ...			 	& $\sim$0.01 		& $-156.44\pm1.26$	& $1099.60\pm1.21$	& $191.53\pm3.81$	& $\sim$1.75\\ \hline
\multicolumn{8}{c}{WISEP J150649.97$+$702736.0; T6 \citep{mace:2013:6}; $W2=11.276\pm0.019$)}\\ \hline 
\textit{WISE}				& $226.70825$		& $70.46003$ 		& $\sim$12 		& $-1187\pm3$		& $1044\pm4$		& $191\pm8$			& $\sim$7\\ 
\citet{marsh:2013:119}		& ... 				& ... 				& $\sim$60--300	& $-1241\pm85$	& $1046\pm64$	& $310\pm42$\tablenotemark{b}	& $\sim$2.5 \\ 
\textit{Gaia} DR2			& ...			 	& ...			 	& $\sim$0.01 		& $-1194.09\pm1.65$& $1044.30\pm1.52$& $193.54\pm4.68$		& $\sim$1.75\\ \hline
\multicolumn{8}{c}{2MASS J04455387$-$3048204; L2 \citep{cruz:2003:2421}; $W2=11.340\pm0.022$}\\ \hline 
\textit{WISE}				& $71.47506$ 		& $-30.80693$ 		& $\sim$12 	& $166\pm4$		& $-421\pm4$		& $70\pm9$			& $\sim$7\\ 
\citet{faherty:2012:56}		& ...				& ...				& $\sim$3 	& $164.0\pm2.8$	& $-415.0\pm2.7$	& $78.5\pm4.9$		& $\sim$3 \\ 
\textit{Gaia} DR2			& ...			 	& ...			 	& $\sim$0.01 	& $161.48\pm0.23$	& $-419.68\pm0.39$	& $61.97\pm0.85$		& $\sim$1.75\\ \hline
\multicolumn{8}{c}{2MASS J09393548$-$2448279; T8 \citep{burgasser:2006:1067}; $W2=11.640\pm0.022$}\\ \hline 
\textit{WISE}				& $144.89964$	& $-24.81069$ 			& $\sim$29 	& $552\pm8$		& $-1026\pm9$		& $195\pm19$			& $\sim$7\\ 
\citet{faherty:2012:56}		& ... 				& ... 				& $\sim$3		& $558.1\pm5.8$	& $-1030.5\pm5.6$	& $196.0\pm10.4$		& $\sim$2.5 \\ \hline 
\multicolumn{8}{c}{2MASS J04390101$-$2353083; L6.5 \citep{cruz:2003:2421}; $W2=11.687\pm0.022$}\\ \hline
\textit{WISE}				& $69.75378$ 		& $-23.88610$ 		& $\sim$14 	& $-120\pm4$		& $-157\pm4$		& $97\pm10$			& $\sim$7\\ 
\citet{faherty:2012:56}		& ...				& ...				& $\sim$3 	& $-116.3\pm3.8$	& $-162.0\pm3.8$	& $110.4\pm4.0$		& $\sim$3 \\ 
\textit{Gaia} DR2			& ...			 	& ...			 	& $\sim$0.01 	& $-113.61\pm0.69$	& $-153.10\pm0.81$	& $80.79\pm2.34$		& $\sim$1.75\\ \hline
\multicolumn{8}{c}{2MASS J23224684$-$3133231; L0$\beta$ \citep{reid:2008:1290, faherty:2012:56}; $W2=11.723\pm0.022$}\\ \hline 
\textit{WISE}				& $350.69440$ 	& $-31.55802$ 		& $\sim$15 	& $-184\pm5$		& $-560\pm4$		& $73\pm11$			& $\sim$7\\ 
\citet{faherty:2012:56}		& ... 				& ... 				& $\sim$3		& $-194.8\pm7.4$	& $-527.3\pm7.5$	& $58.6\pm5.6$		& $\sim$1.5 \\ 
\textit{Gaia} DR2			& ...			 	& ...			 	& $\sim$0.01 	& $-203.20\pm0.53$	& $-540.48\pm0.55$	& $50.32\pm1.37$		& $\sim$1.75\\ \hline
\multicolumn{8}{c}{UGPS J072227.51$-$054031.2; T9 \citep{cushing:2011:50}; $W2=12.200\pm0.023$}\\ \hline  
\textit{WISE}				& $110.61366$		& $-5.67499$ 		& $\sim$40 	& $-916\pm12$		& $353\pm13$		& $251\pm26$			& $\sim$7\\ 
\citet{leggett:2012:74}		& ... 				& ... 				& $\sim$6.2	& $-904.14\pm1.71$	& $352.025\pm1.21$	& $242.8\pm2.4$		& $\sim$5 \\ \hline 
\multicolumn{8}{c}{WISEA J025409.55$+$022358.5; T8 \citep{kirkpatrick:2011:19}; $W2=12.758\pm0.026$}\\ \hline 
\textit{WISE}				& $43.53936$		& $2.39960$ 		& $\sim$66 	& $2584\pm16$	& $214\pm18$		& $139\pm40$			& $\sim$7\\ 
\citet{dupuy:2013:1492}		& ... 				& ... 				& $\sim$30	& $2588\pm27$	& $273\pm27$		& $135\pm15$			& $\sim$2 \\ \hline 
\multicolumn{8}{c}{2MASS J07290002$-$3954043; T8 \citep{looper:2007:1162}; $W2=12.972\pm0.024$}\\ \hline 
\textit{WISE}				& $112.24790$	 	& $-39.89604$ 		& $\sim$48 	& $-601\pm11$		& $1670\pm11$		& $105\pm24$			& $\sim$7\\ 
\citet{faherty:2012:56}		& ... 				& ... 				& $\sim$3		& $-566.6\pm5.3$	& $1643.4\pm5.5$	& $126.3\pm8.3$		& $\sim$3.7 \\ \hline 
\multicolumn{8}{c}{2MASS J22282889$-$4310262; T6 \citep{burgasser:2006:1067}; $W2=13.326\pm0.030$}\\ \hline 
\textit{WISE}				& $337.12086$ 	& $-43.17486$ 		& $\sim$73 	& $107\pm21$		& $-304\pm22$		& $109\pm46$			& $\sim$7\\ 
\citet{faherty:2012:56}		& ... 				& ... 				& $\sim$3		& $102.3\pm5.8$	& $-324.4\pm5.1$	& $94.0\pm7.0$		& $\sim$1.5 \\ \hline 
\multicolumn{8}{c}{WISE J085510.83$-$071442.5; $\geqslant$Y2 \citep{tinney:2014:39}; $W2=14.016\pm0.048$ \citep{wright:2014:82}}\\ \hline 
\textit{WISE}\tablenotemark{c}	& $133.78619$		& $-7.24432$		& $\sim$140 		& $-8055\pm56$	& $677\pm58$	& $520\pm67$			& $\sim$3.5\\ 
\citet{wright:2014:82}			& ... 				& ... 				& $\sim$40--500 	& $-8051\pm47$	& $657\pm50$	& $448\pm33$			& $\sim$4 \\ \hline 
\multicolumn{8}{c}{WISEP J041022.71$+$150248.5; Y0 \citep{mace:2013:6}; $W2=14.113\pm0.047$}\\ \hline 
\textit{WISE}				& $62.59465$		& $15.04681$		& $\sim$196 		& $968\pm55$	& $-2239\pm59$	& $209\pm113$				& $\sim$7\\ 
\citet{marsh:2013:119}		& ... 				& ... 				& $\sim$50--200 	& $974\pm79$	& $-2144\pm72$	& $233\pm56$\tablenotemark{d}& $\sim$2.5 \\ \hline 
\multicolumn{8}{c}{WISEPA J182831.08$+$265037.8; $\geqslant$Y2 \citep{kirkpatrick:2012:156}; $W2=14.353\pm0.045$}\\ \hline  
\textit{WISE}				& $277.12949$ 	& $26.84381$ 		& $\sim$159 		& $1058\pm47$	& $111\pm47$	& $131\pm88$			& $\sim$7\\ 
\citet{beichman:2013:101}		& ... 				& ... 				& $\sim$50 		& $1069\pm11$		& $153\pm13$	& $90\pm10$			& $\sim$2.5 \\
\enddata
\tablenotetext{a}{No uncertainties reported.}
\tablenotetext{b}{\citet{kirkpatrick:2012:156} quote a value of $\pi=193\pm26$ mas and cite the measurement to a pre-published version of \citet{marsh:2013:119}.}
\tablenotetext{c}{Measurements made using only NEOWISE(-R) data.}
\tablenotetext{d}{\citet{kirkpatrick:2012:156} quote a value of $\pi=164\pm24$ mas and cite the measurement to a pre-published version of \citet{marsh:2013:119}.}
\end{deluxetable*}
	 
\begin{figure*}
\centering
\includegraphics[width=\linewidth]{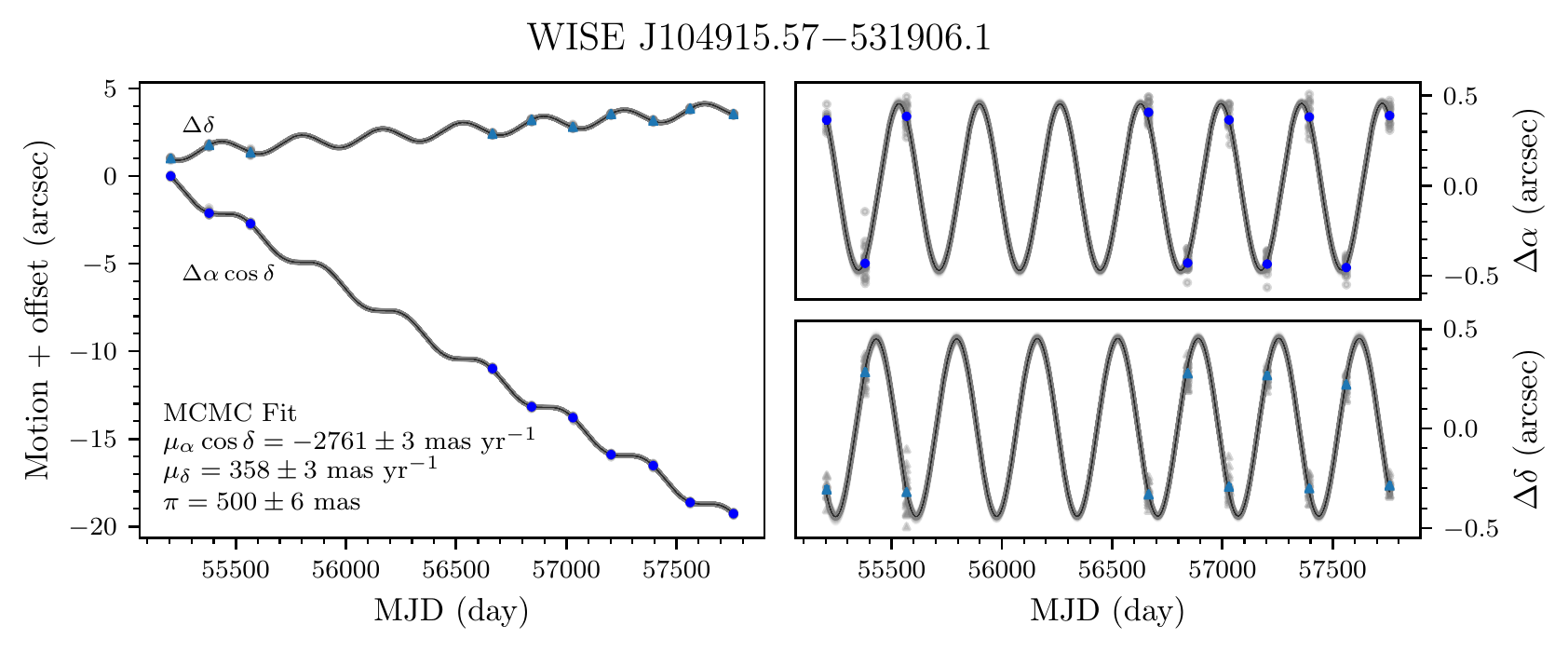}
\caption{
\textit{Left}: Astrometric solution for WISE J104915.57$-$531906.1 (solid lines). The $\alpha$ and $\delta$ solutions are offset for visibility. Individual positions for each exposure are shown as translucent gray points, with blue points and cyan triangles indicating the uncertainty weighted mean positions for each epoch in $\alpha$ and $\delta$, respectively. Errorbars are plotted, but are typically smaller than the plotted symbols. The gray bands show 300 random realizations from the MCMC posterior distributions.
\textit{Right}: Astrometric solution with the proper motions removed. The dark gray band indicates the uncertainty in the parallax.
The complete figure set (20 images) is available in the online journal.
\label{fig:new}}
\end{figure*}

\begin{figure*}
\centering
\includegraphics[width=\linewidth]{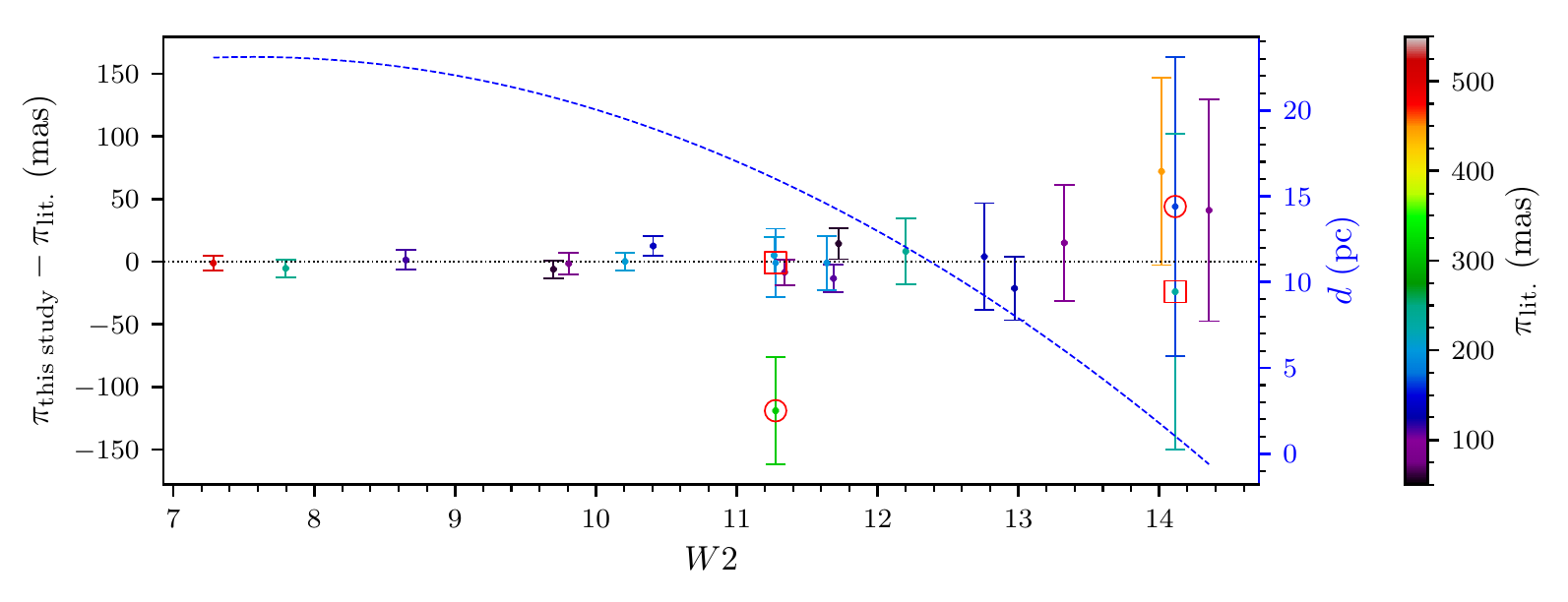}
\caption{
Residuals of parallaxes derived in this study using \textit{WISE} data to literature parallax measurements (not including measurements from \textit{Gaia} DR2). The symbol color and colorbar indicate the literature value of the parallax. The four points with red symbols around them indicate the measurements from \citet[circles;][]{marsh:2013:119}, and the measurements \citet[squares;][]{kirkpatrick:2012:156} cite to a pre-published version of \citet{marsh:2013:119}.
The blue dashed line shows the approximate distance limit (right y-axis) as a function of $W2$ magnitude for 15\% uncertainties using this method. 
\label{fig:residuals}}
\end{figure*}

	 For all of the comparisons in Table~\ref{tbl:parallax}, the computed \textit{WISE} parallax values from this study are within the 2$\sigma$ combined uncertainty of the highest precision literature value (18 of 20 within 1$\sigma$), excluding the \textit{Gaia} DR2\footnote{\textit{Gaia} DR2 was released while this manuscript was under review} measurements. The full astrometric solutions are shown in Figure Set~\ref{fig:new}. Figure~\ref{fig:residuals} shows the residuals between the parallax value derived using \textit{WISE} and the highest precision literature parallax, as a function of $W2$ magnitude. The astrometric precision severely deteriorates for sources fainter than $W2\approx14$, setting the approximate limit for where this method is valid.

\figsetstart
\figsetnum{5}
\figsettitle{\textit{WISE} Astrometric Solutions for Literature Sources} 

\figsetgrpstart
\figsetgrpnum{5.1}
\figsetgrptitle{\textit{WISE} astrometric solution for WISE J10491557$-$5319061.}
\figsetplot{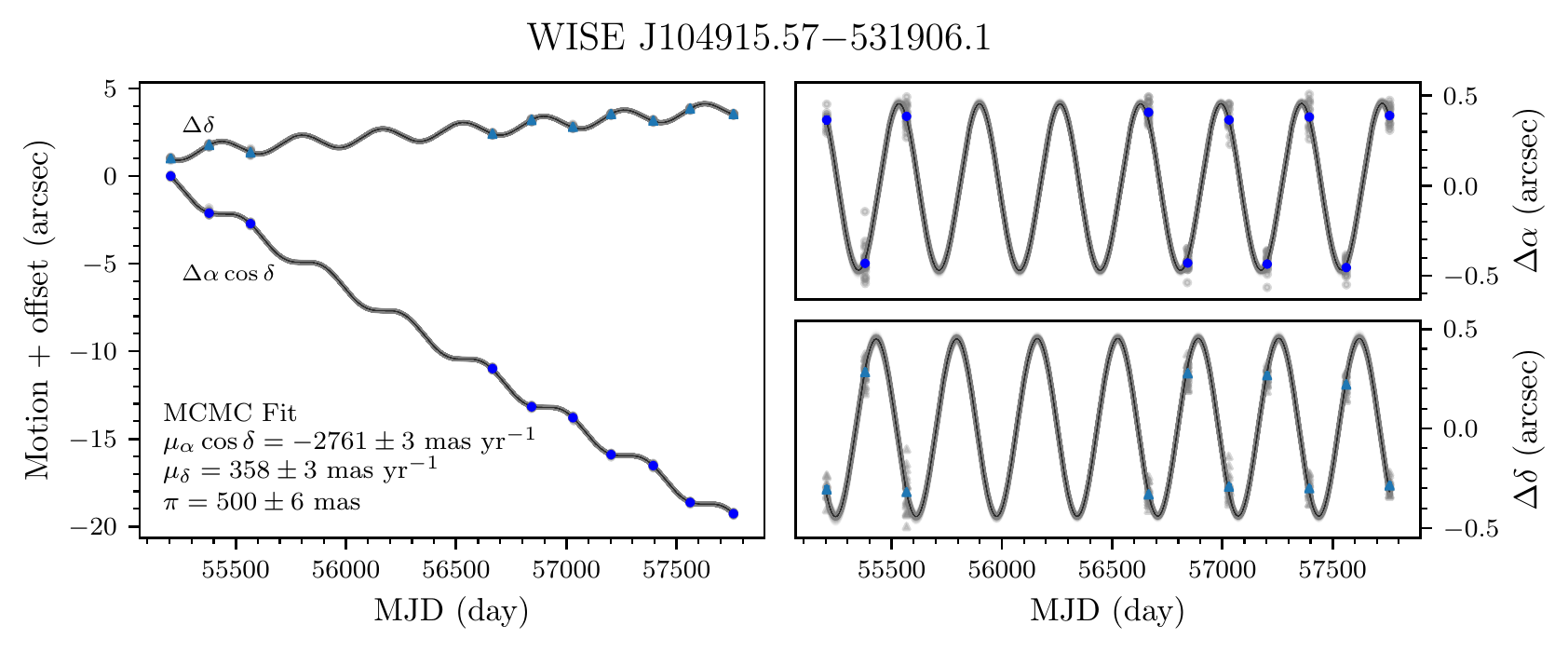}
\figsetgrpnote{}
\figsetgrpend
 
\figsetgrpstart
\figsetgrpnum{5.2}
\figsetgrptitle{\textit{WISE} astrometric solution for 2MASS J10481463$-$3956062.}
\figsetplot{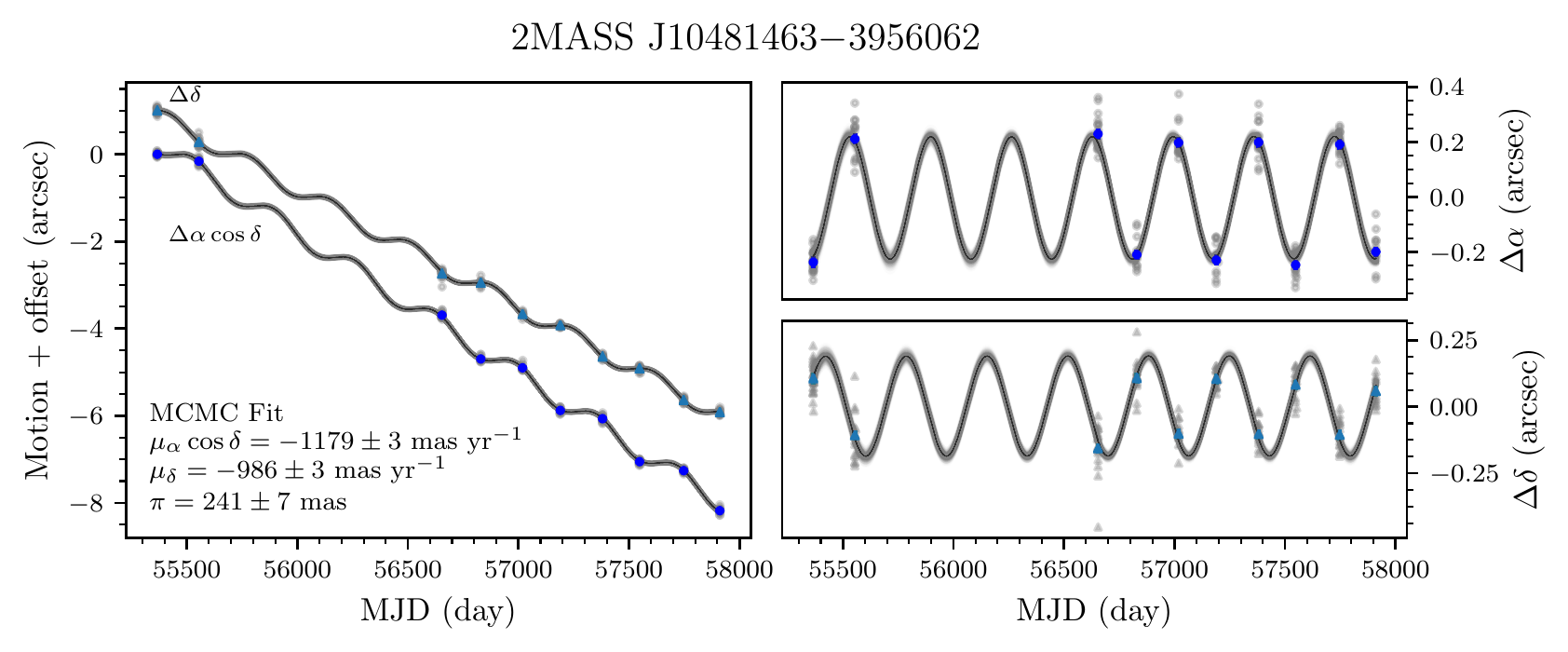}
\figsetgrpnote{}
\figsetgrpend

\figsetgrpstart
\figsetgrpnum{5.3}
\figsetgrptitle{\textit{WISE} astrometric solution for 2MASS J00113182$+$5908400.}
\figsetplot{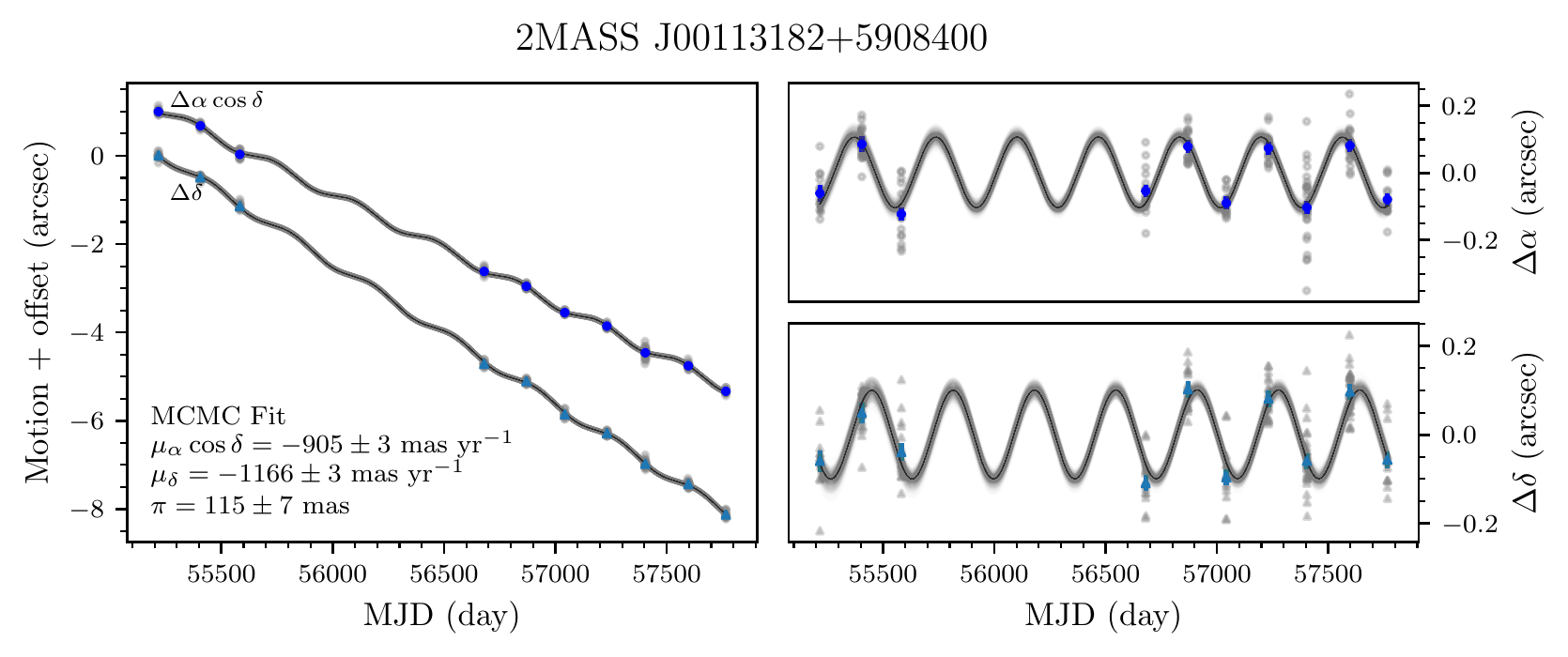}
\figsetgrpnote{}
\figsetgrpend

\figsetgrpstart
\figsetgrpnum{5.4}
\figsetgrptitle{\textit{WISE} astrometric solution for 2MASS J02461477$-$0459182.}
\figsetplot{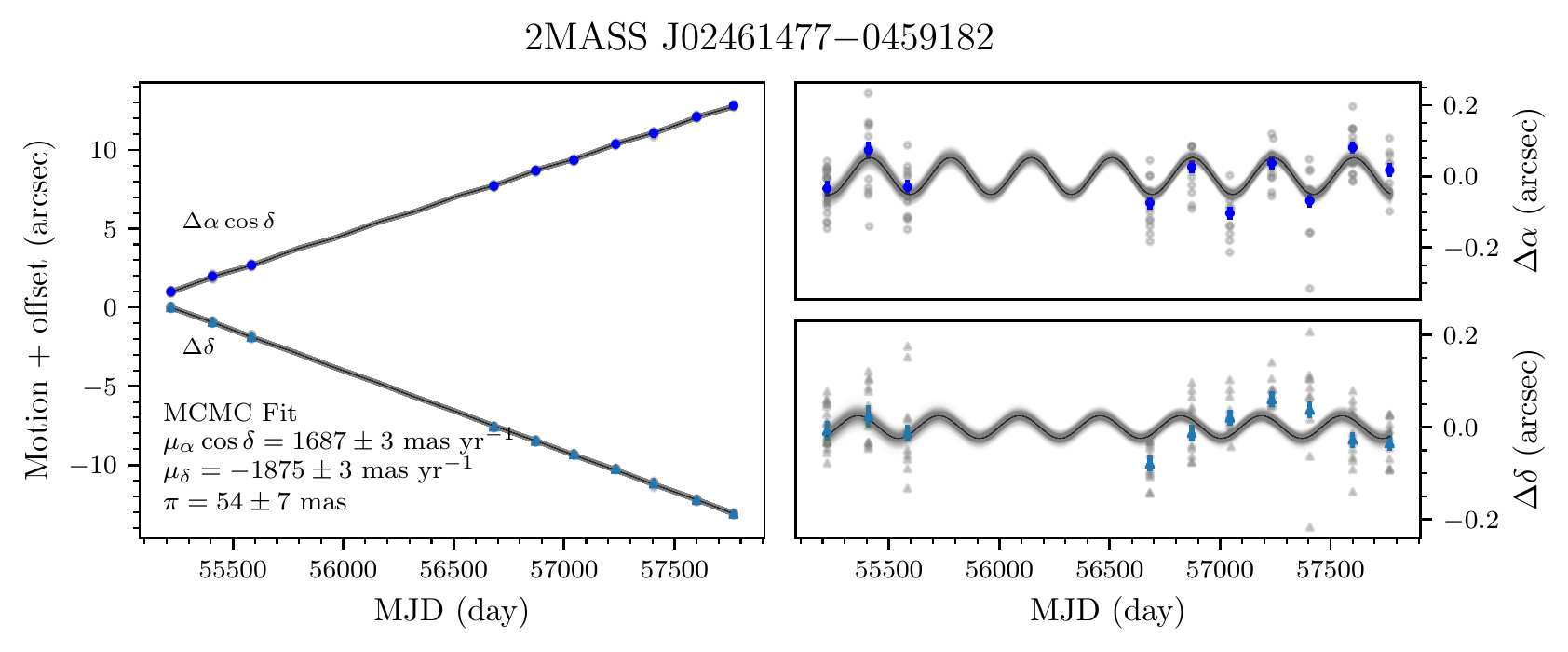}
\figsetgrpnote{}
\figsetgrpend

\figsetgrpstart
\figsetgrpnum{5.5}
\figsetgrptitle{\textit{WISE} astrometric solution for 2MASS J23062928$-$0502285.}
\figsetplot{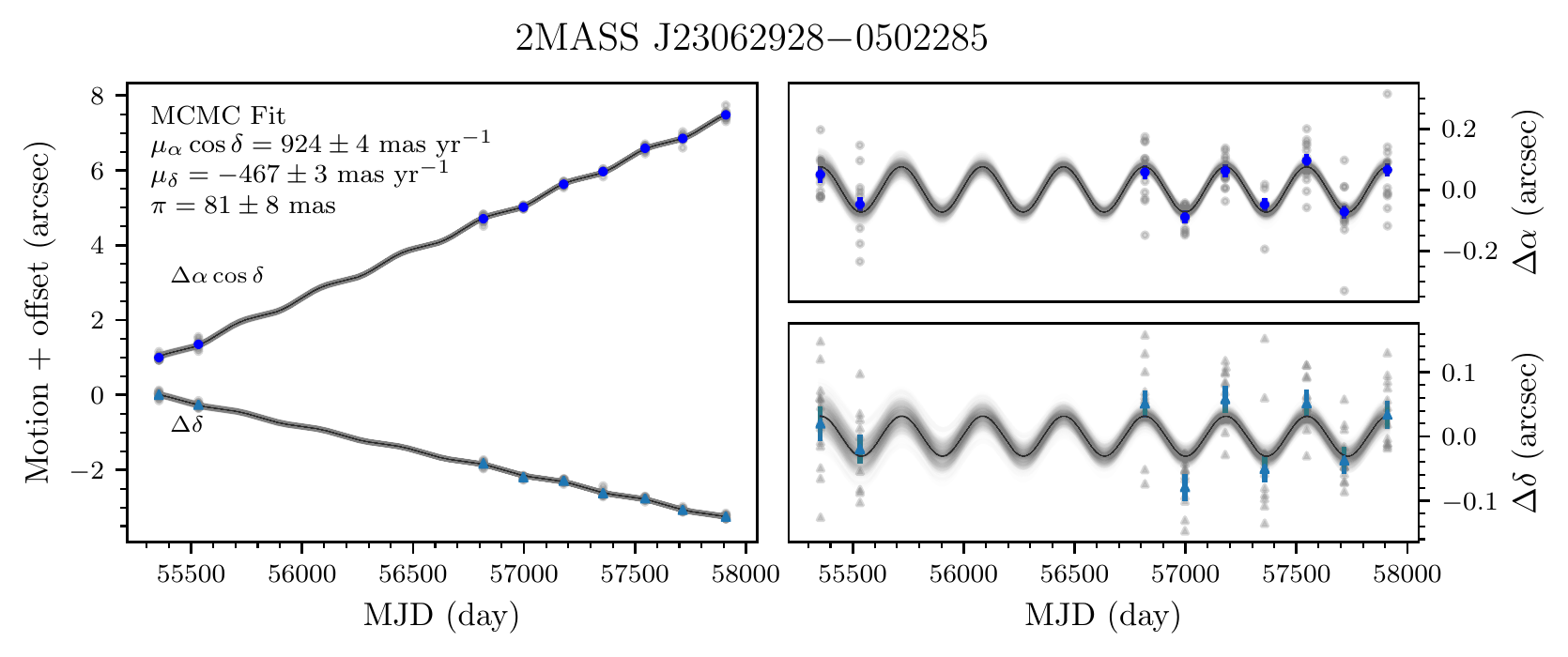}
\figsetgrpnote{}
\figsetgrpend

\figsetgrpstart
\figsetgrpnum{5.6}
\figsetgrptitle{\textit{WISE} astrometric solution for 2MASS J02550357$-$4700509.}
\figsetplot{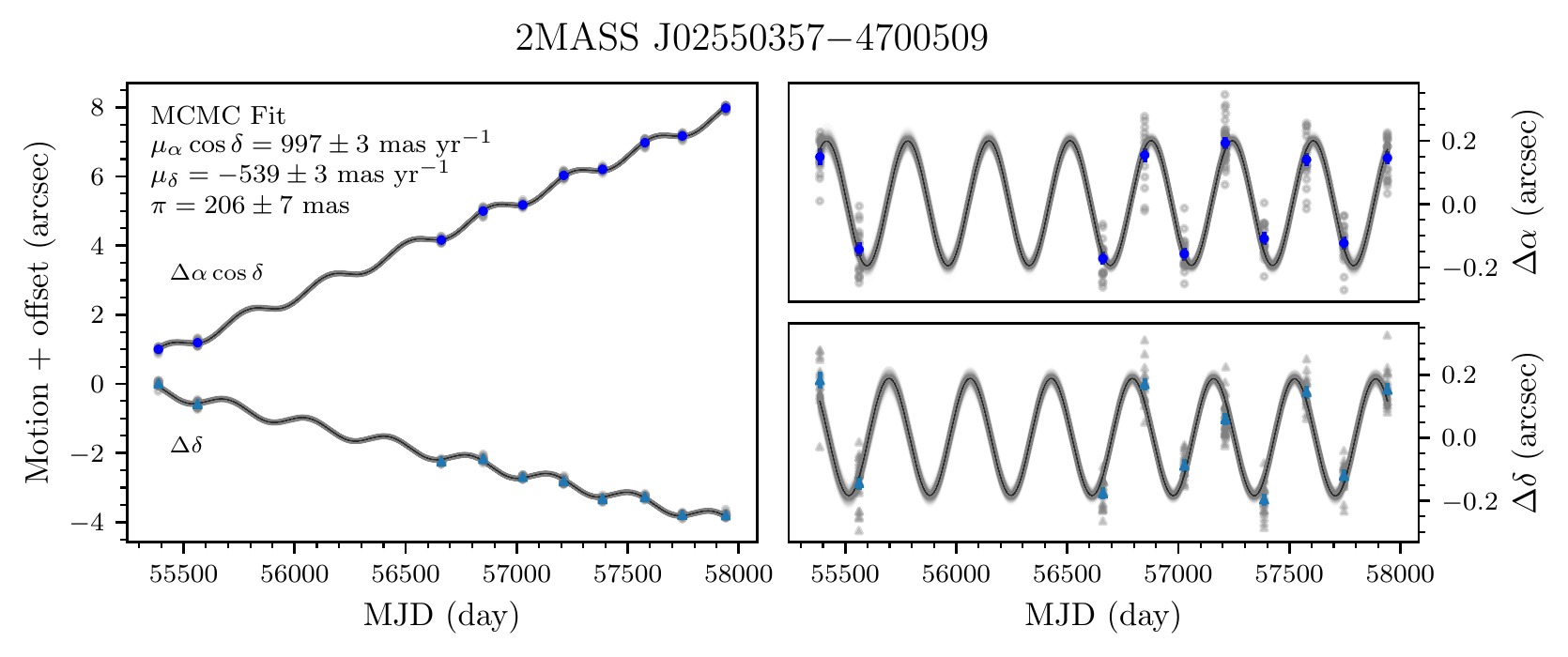}
\figsetgrpnote{}
\figsetgrpend

\figsetgrpstart
\figsetgrpnum{5.7}
\figsetgrptitle{\textit{WISE} astrometric solution for 2MASS J08354193$-$0819227.}
\figsetplot{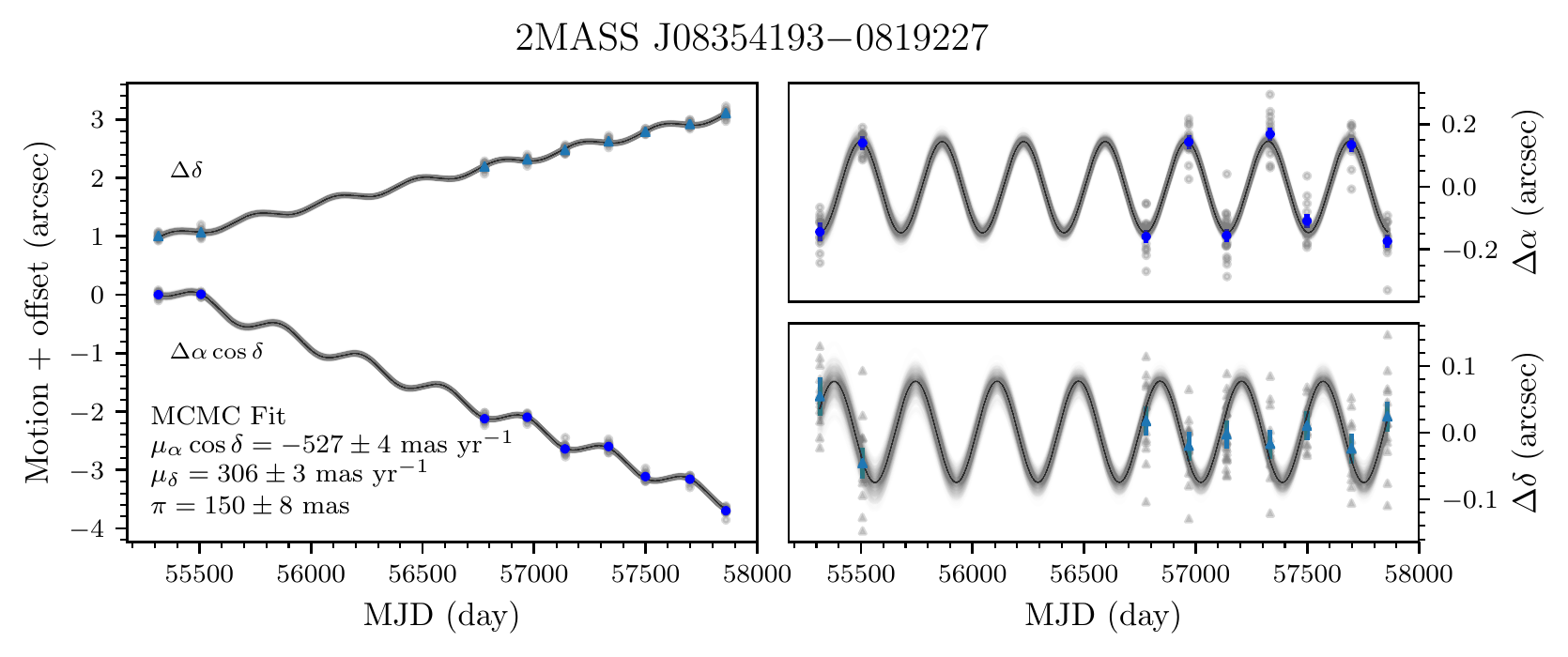}
\figsetgrpnote{}
\figsetgrpend

\figsetgrpstart
\figsetgrpnum{5.8}
\figsetgrptitle{\textit{WISE} astrometric solution for 2MASS J08173001$-$6155158.}
\figsetplot{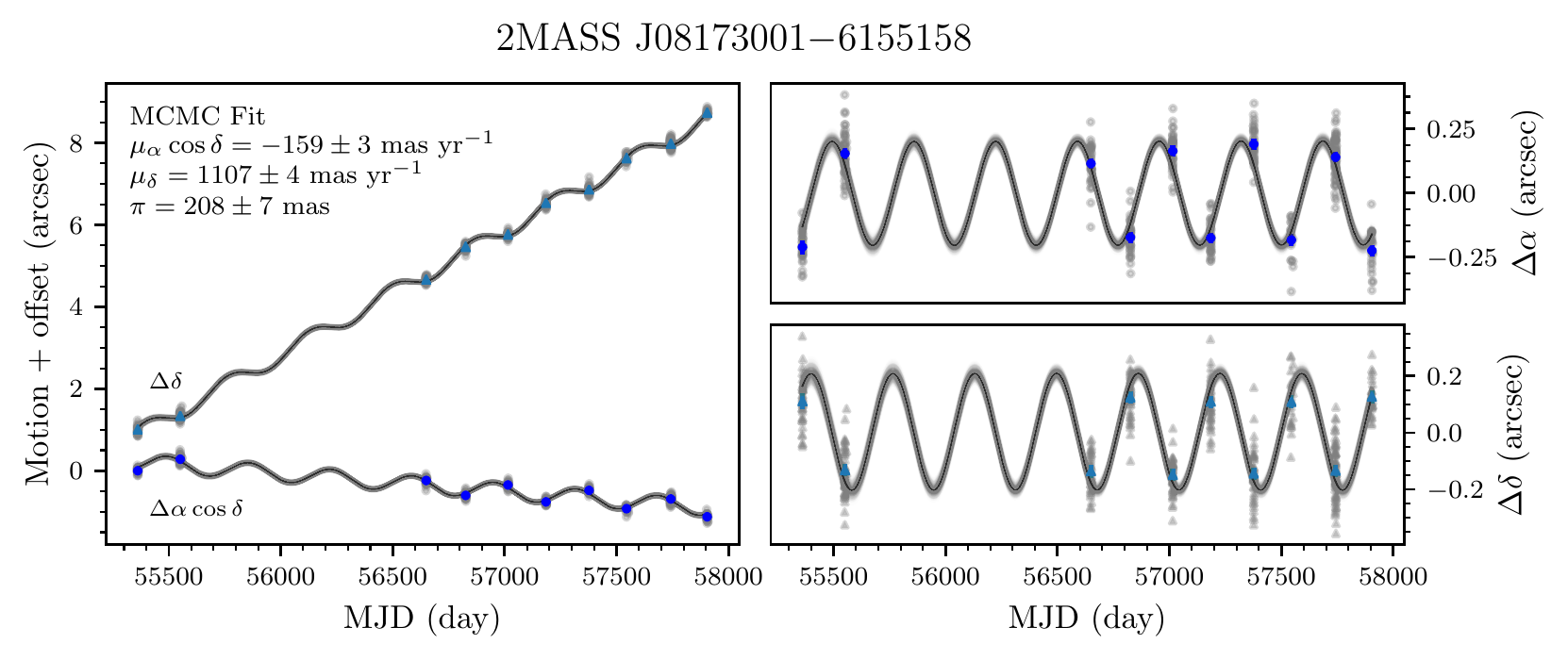}
\figsetgrpnote{}
\figsetgrpend

\figsetgrpstart
\figsetgrpnum{5.9}
\figsetgrptitle{\textit{WISE} astrometric solution for WISEP J150649.97$+$702736.0.}
\figsetplot{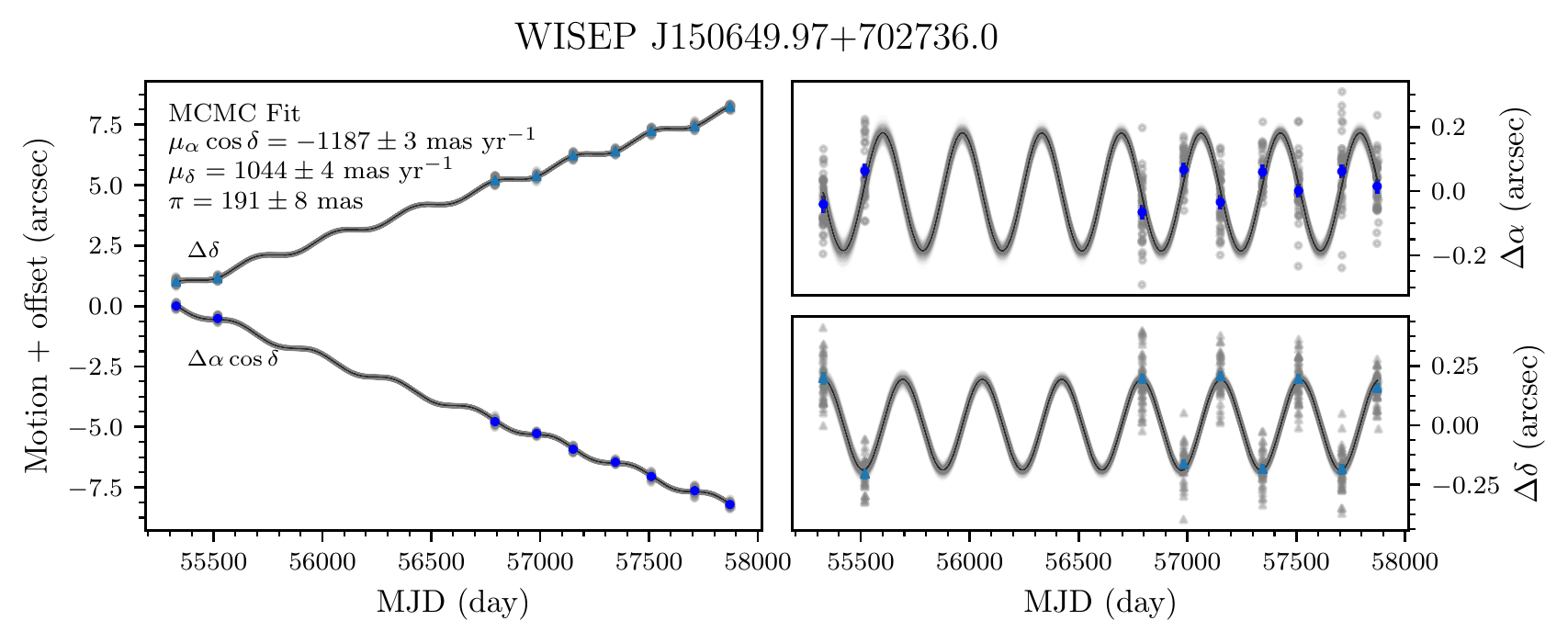}
\figsetgrpnote{}
\figsetgrpend

\figsetgrpstart
\figsetgrpnum{5.10}
\figsetgrptitle{\textit{WISE} astrometric solution for 2MASS J04455387$-$3048204.}
\figsetplot{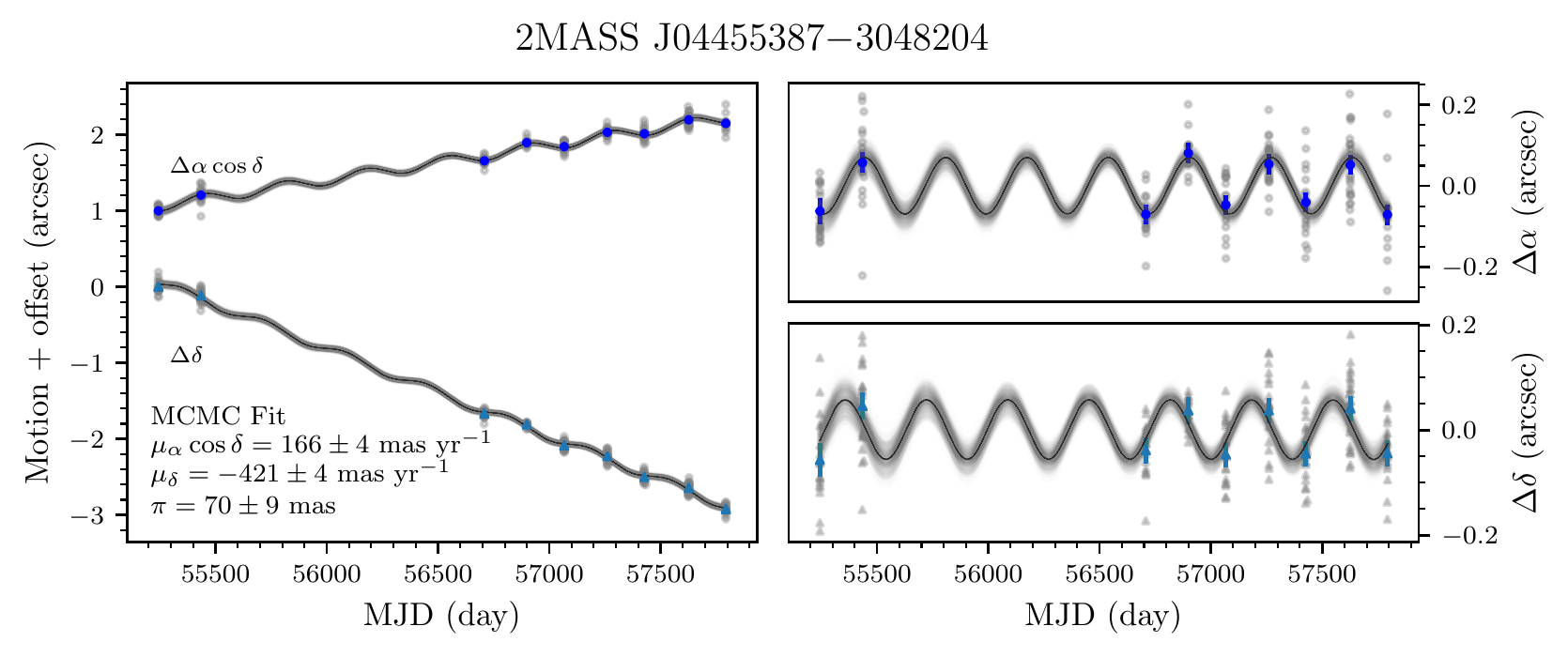}
\figsetgrpnote{}
\figsetgrpend

\figsetgrpstart
\figsetgrpnum{5.11}
\figsetgrptitle{\textit{WISE} astrometric solution for 2MASS J09393548$-$2448279.}
\figsetplot{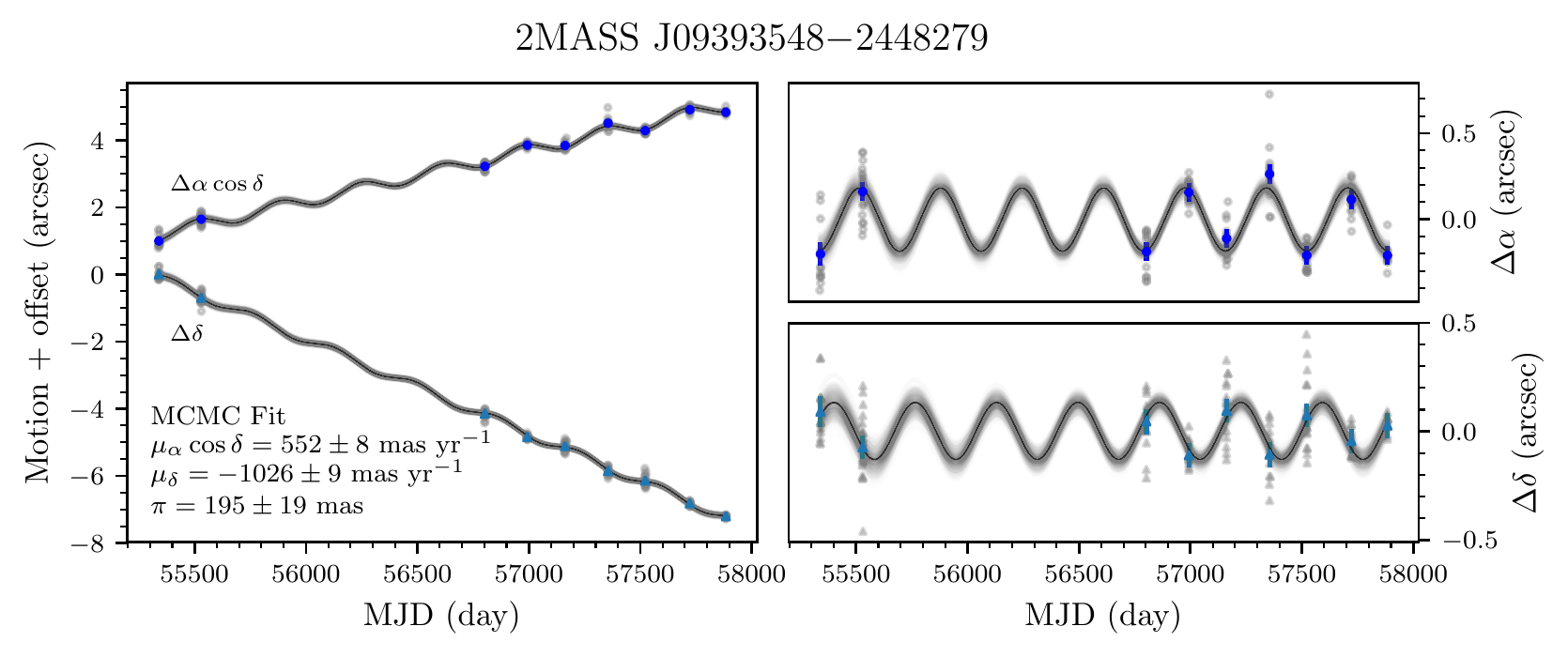}
\figsetgrpnote{}
\figsetgrpend

\figsetgrpstart
\figsetgrpnum{5.12}
\figsetgrptitle{\textit{WISE} astrometric solution for 2MASS J04390101$-$2353083.}
\figsetplot{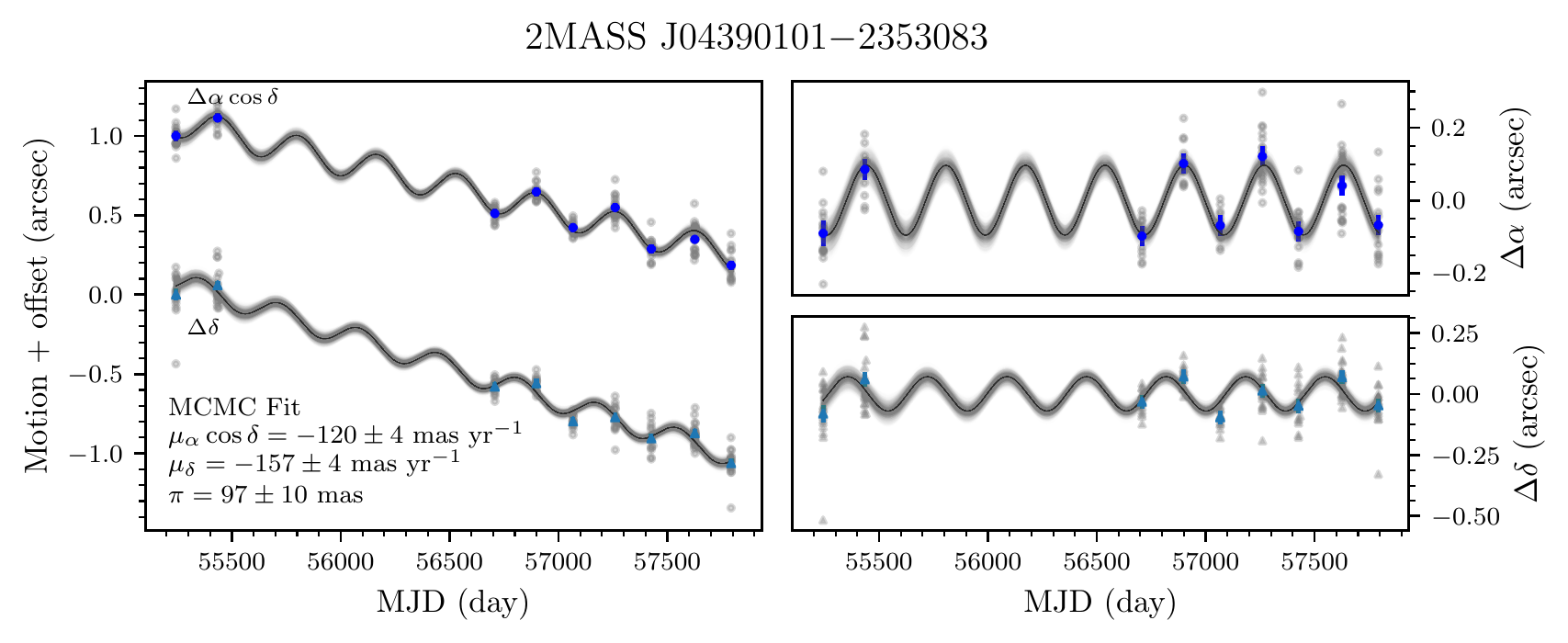}
\figsetgrpnote{}
\figsetgrpend

\figsetgrpstart
\figsetgrpnum{5.13}
\figsetgrptitle{\textit{WISE} astrometric solution for 2MASS J23224684$-$3133231.}
\figsetplot{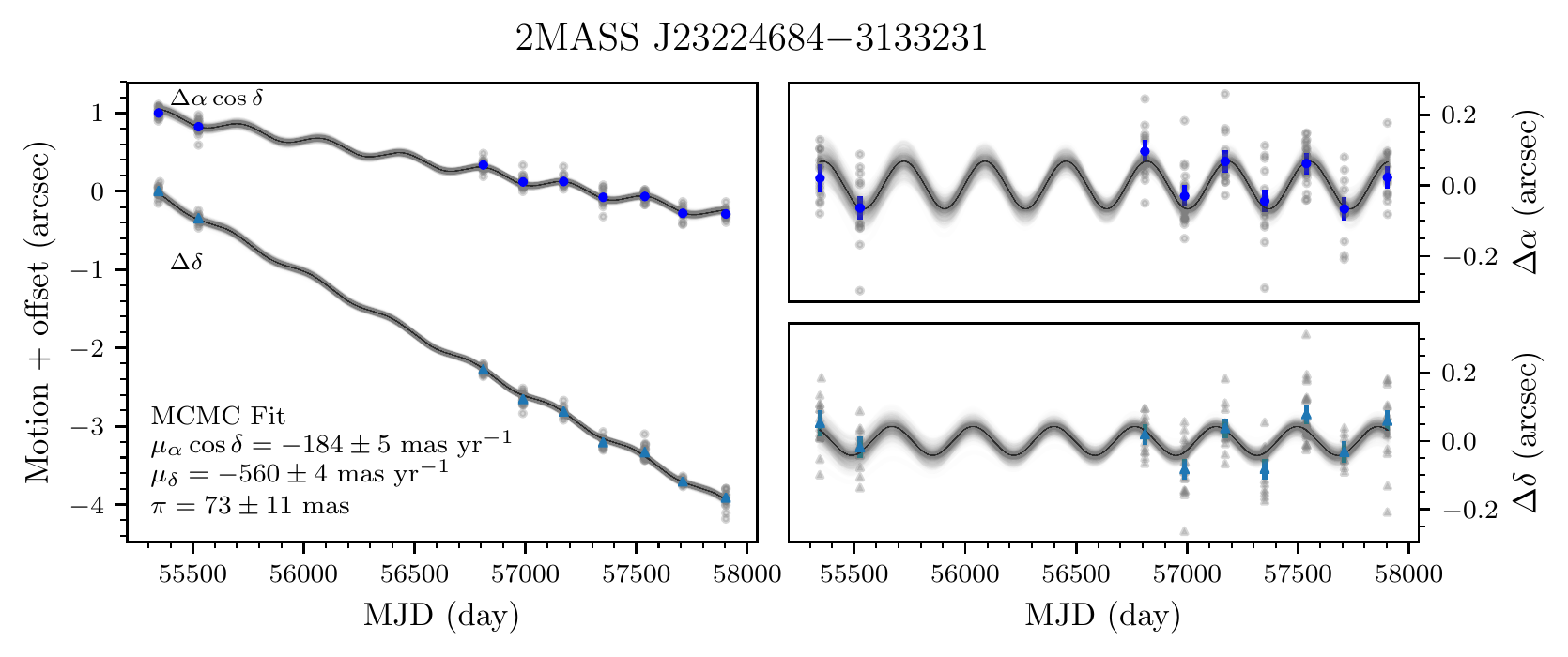}
\figsetgrpnote{}
\figsetgrpend

\figsetgrpstart
\figsetgrpnum{5.14}
\figsetgrptitle{\textit{WISE} astrometric solution for UGPS J072227.51$-$054031.2.}
\figsetplot{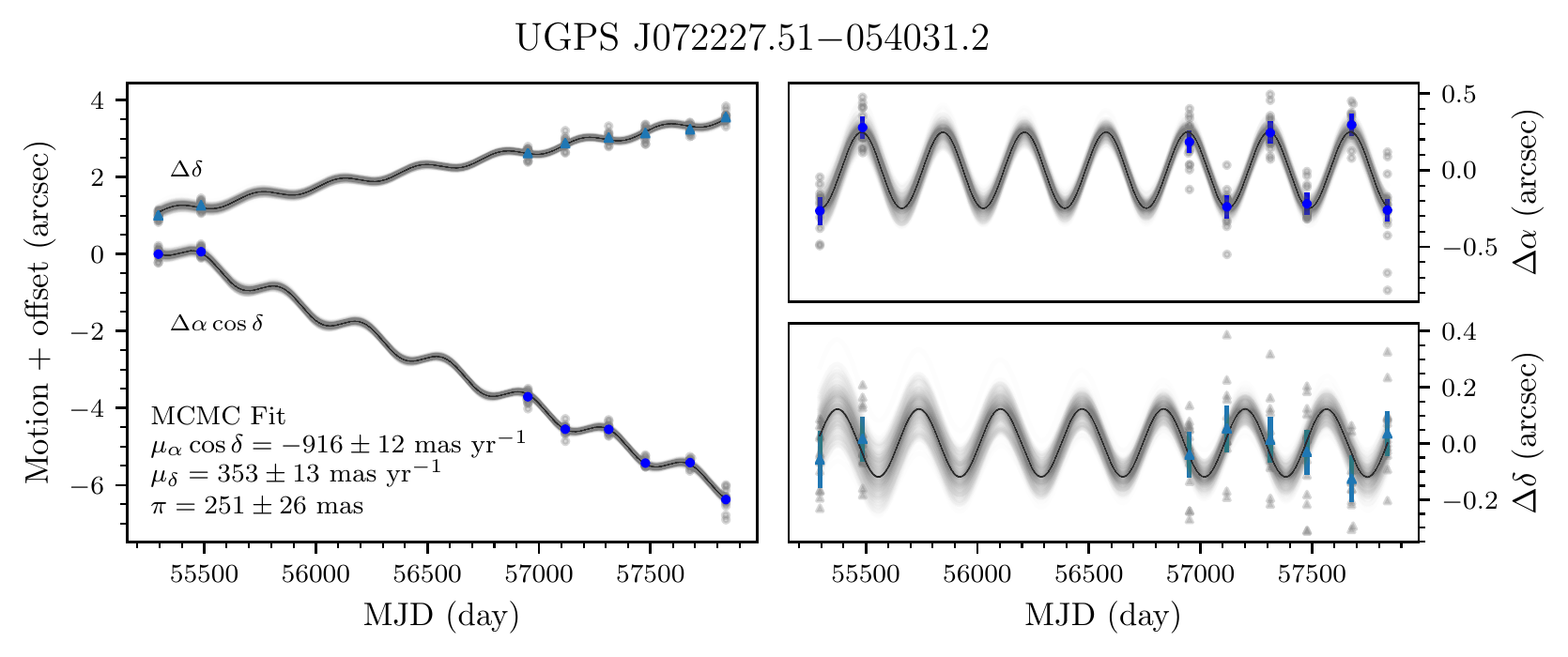}
\figsetgrpnote{}
\figsetgrpend

\figsetgrpstart
\figsetgrpnum{5.15}
\figsetgrptitle{\textit{WISE} astrometric solution for WISEA J025409.55$+$022358.5.}
\figsetplot{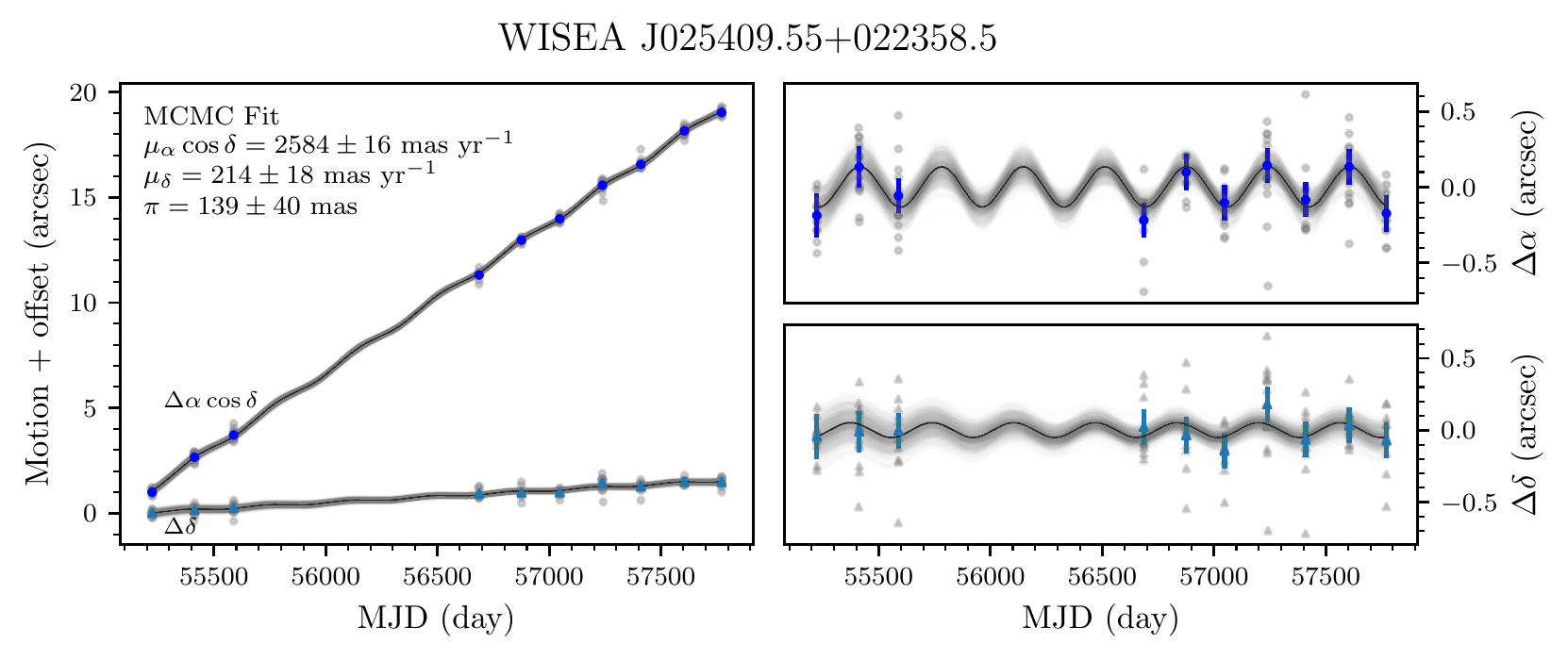}
\figsetgrpnote{}
\figsetgrpend

\figsetgrpstart
\figsetgrpnum{5.16}
\figsetgrptitle{\textit{WISE} astrometric solution for 2MASS J07290002$-$3954043.}
\figsetplot{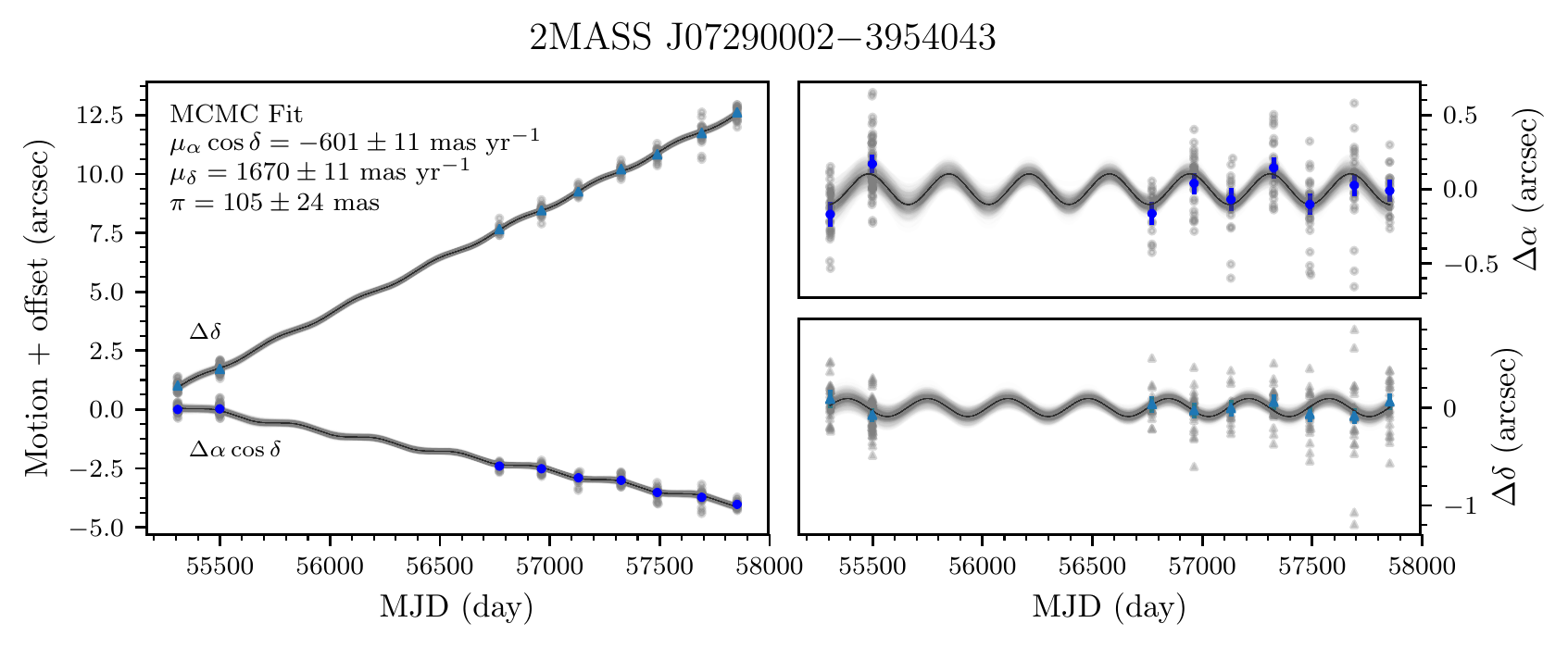}
\figsetgrpnote{}
\figsetgrpend

\figsetgrpstart
\figsetgrpnum{5.17}
\figsetgrptitle{\textit{WISE} astrometric solution for 2MASS J22282889$-$4310262.}
\figsetplot{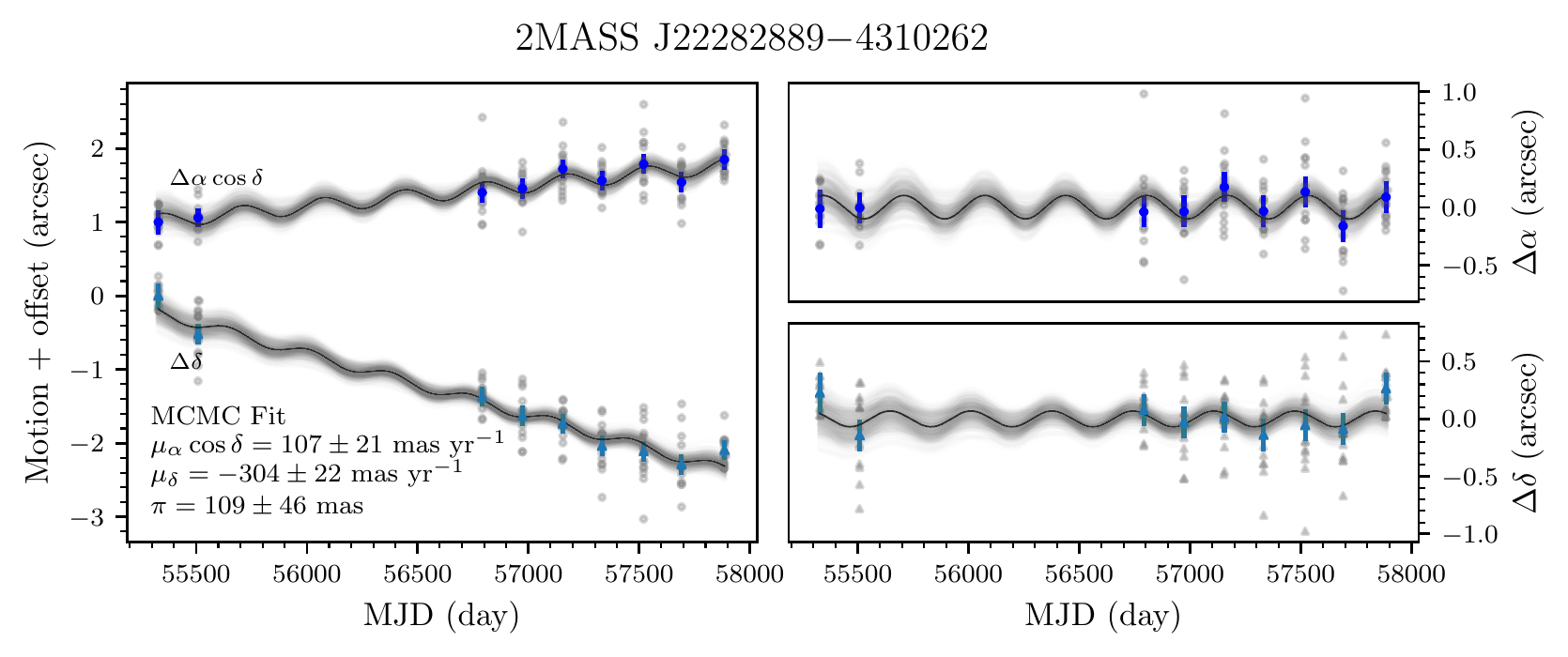}
\figsetgrpnote{}
\figsetgrpend

\figsetgrpstart
\figsetgrpnum{5.18}
\figsetgrptitle{\textit{WISE} astrometric solution for WISE J085510.83$-$071442.5.}
\figsetplot{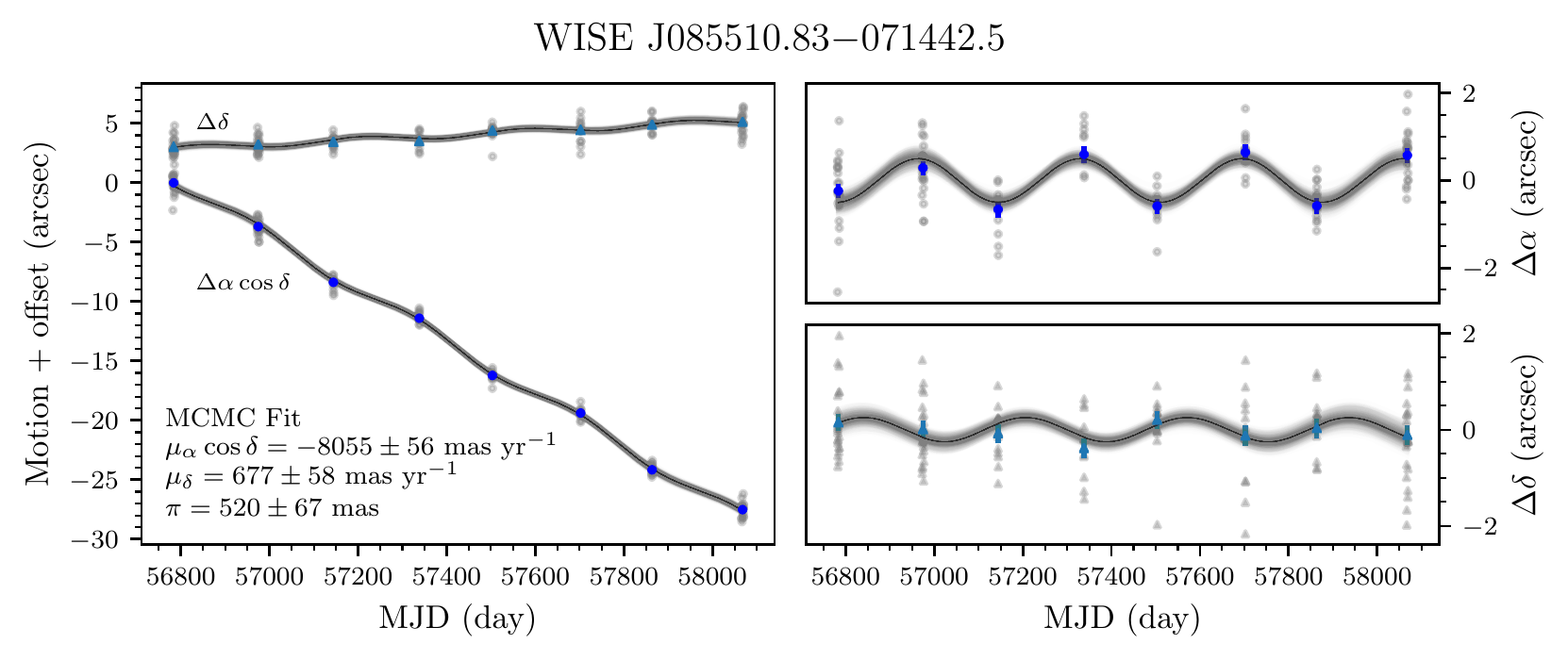}
\figsetgrpnote{}
\figsetgrpend

\figsetgrpstart
\figsetgrpnum{5.19}
\figsetgrptitle{\textit{WISE} astrometric solution for WISEP J041022.71$+$150248.5.}
\figsetplot{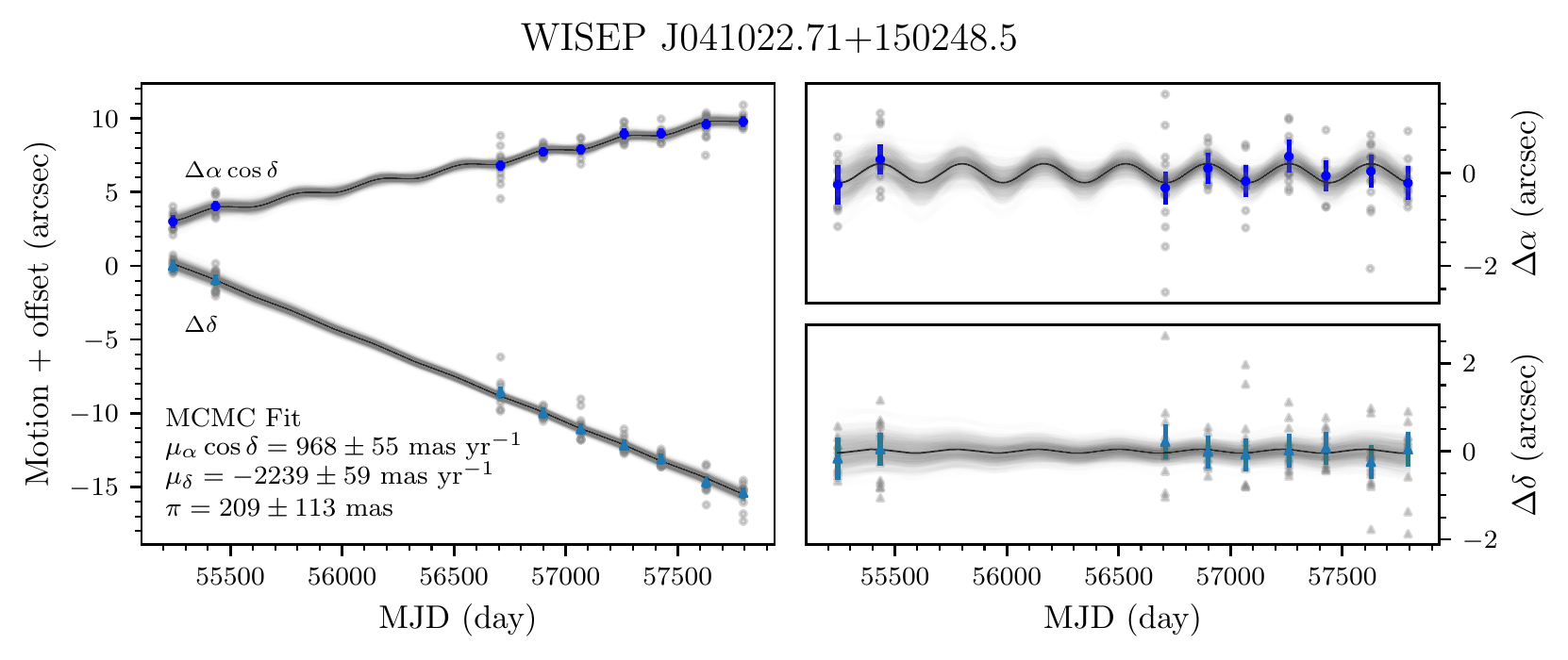}
\figsetgrpnote{}
\figsetgrpend

\figsetgrpstart
\figsetgrpnum{5.20}
\figsetgrptitle{\textit{WISE} astrometric solution for WISEPA J182831.08$+$265037.8.}
\figsetplot{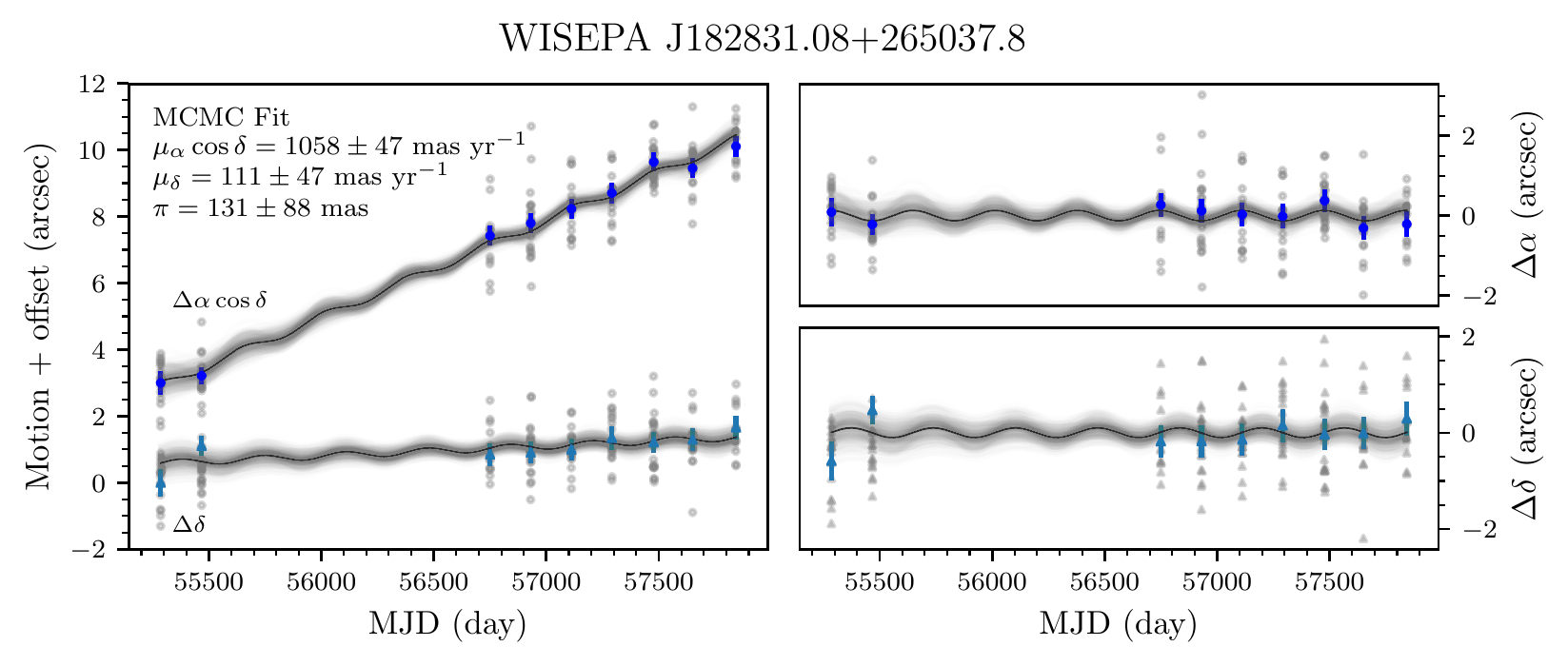}
\figsetgrpnote{}
\figsetgrpend
 
\figsetend

\begin{figure}
\centering
\includegraphics[width=\linewidth]{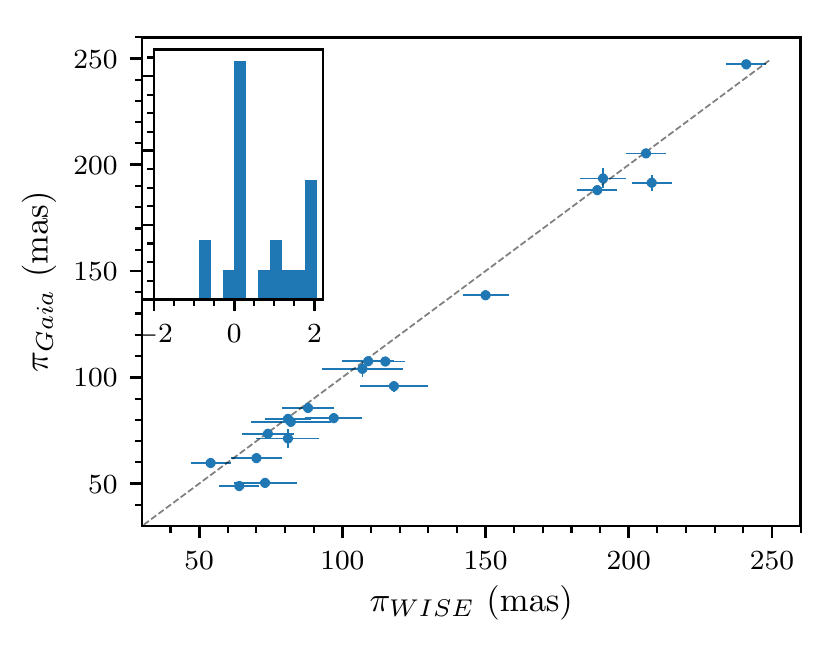}
\caption{
Comparison between parallaxes derived in this study using \textit{WISE} data to \textit{Gaia} DR2 parallaxes. 
Errorbars show the 1$\sigma$ uncertainties on each measurement. 
The inset plot shows the distribution of the quantity $(\pi_{WISE} - \pi_{Gaia}) / \sqrt{\sigma_{\pi_{WISE}}^2 + \sigma_{\pi_{Gaia}}^2}$, which has $\mu = 0.65$ and $\sigma = 0.90$. 
\label{fig:GaiaComparison}}
\end{figure}

	With the recent release of \textit{Gaia} DR2, it is essential to compare results obtained here with results from DR2. Of the 20 objects in Table~\ref{tbl:parallax}, 11 have cross-matches within \textit{Gaia} DR2, and measurements are listed in Table~\ref{tbl:parallax}. Many of these sources have poor \textit{Gaia} goodness-of-fit statistics, and large excess astrometric noise parameters \citep{lindegren:2012:a78,lindegren:2018:}. To account for possibly underestimated uncertainties on the measured parallax values, the \textit{Gaia} uncertainties reported in this study are the quadrature sum of the quoted DR2 parallax uncertainty and the excess astrometric noise parameter. Figure~\ref{fig:GaiaComparison} shows the comparison between values in this study and \textit{Gaia} DR2. All computed values are within the 3$\sigma$ combined uncertainty (85\% within 2$\sigma$).
	
	To evaluate the uncertainties of the measurements in this study to those from \textit{Gaia} DR2, the quantity $(\pi_{WISE} - \pi_{Gaia}) / \sqrt{\sigma_{\pi_{WISE}}^2 + \sigma_{\pi_{Gaia}}^2}$ was computed. If the uncertainties are not over/underestimated, and the measurements are unbiased, this quantity should follow a normal distribution with $\mu=0$ and $\sigma=1$. The inset plot of Figure~\ref{fig:GaiaComparison} shows the distribution of the above quantity, which has $\mu = 0.65$ mas and $\sigma = 0.90$ mas. Further comparisons are needed as the sample size in Figure~\ref{fig:GaiaComparison} is small, and statistical comparison using tests such as the Anderson-Darling test \citep{anderson:1952:193} provide limited information. It appears that \textit{WISE} parallax measurements are slightly overestimated as compared to \textit{Gaia} DR2, however, this is not the case for comparison to other literature values. Further investigation is warranted using a larger comparison sample and future \textit{Gaia} data releases that may account for possible systematics not yet discovered in the data.

	In principle, this method can be applied to any source bright enough to be extracted within a single \textit{WISE} L1b frame. Saturated photometry may cause an issue with centroiding. Crowded fields also pose a challenge due to multiple nearby objects causing source confusion and poor centroiding. It is unlikely that robust parallaxes ($\lesssim15\%$ uncertainty) can be measured farther than $\sim$20~pc using \textit{WISE} data alone, assuming an average parallax precision of 8~mas. However, this distance limit is highly dependent on $W2$ magnitude. The first 11 objects listed in Table~\ref{tbl:parallax} (excluding the binary Luhman 16AB and T8 dwarf 2MASS~J09393548$-$2448279) are contained within \textit{Gaia} DR2, which is roughly consistent with the $G<19$ ($W2<12$) limit discussed in Section~\ref{gaia}.

\section{New and Improved Astrometric Measurements}
\label{new}
	
\begin{figure*}
\includegraphics[width=\linewidth]{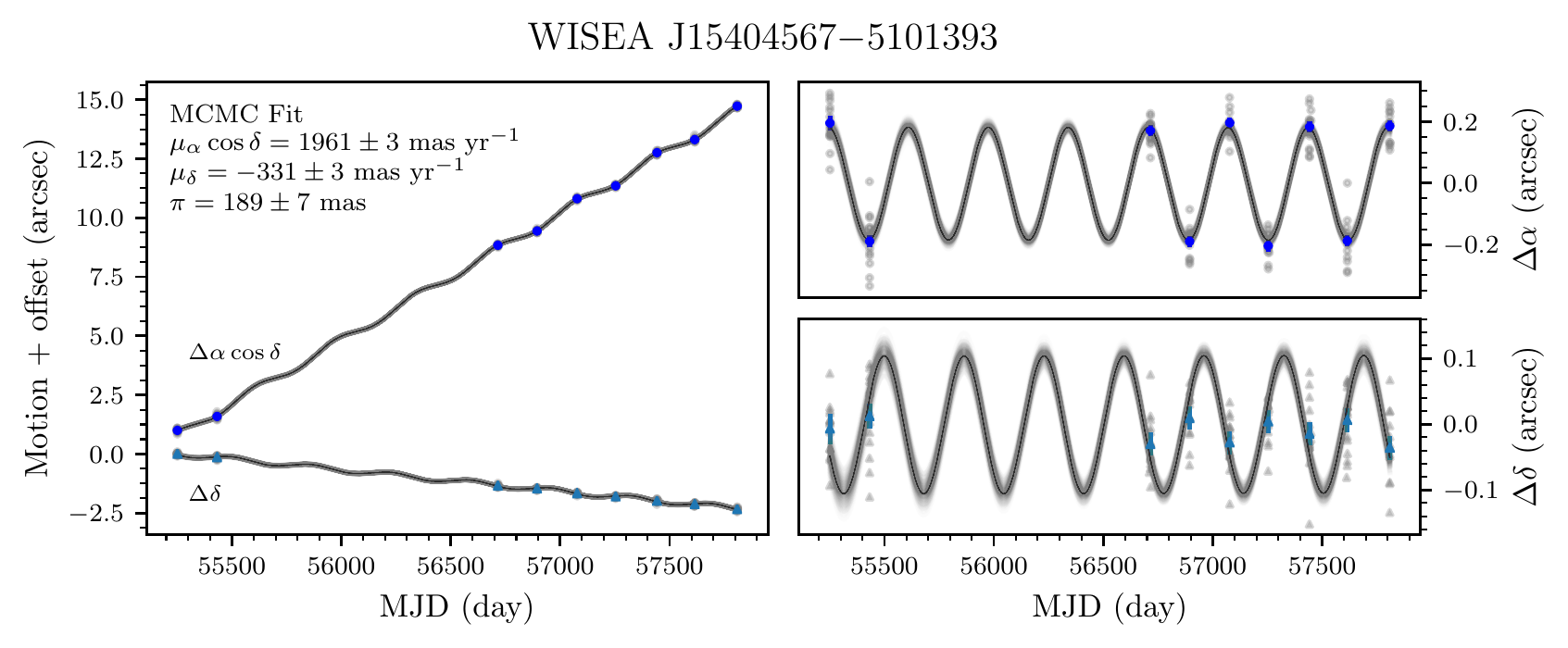}
\caption{Astrometric solution for WISEA J154045.67$-$510139.3, similar to Figure~\ref{fig:new}. 
The complete figure set (23 images) is available in the online journal.
\label{fig:updated}}
\end{figure*}

	There are many known low-mass objects estimated to be within 20~pc based on spectro-photometric parallax relationships, that have either no trigonometric parallax measurement, or measurements with large uncertainties ($>$20\%). Here, 23 such cases are investigated, sourced from the literature to cover a range of spectral types, distances, and $W2$ magnitudes, nine of which have parallax measurements within \textit{Gaia} DR2 (Table~\ref{tbl:parallax2}). The newly computed astrometric solutions are shown in Figure Set~\ref{fig:updated}.

\figsetstart
\figsetnum{8}
\figsettitle{Updated \textit{WISE} Astrometric Solutions}

\figsetgrpstart
\figsetgrpnum{8.1}
\figsetgrptitle{\textit{WISE} astrometric solution for WISEA J154045.67$-$510139.3.}
\figsetplot{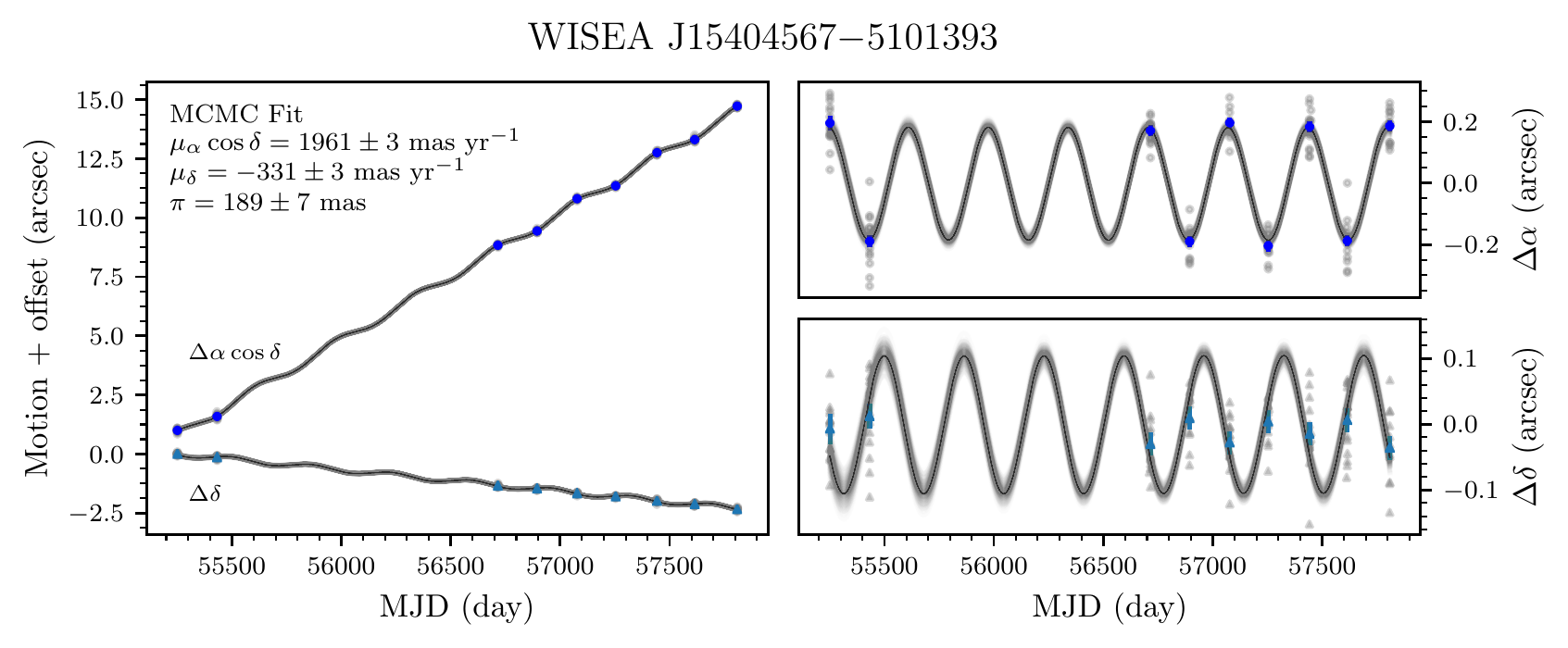}
\figsetgrpnote{}
\figsetgrpend

\figsetgrpstart
\figsetgrpnum{8.2}
\figsetgrptitle{\textit{WISE} astrometric solution for SDSS J122150.17$+$463244.4.}
\figsetplot{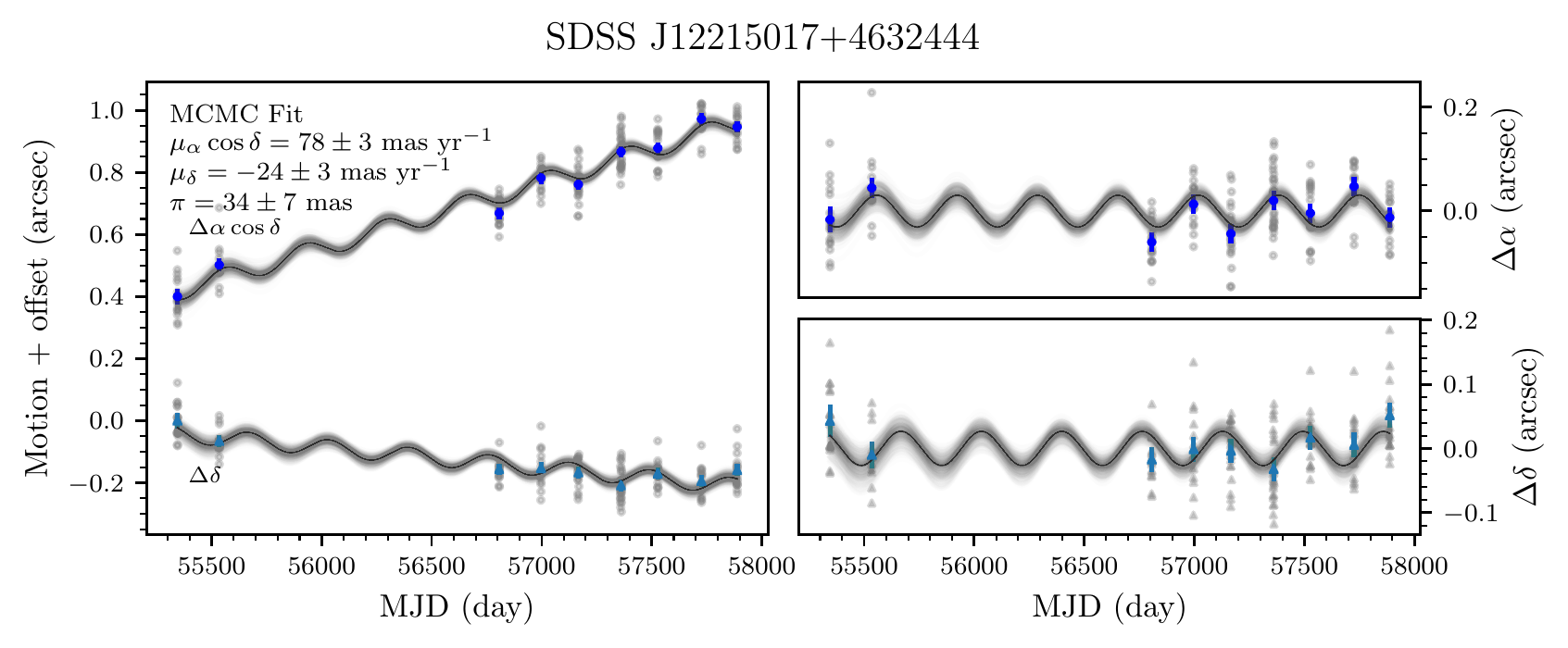}
\figsetgrpnote{}
\figsetgrpend

\figsetgrpstart
\figsetgrpnum{8.3}
\figsetgrptitle{\textit{WISE} astrometric solution for 2MASS J12351726$+$1318054.}
\figsetplot{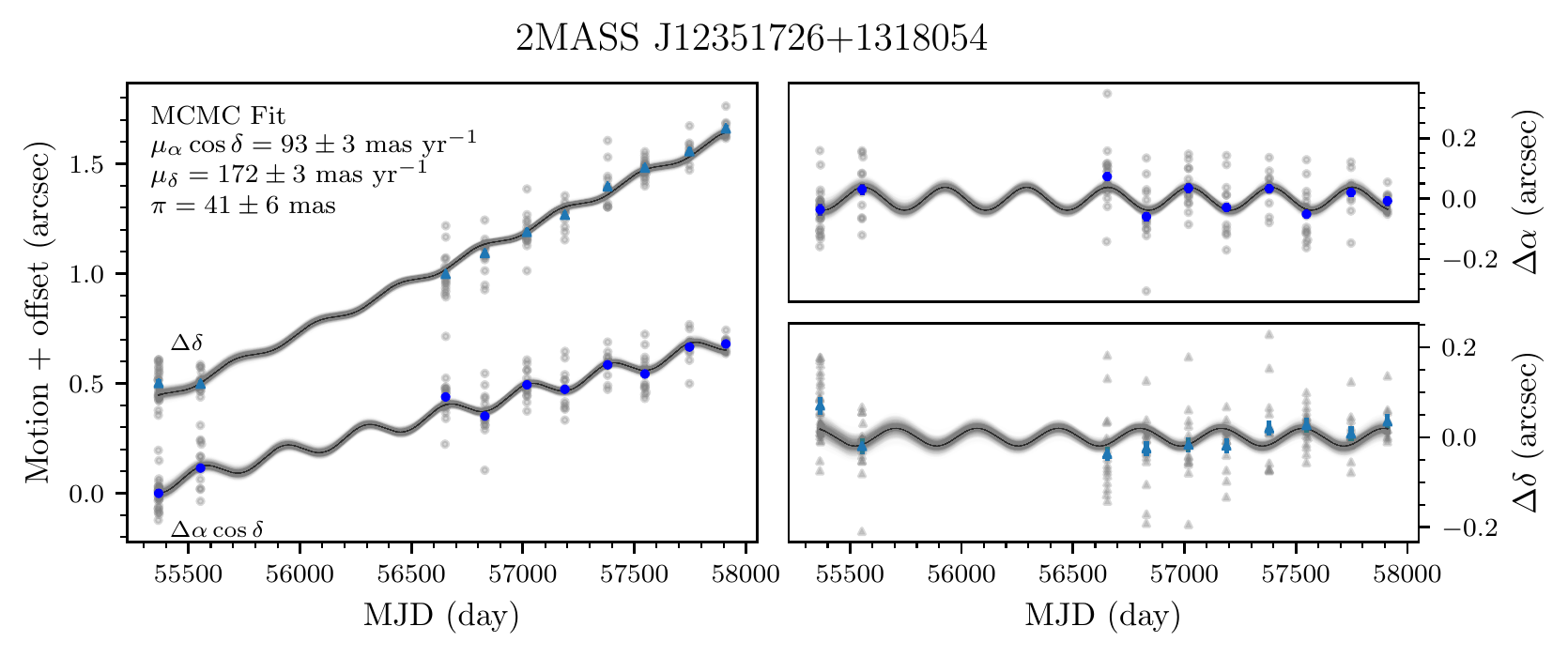}
\figsetgrpnote{}
\figsetgrpend

\figsetgrpstart
\figsetgrpnum{8.4}
\figsetgrptitle{\textit{WISE} astrometric solution for 2MASS J03140344$+$1603056.}
\figsetplot{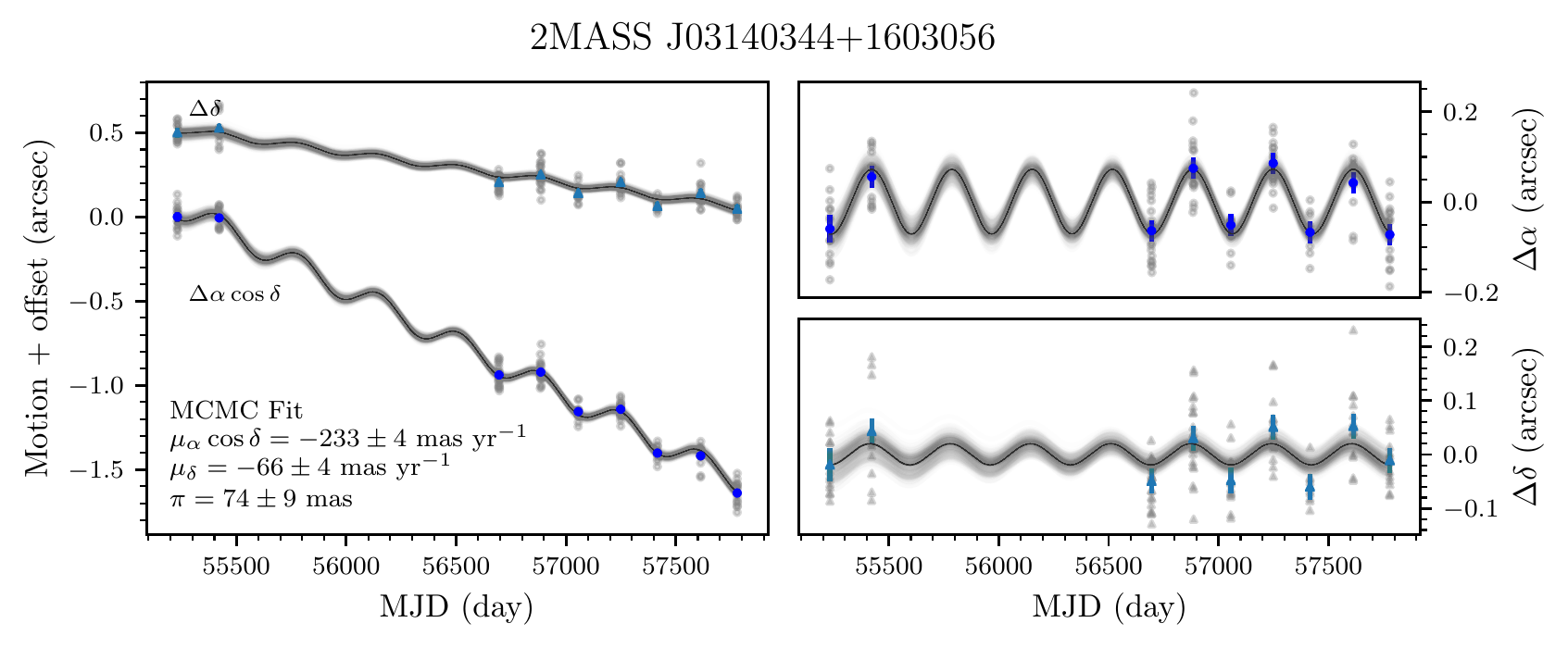}
\figsetgrpnote{}
\figsetgrpend

\figsetgrpstart
\figsetgrpnum{8.5}
\figsetgrptitle{\textit{WISE} astrometric solution for 2MASS J15065441$+$1321060.}
\figsetplot{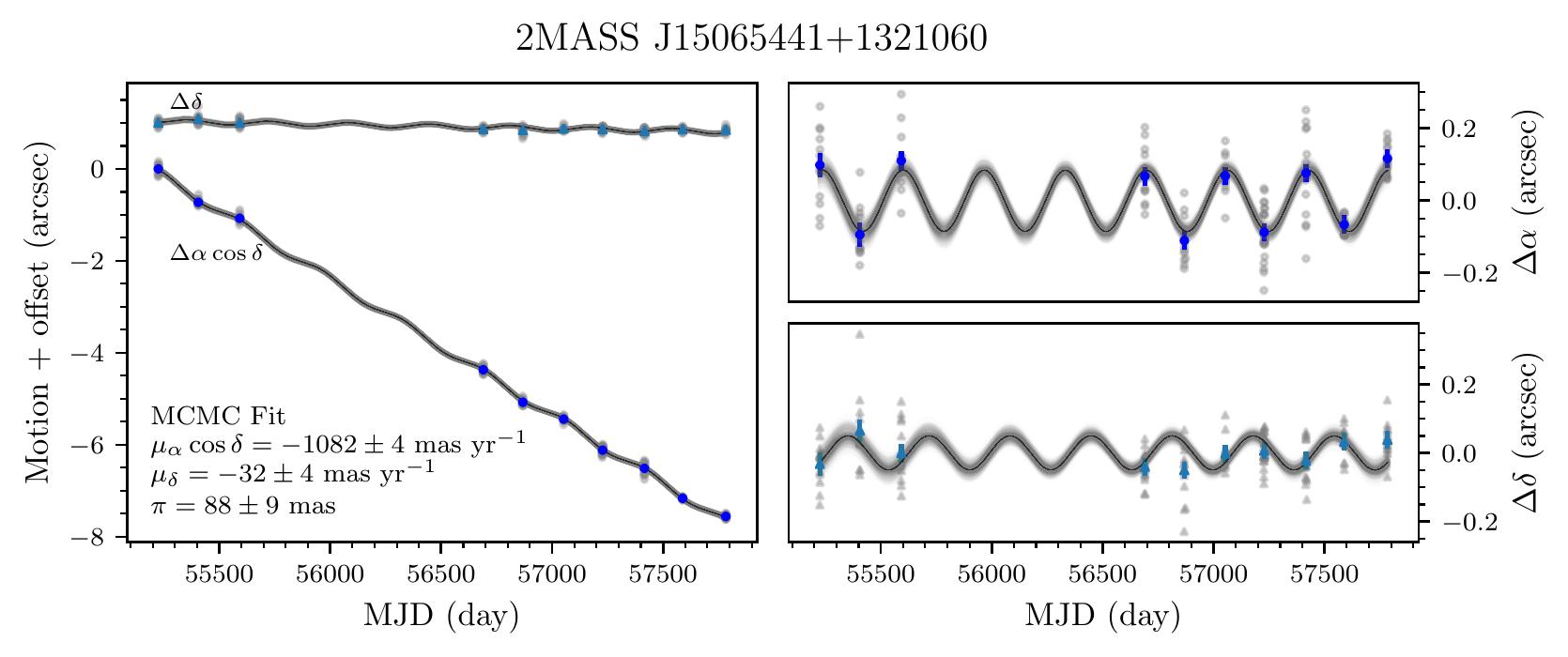}
\figsetgrpnote{}
\figsetgrpend

\figsetgrpstart
\figsetgrpnum{8.6}
\figsetgrptitle{\textit{WISE} astrometric solution for SDSS J141624.08$+$134826.7.}
\figsetplot{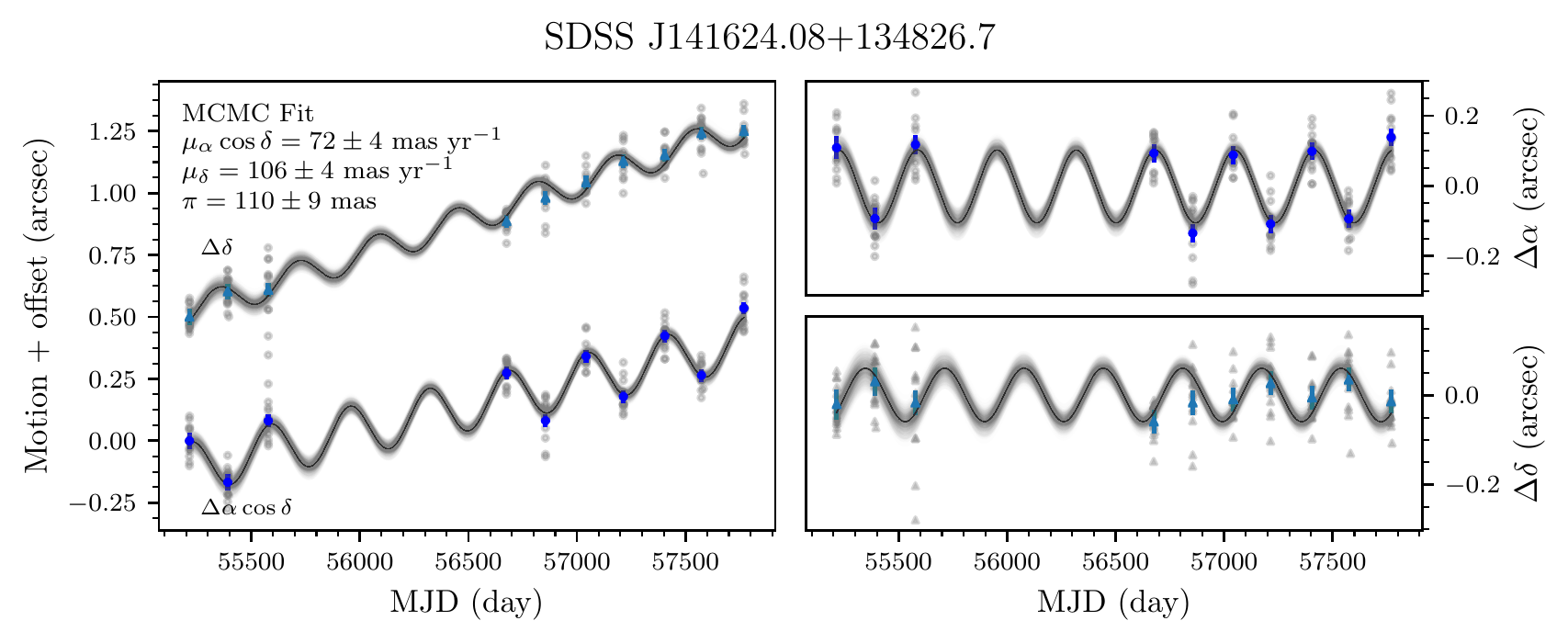}
\figsetgrpnote{}
\figsetgrpend

\figsetgrpstart
\figsetgrpnum{8.7}
\figsetgrptitle{\textit{WISE} astrometric solution for 2MASS J15150083$+$4847416.}
\figsetplot{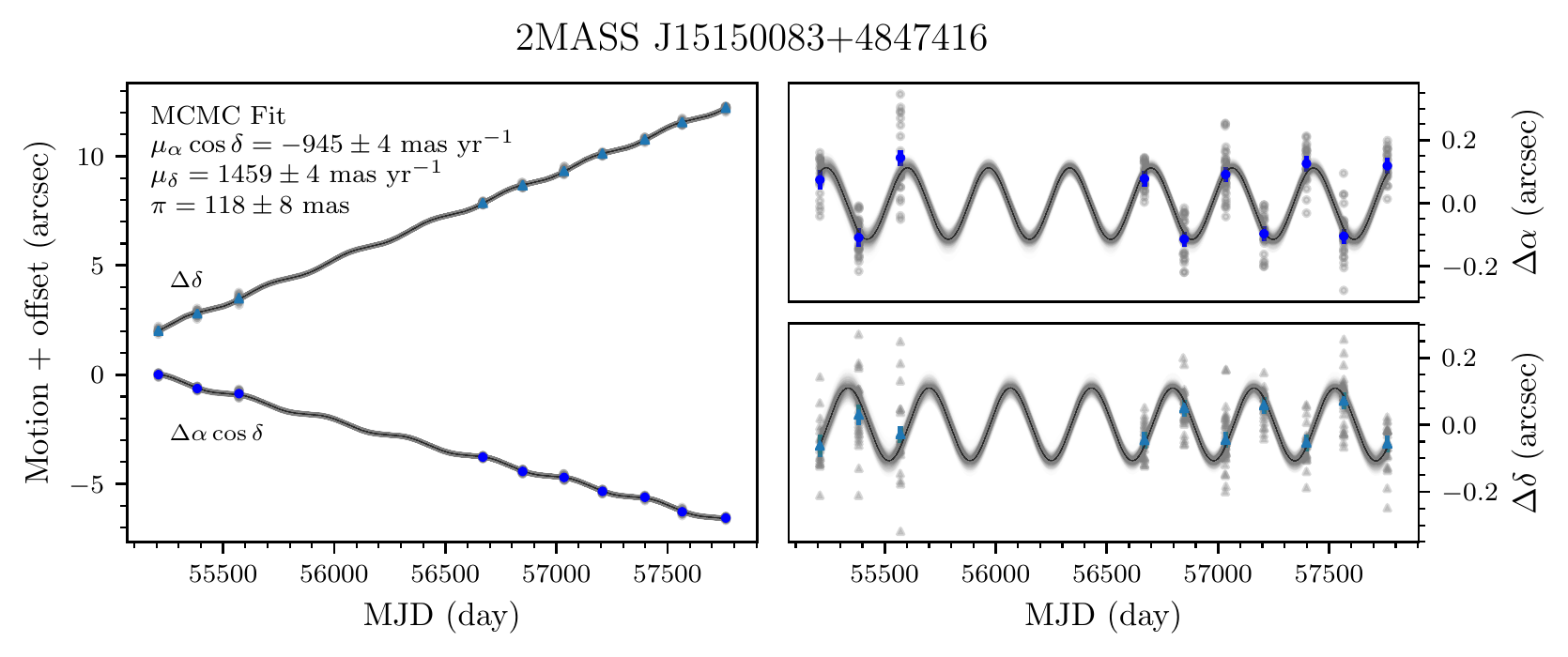}
\figsetgrpnote{}
\figsetgrpend

\figsetgrpstart
\figsetgrpnum{8.8}
\figsetgrptitle{\textit{WISE} astrometric solution for 2MASS J01443536$-$0716142.}
\figsetplot{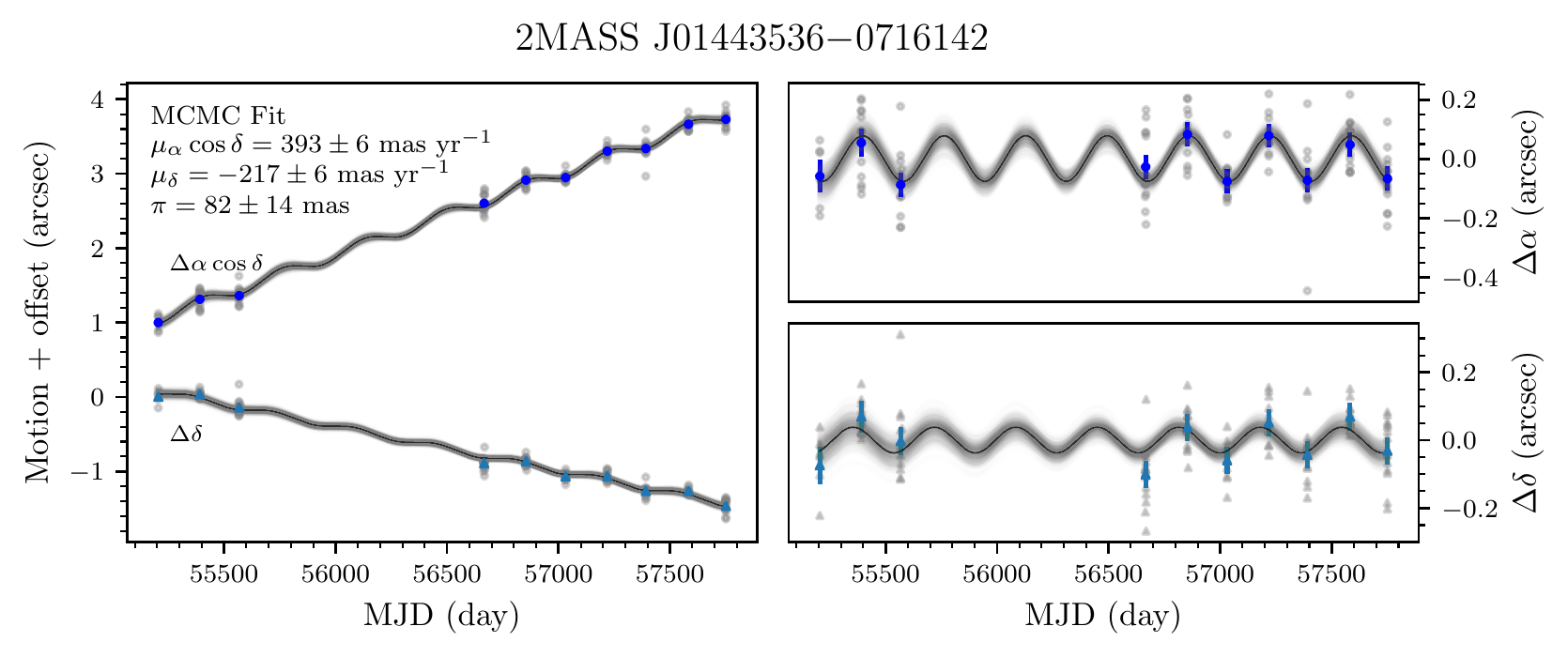}
\figsetgrpnote{}
\figsetgrpend

\figsetgrpstart
\figsetgrpnum{8.9}
\figsetgrptitle{\textit{WISE} astrometric solution for 2MASS J08575849$+$5708514.}
\figsetplot{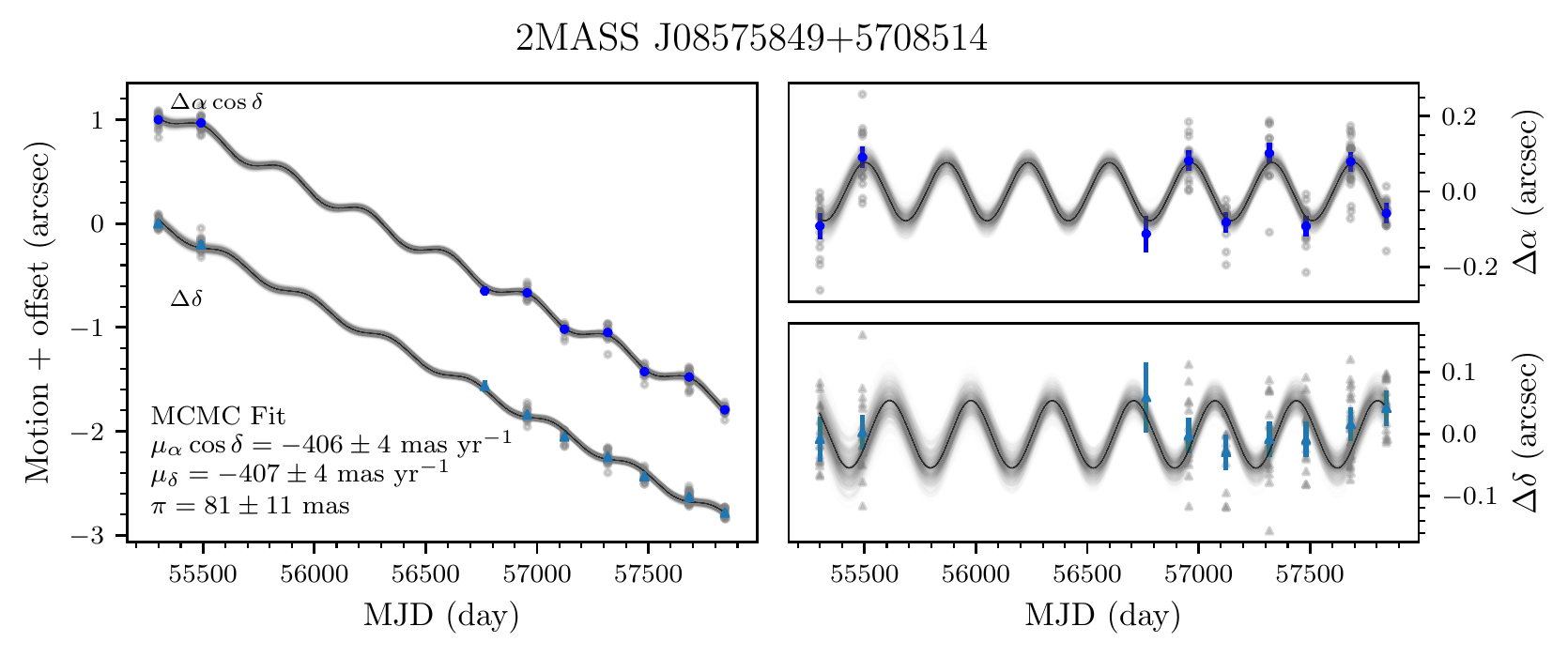}
\figsetgrpnote{}
\figsetgrpend

\figsetgrpstart
\figsetgrpnum{8.10}
\figsetgrptitle{\textit{WISE} astrometric solution for SDSS J090837.91$+$503207.5.}
\figsetplot{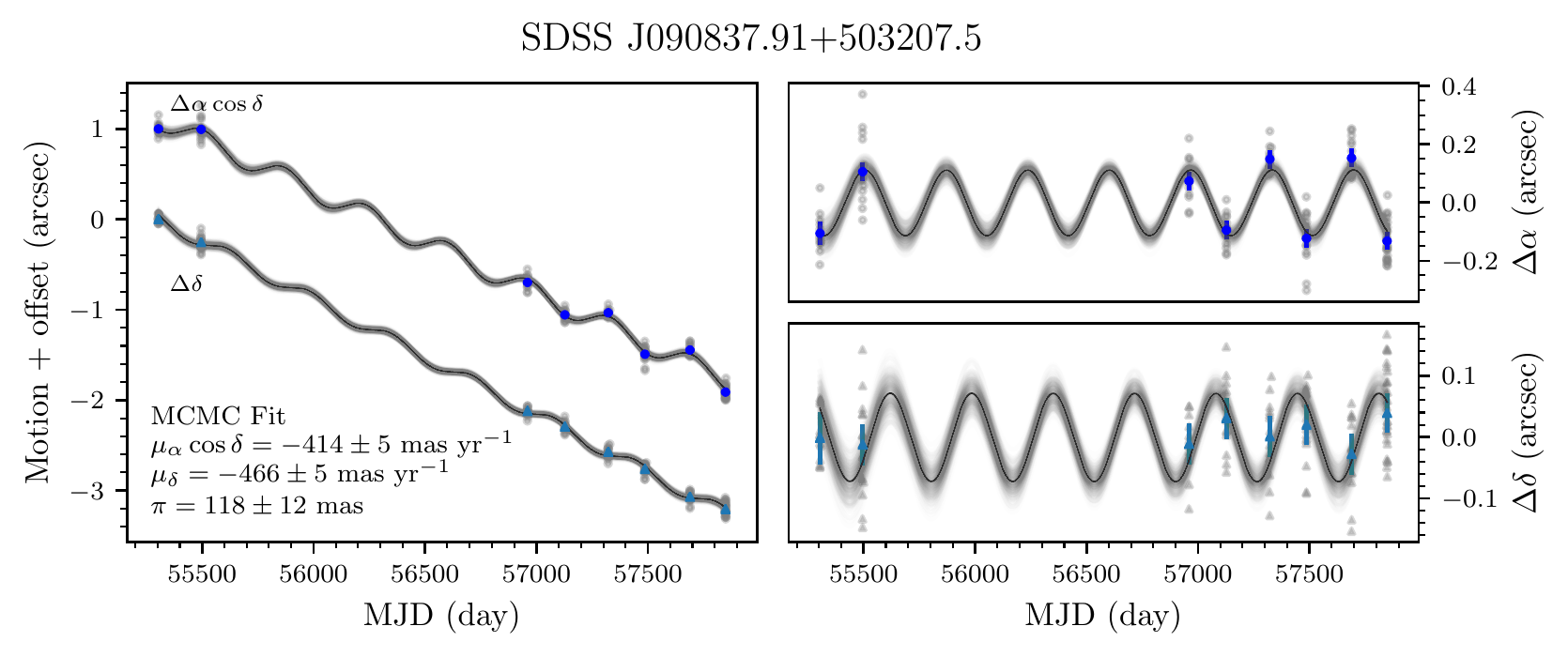}
\figsetgrpnote{}
\figsetgrpend

\figsetgrpstart
\figsetgrpnum{8.11}
\figsetgrptitle{\textit{WISE} astrometric solution for WISE J003110.04$+$574936.3.}
\figsetplot{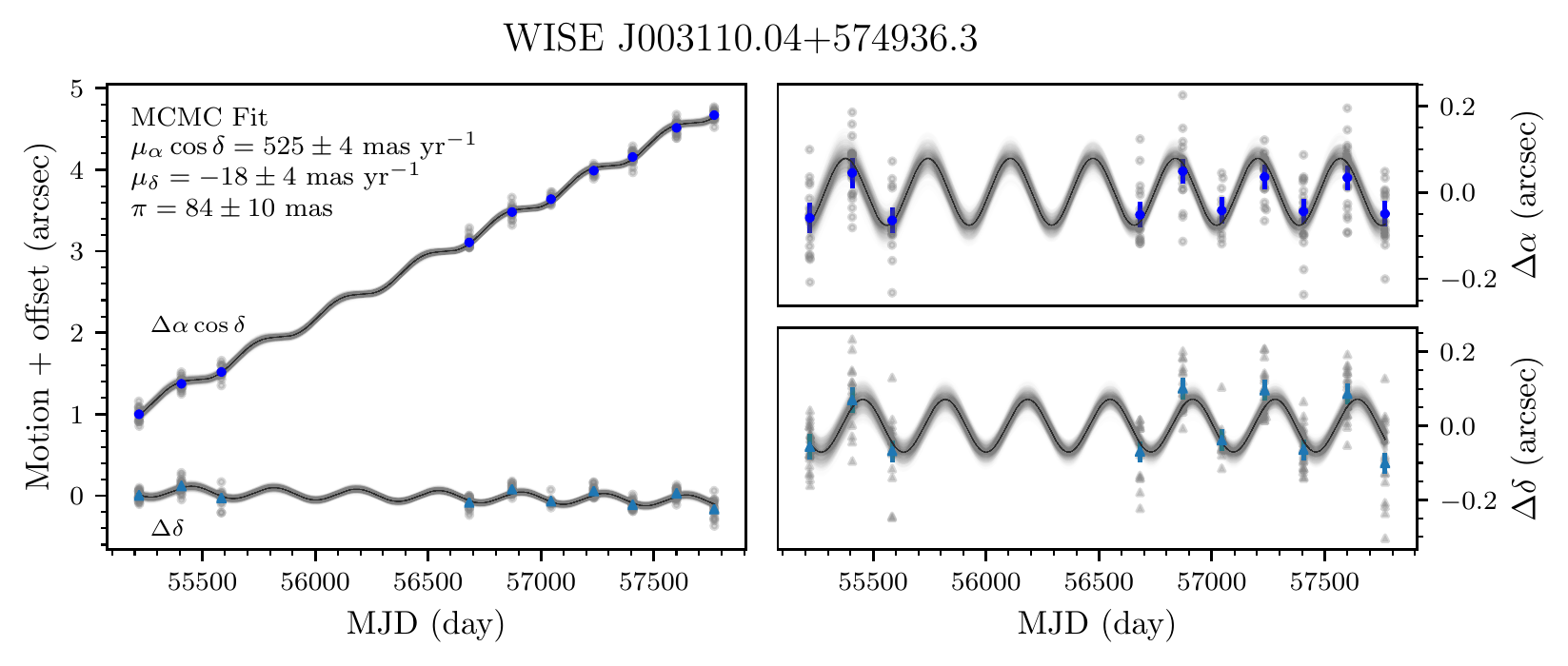}
\figsetgrpnote{}
\figsetgrpend

\figsetgrpstart
\figsetgrpnum{8.12}
\figsetgrptitle{\textit{WISE} astrometric solution for WISE J203042.79$+$074934.7.}
\figsetplot{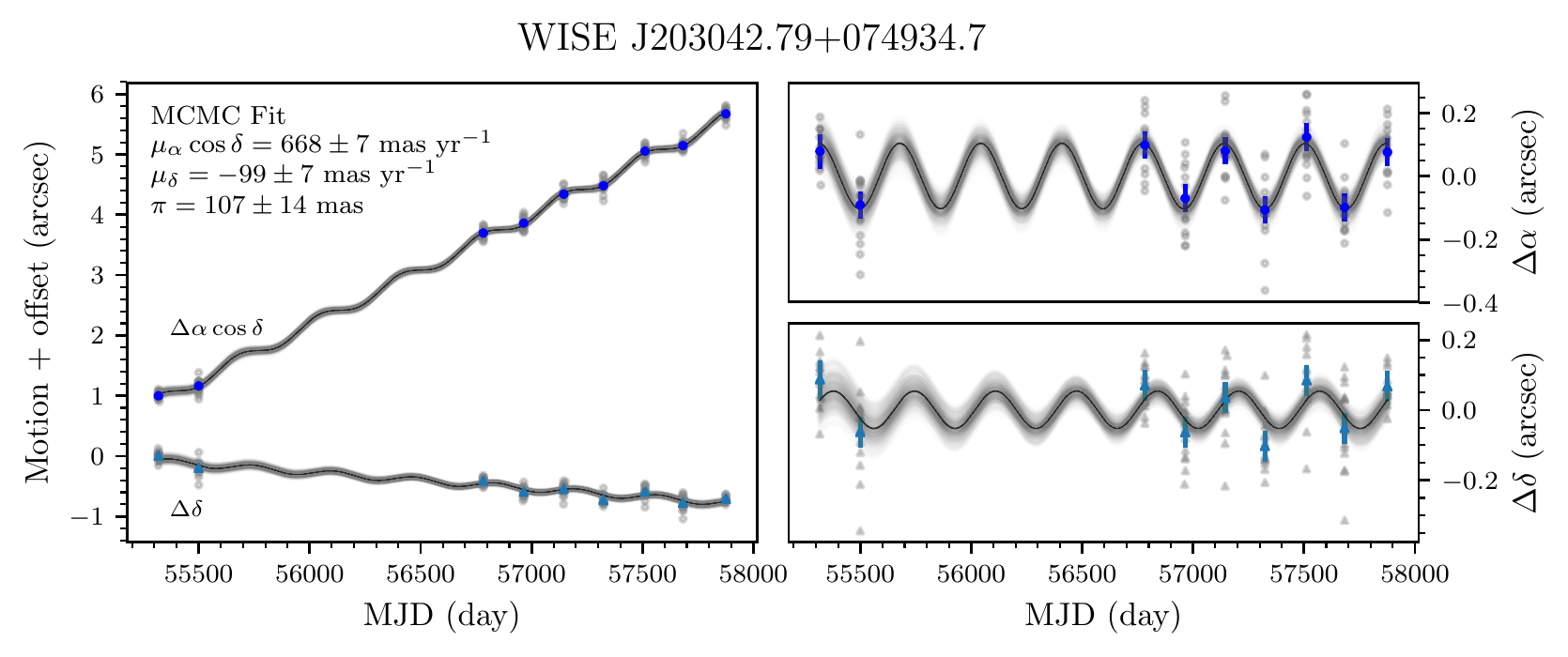}
\figsetgrpnote{}
\figsetgrpend

\figsetgrpstart
\figsetgrpnum{8.13}
\figsetgrptitle{\textit{WISE} astrometric solution for SDSS J075840.33$+$324723.4.}
\figsetplot{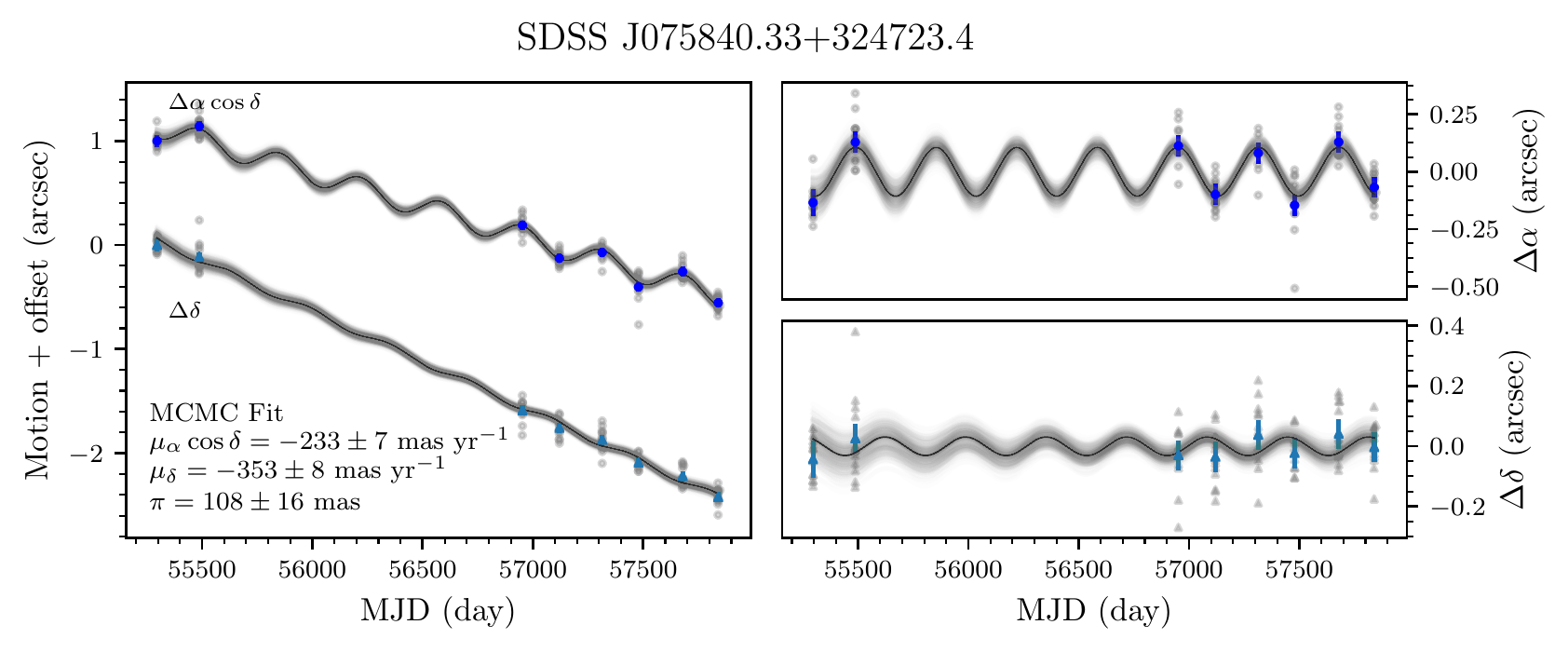}
\figsetgrpnote{}
\figsetgrpend

\figsetgrpstart
\figsetgrpnum{8.14}
\figsetgrptitle{\textit{WISE} astrometric solution for WISE J185101.83$+$593508.6.}
\figsetplot{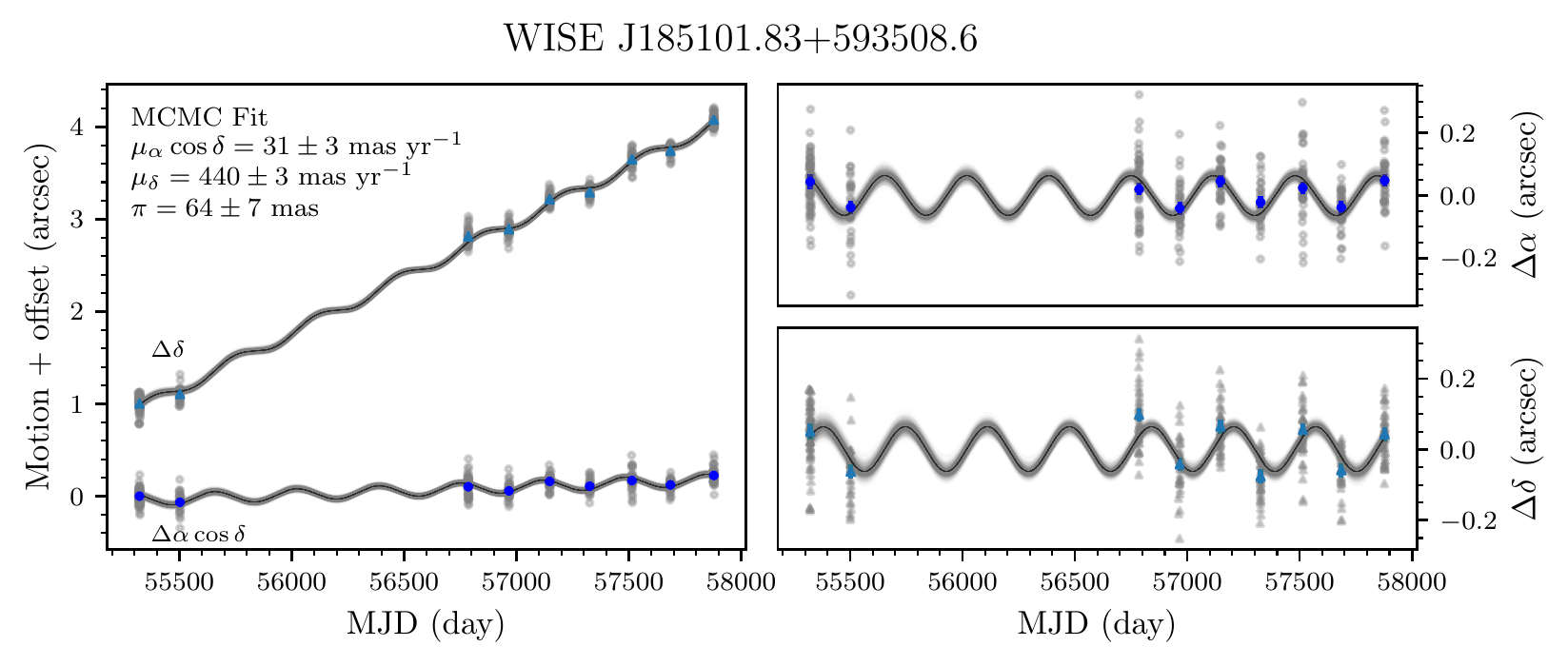}
\figsetgrpnote{}
\figsetgrpend

\figsetgrpstart
\figsetgrpnum{8.15}
\figsetgrptitle{\textit{WISE} astrometric solution for 2MASS J13243553$+$6358281.}
\figsetplot{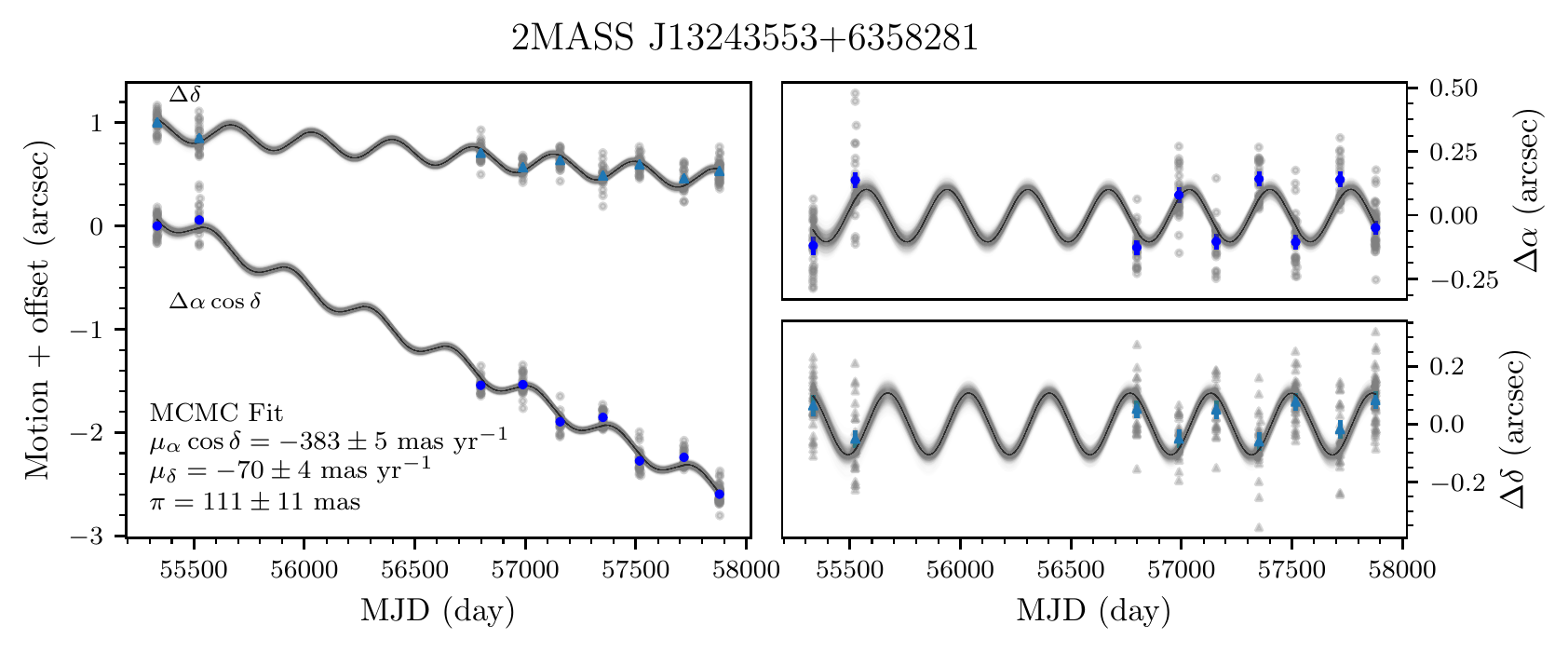}
\figsetgrpnote{}
\figsetgrpend

\figsetgrpstart
\figsetgrpnum{8.16}
\figsetgrptitle{\textit{WISE} astrometric solution for 2MASS J11061197$+$2754225.}
\figsetplot{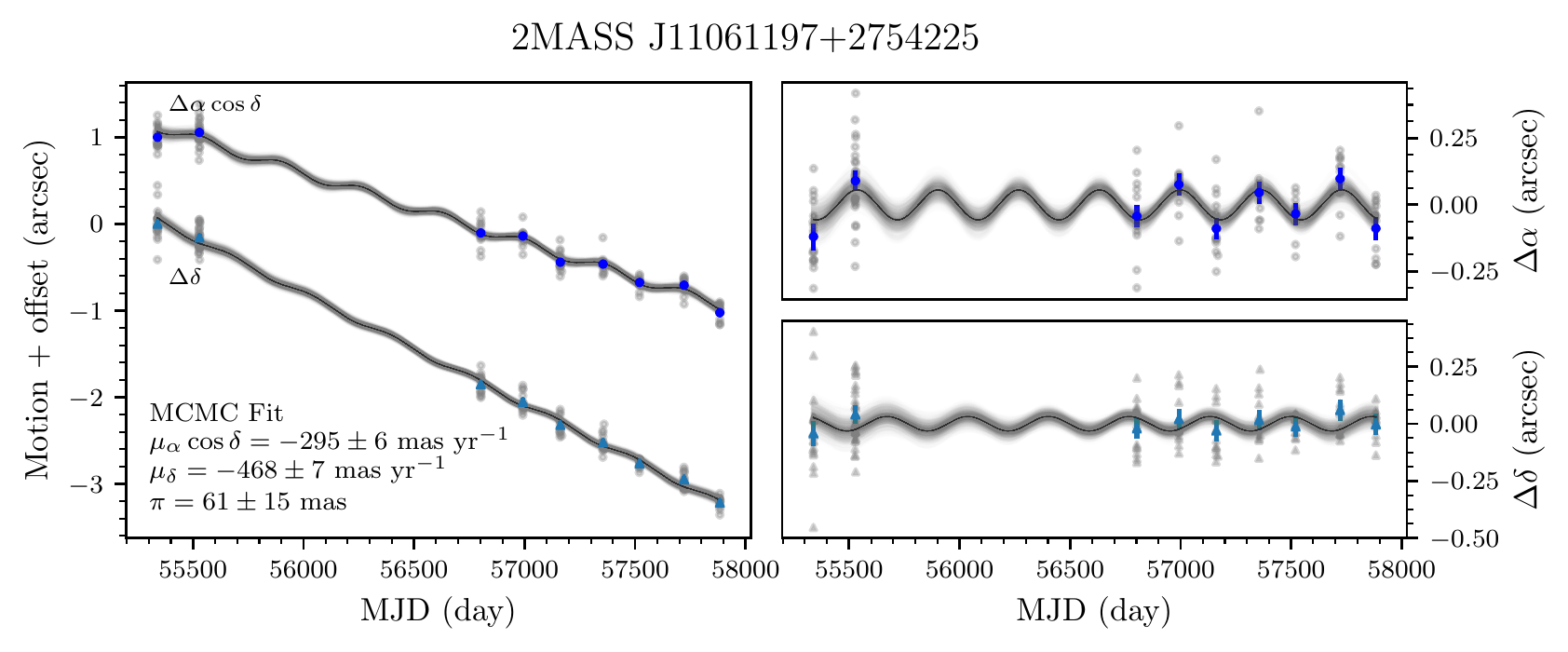}
\figsetgrpnote{}
\figsetgrpend

\figsetgrpstart
\figsetgrpnum{8.17}
\figsetgrptitle{\textit{WISE} astrometric solution for PSO J140.2308$+$45.6487.}
\figsetplot{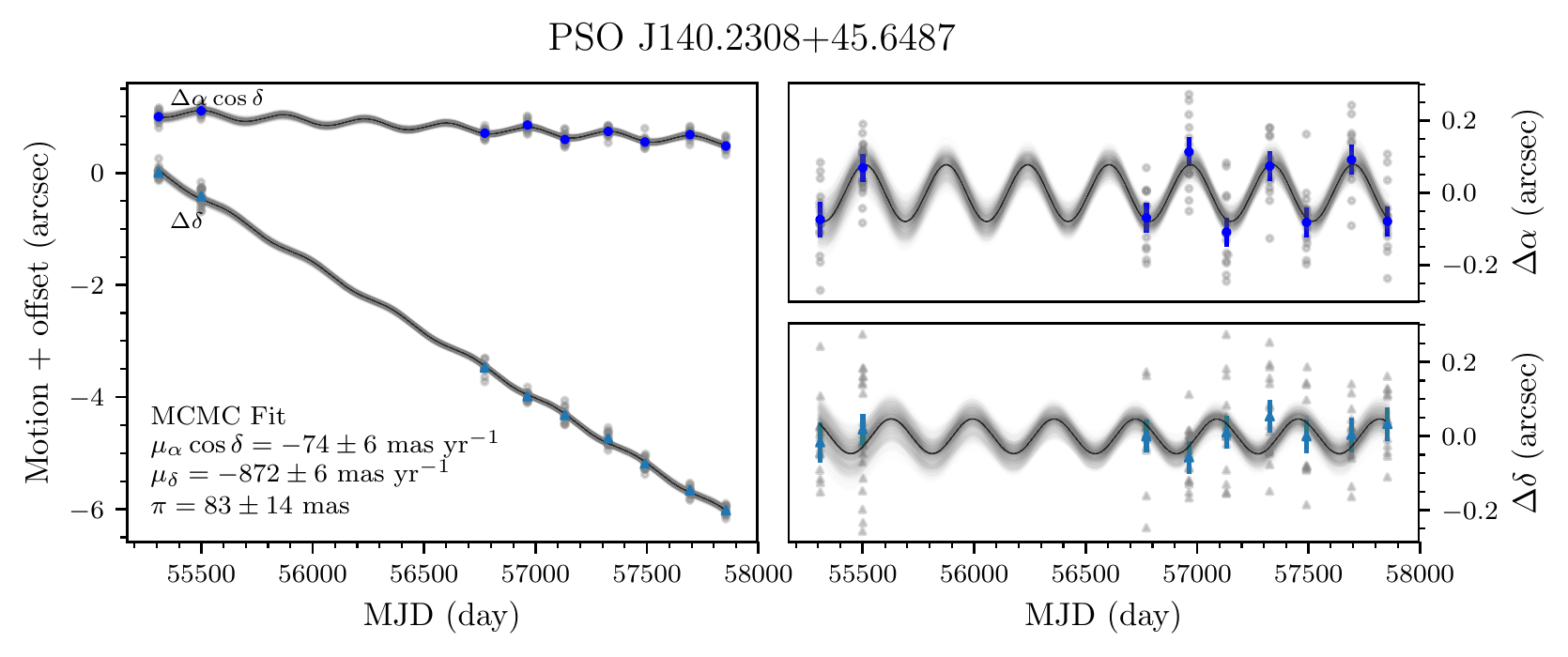}
\figsetgrpnote{}
\figsetgrpend

\figsetgrpstart
\figsetgrpnum{8.18}
\figsetgrptitle{\textit{WISE} astrometric solution for WISE J223617.59$+$510551.9.}
\figsetplot{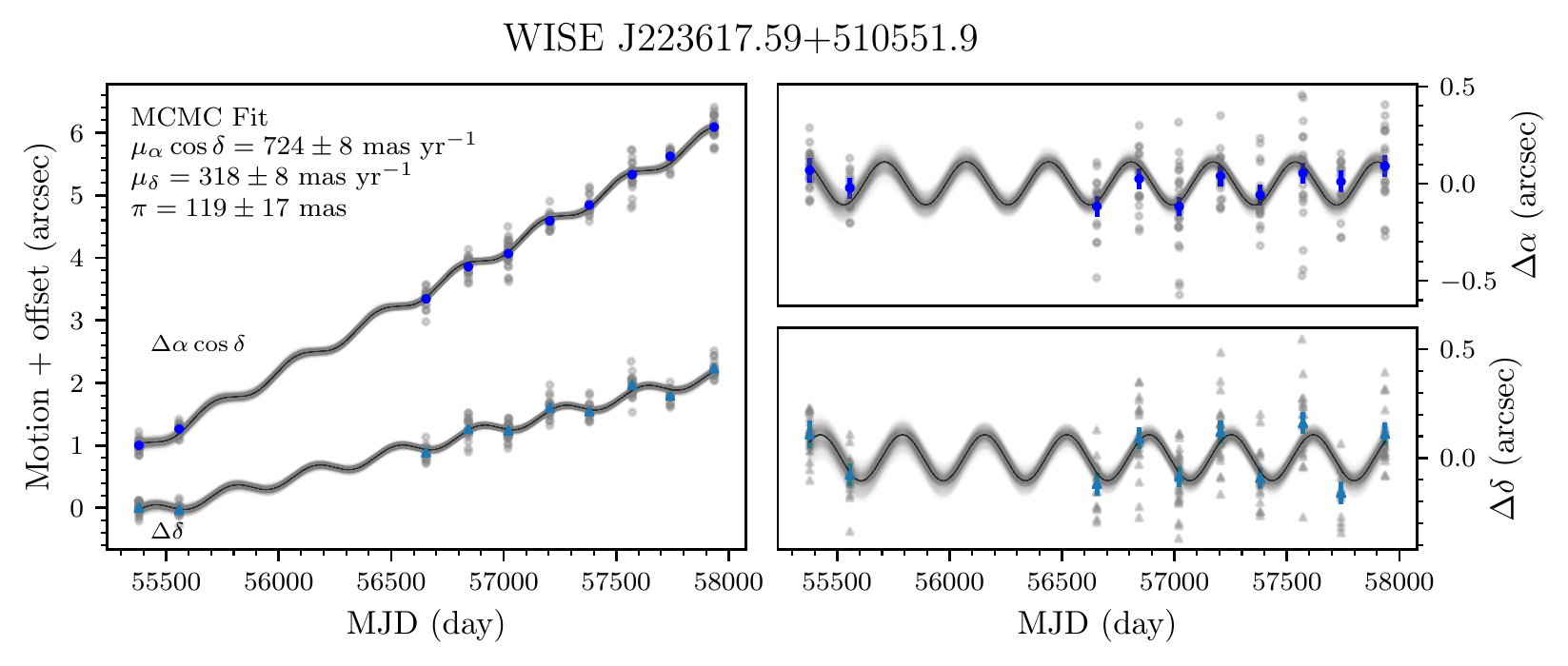}
\figsetgrpnote{}
\figsetgrpend

\figsetgrpstart
\figsetgrpnum{8.19}
\figsetgrptitle{\textit{WISE} astrometric solution for 2MASS J03480772$-$6022270.}
\figsetplot{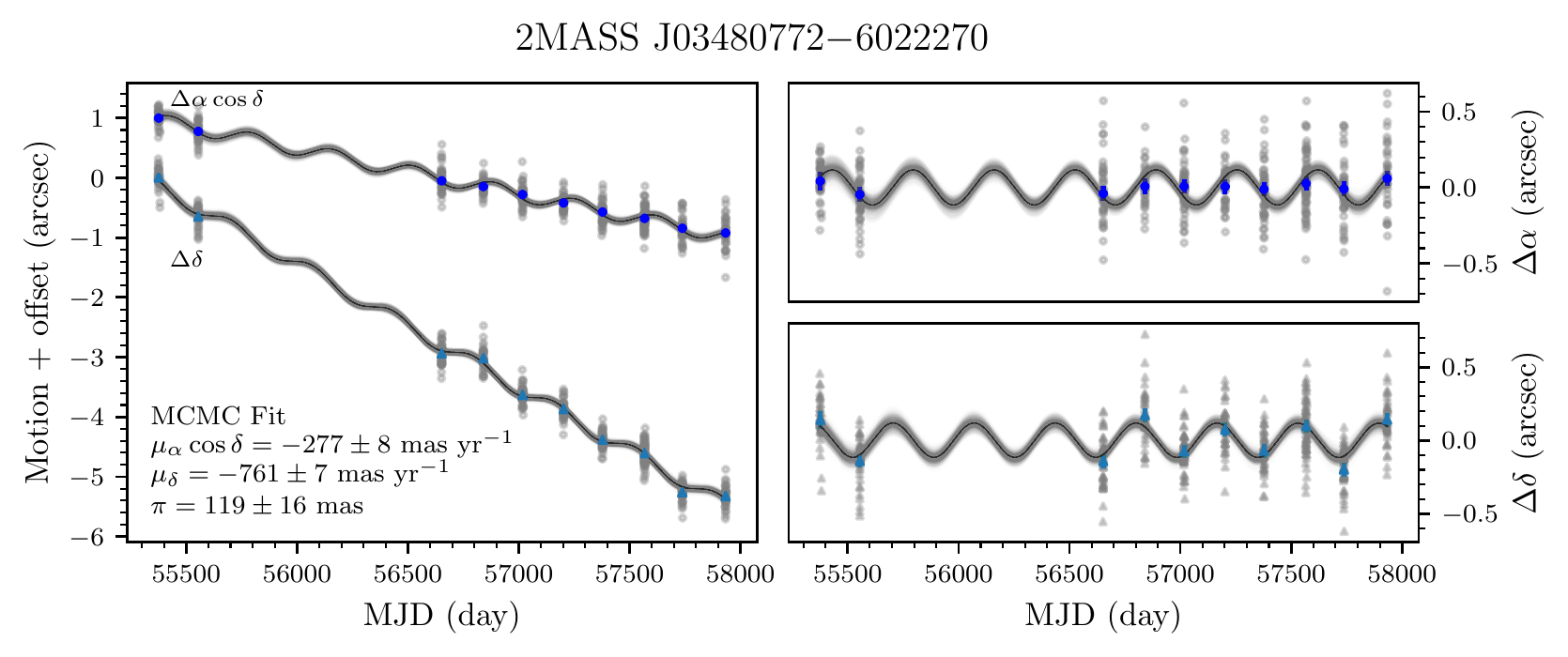}
\figsetgrpnote{}
\figsetgrpend

\figsetgrpstart
\figsetgrpnum{8.20}
\figsetgrptitle{\textit{WISE} astrometric solution for WISE J180952.53$-$044812.5.}
\figsetplot{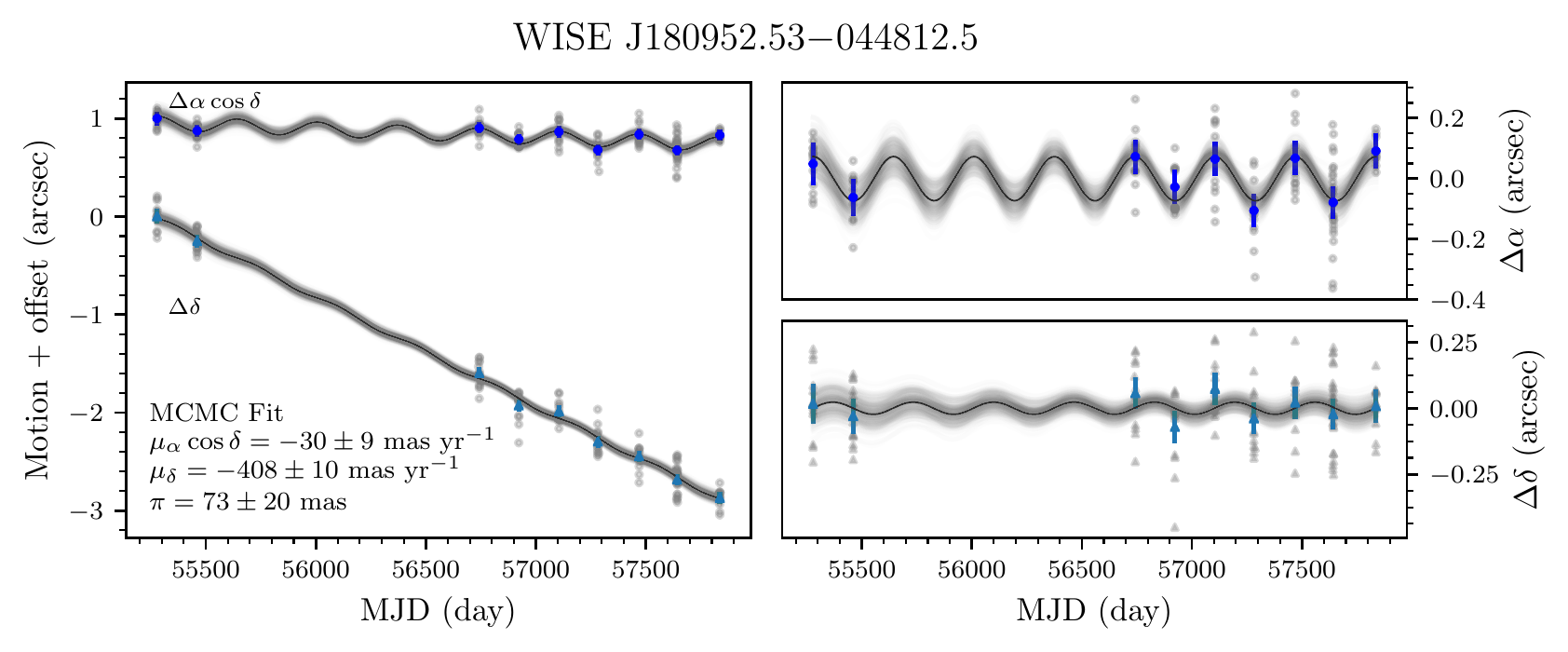}
\figsetgrpnote{}
\figsetgrpend

\figsetgrpstart
\figsetgrpnum{8.21}
\figsetgrptitle{\textit{WISE} astrometric solution for 2MASS J12314753$+$0847331.}
\figsetplot{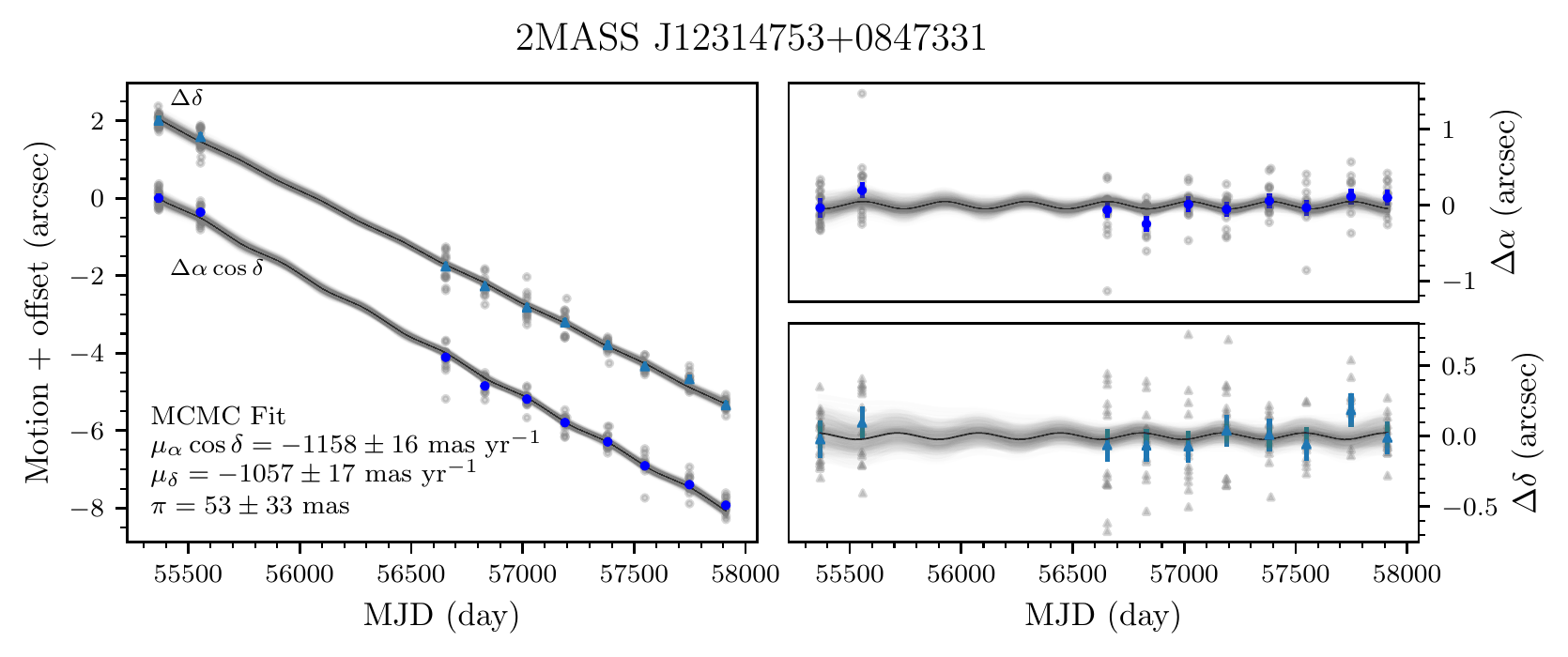}
\figsetgrpnote{}
\figsetgrpend

\figsetgrpstart
\figsetgrpnum{8.22}
\figsetgrptitle{\textit{WISE} astrometric solution for 2MASSI J2254188$+$312349.}
\figsetplot{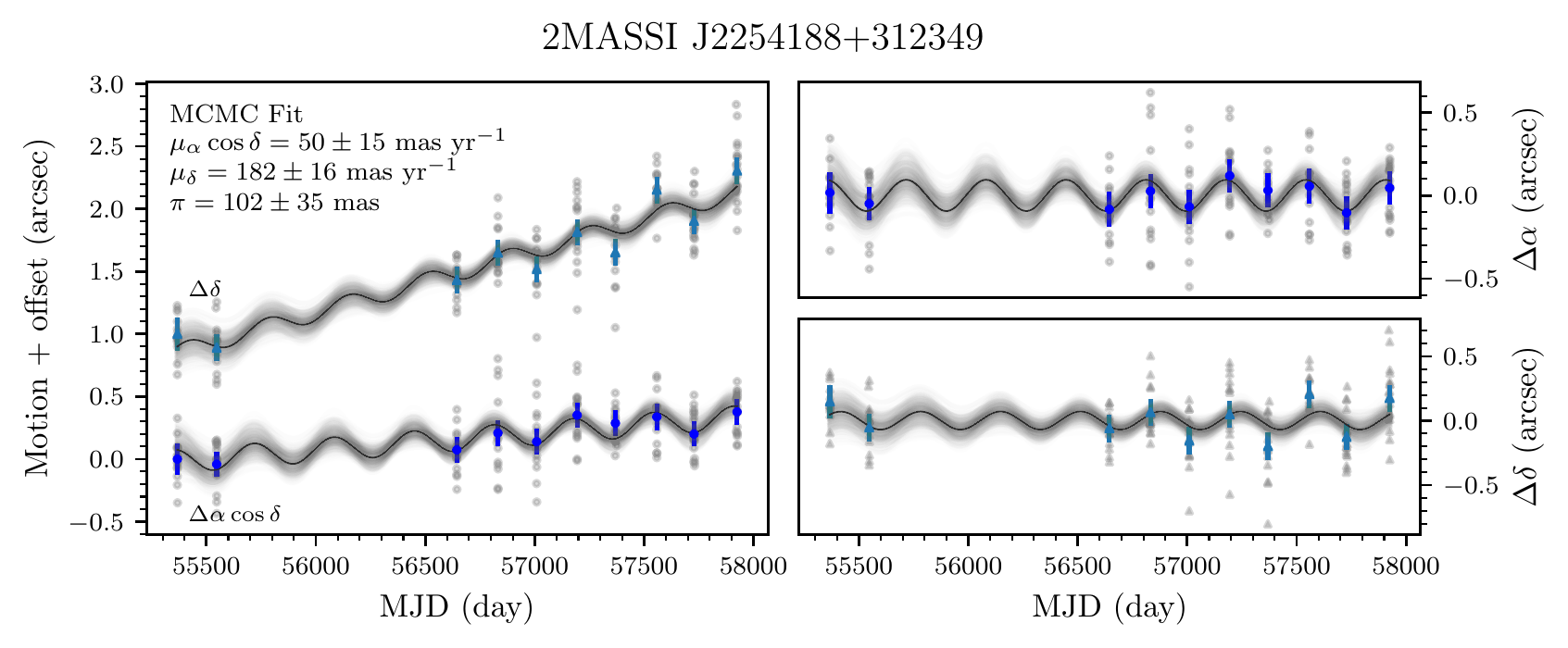}
\figsetgrpnote{}
\figsetgrpend

\figsetgrpstart
\figsetgrpnum{8.23}
\figsetgrptitle{\textit{WISE} astrometric solution for 2MASS J21543318$+$5942187.}
\figsetplot{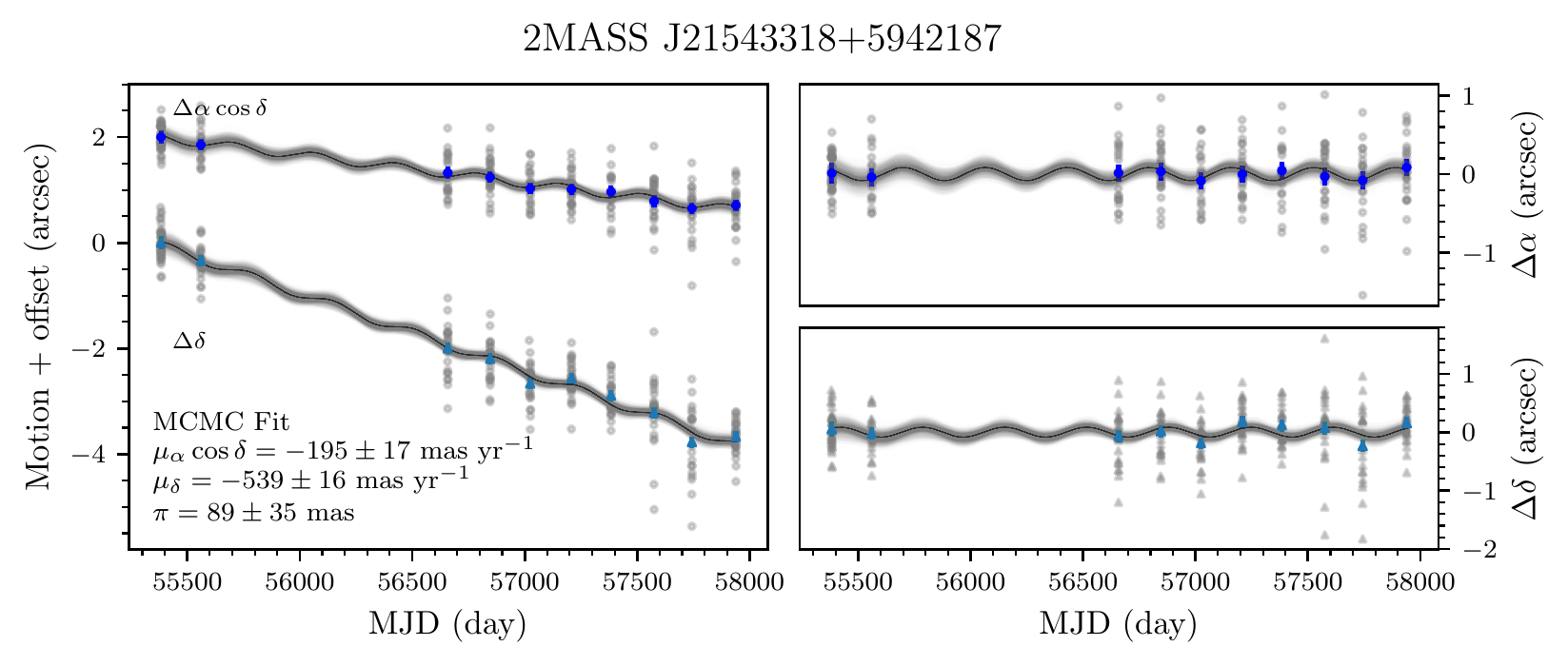}
\figsetgrpnote{}
\figsetgrpend

\figsetend

\begin{deluxetable*}{lccccccccccc}
\tabletypesize{\footnotesize}
\tablecolumns{12}
\tablecaption{New and Updated Astrometric Measurements\label{tbl:parallax2}}
\phs
\tablehead{
\colhead{Source} & \colhead{SpT} & \colhead{SpT} & $W2$ & \colhead{$\mu_\alpha \cos \delta$} & \colhead{$\mu_\delta$} & \colhead{$\pi$} & \colhead{$\pi_\mathrm{lit.}$\tablenotemark{a}} & \colhead{$\pi_\mathrm{lit.}$} & \colhead{\textit{Gaia}} \\
 & & \colhead{Ref.} & & \colhead{(mas yr$^{-1}$)} & \colhead{(mas yr$^{-1}$)} & \colhead{(mas)} & \colhead{(mas)} & \colhead{Ref.} & \colhead{DR2}
}
\startdata
WISEA J154045.67$-$510139.3 	& M6			& 9 	& $7.465\pm0.022$		& $1961\pm3$		& $-331\pm3$		& $189\pm7$	& $165\pm41$		& 9	& $188.04\pm0.37$	\\ 
SDSS J122150.17$+$463244.4 	& M6\tablenotemark{b}	& ...	& $9.797\pm0.020$	& $78\pm3$	& $-24\pm3$		& $34\pm7$	& ($69\pm7$)		& 19	& Y\tablenotemark{c} \\ 
2MASS J12351726$+$1318054 	& M6			& 14	& $9.893\pm0.021$		& $93\pm3$		& $172\pm3$		& $41\pm6$	& ($67\pm7$)		& 19	& Y\tablenotemark{c}\\
2MASS J03140344$+$1603056 	& L0			& 15	& $10.649\pm0.021$		& $-233\pm4$		& $-66\pm4$		& $74\pm9$	& ($69\pm4$)		& 15	& $73.42\pm1.06$	\\ 
2MASS J15065441$+$1321060 	& L3			& 8	& $10.872\pm0.021$		& $-1082\pm4$		& $-32\pm4$		& $88\pm9$	& ($83\pm14$)		& 8	& $85.58\pm1.00$	\\ 
SDSS J141624.08$+$134826.7 	& L6			& 17	& $11.026\pm0.020$		& $72\pm4$		& $106\pm4$		& $110\pm9$	& $127\pm27$		& 18	& $107.55\pm1.67$	\\ 
2MASS J15150083$+$4847416 	& L6			& 16	& $11.332\pm0.021$		& $-945\pm4$		& $1459\pm4$		& $118\pm8$	& ($145\pm27$)	& 17	& Y\tablenotemark{c}\\
2MASS J01443536$-$0716142		& L5			& 11	& $11.371\pm0.021$		& $363\pm6$		& $-217\pm6$		& $82\pm14$	& ($75\pm8$)		& 5	& $79.03\pm2.38$	\\
2MASS J08575849$+$5708514	& L8\tablenotemark{d}	& 7	& $11.439\pm0.021$		& $-406\pm4$		& $-407\pm4$		& $81\pm11$	& $98.0\pm2.6$\tablenotemark{e}	& 21	& $71.23\pm4.65$	\\
SDSS J090837.91$+$503207.5	& L8			& 17	& $11.651\pm0.023$		& $-414\pm5$		& $-466\pm5$		& $118\pm12$	& ($122\pm24$)	& 17	& $95.82\pm2.63$	\\ 
WISE J003110.04$+$574936.3		& L9			& 1	& $11.843\pm0.021$		& $525\pm4$		& $-18\pm4$		& $84\pm10$	& ($91\pm8$)		& 20	& N	\\ 
WISE J203042.79$+$074934.7		& T1.5		& 13	& $12.129\pm0.024$		& $668\pm7$		& $-99\pm7$		& $107\pm14$	& ($92\pm8$)		& 1	& $103.97\pm3.61$	\\ 
SDSS J075840.33$+$324723.4	& T2			& 10	& $12.170\pm0.024$		& $-233\pm7$		& $-353\pm8$		& $108\pm16$	& ($91\pm8$) 		& 6	& Y\tablenotemark{f} \\
WISE J185101.83$+$593508.6		& L7$+$T2	& 20	& $12.178\pm0.022$		& $31\pm3$		& $440\pm3$		& $64\pm7$	& ($91\pm8$)		& 20	& $48.85\pm2.45$ \\ 
2MASS J13243553$+$6358281	& L8$+$T3.5 (T2)	& 4 (12)	& $12.294\pm0.022$		& $-383\pm5$		& $-70\pm4$		& $111\pm11$	& ($77\pm6$)		& 6	& N \\ 
2MASS J11061197$+$2754225	& T0$+$T4.5	& 4	& $12.361\pm0.024$		& $-295\pm6$		& $-468\pm7$		& $61\pm15$	& ($91\pm8$) 		& 6	& Y\tablenotemark{f} \\ 
PSO J140.2308$+$45.6487		& L9			& 13	& $12.439\pm0.023$		& $-74\pm6$		& $-872\pm6$		& $83\pm14$	& ($70\pm7$)		& 2	& Y\tablenotemark{f} \\ 
WISE J223617.59$+$510551.9		& T5.5		& 13	& $12.499\pm0.025$		& $724\pm8$		& $318\pm8$		& $119\pm17$	& ($106\pm9$)		& 1	& Y\tablenotemark{f} \\
2MASS J03480772$-$6022270		& T7			& 3	& $12.550\pm0.022$		& $-277\pm8$		& $-761\pm7$		& $119\pm16$	& ($111\pm12$)	& 6	& N \\
WISE J180952.53$-$044812.5		& T1			& 2	& $12.745\pm0.028$		& $-30\pm9$		& $-408\pm10$		& $73\pm20$	& ($67\pm7$)		& 2	& Y\tablenotemark{f} \\ 
2MASS J12314753$+$0847331	& T5.5		& 3	& $13.083\pm0.031$		& $-1158\pm16$	& $-1057\pm17$	& $53\pm33$	& ($83\pm7$)		& 6	& N \\
2MASSI J2254188$+$312349		& T4			& 3	& $13.288\pm0.030$		& $57\pm16$		& $188\pm16$		& $108\pm33$	& ($71\pm10$)		& 6	& Y\tablenotemark{f} \\
2MASS J21543318$+$5942187	& T6			& 12	& $13.579\pm0.029$		& $-195\pm17$		& $-539\pm16$		& $89\pm35$	& ($100\pm10$)	& 6	& N 
\enddata
\tablecomments{
(1) \citealt{best:2013:84}; 
(2) \citealt{best:2015:118};
(3) \citealt{burgasser:2006:1067};
(4) \citealt{burgasser:2010:1142};
(5) \citealt{cruz:2003:2421}; 
(6) \citealt{faherty:2009:1};
(7) \citealt{geballe:2002:466}; 
(8) \citealt{gizis:2000:1085}; 
(9) \citealt{kirkpatrick:2014:122}; 
(10) \citealt{knapp:2004:3553};
(11) \citealt{liebert:2003:343}; 
(12) \citealt{looper:2007:1162};
(13) \citealt{mace:2013:6}; 
(14) \citealt{reid:2007:2825};
(15) \citealt{reid:2008:1290};
(16) \citealt{schmidt:2007:2258};
(17) \citealt{schmidt:2010:1808}; 
(18) \citealt{scholz:2010:l8}; 
(19) \citealt{theissen:2017:92};
(20) \citealt{thompson:2013:809};
(21) \citealt{wang:2018:064402}.
}
\tablenotetext{a}{Values in parentheses indicate spectrophotometric distance estimates.}
\tablenotetext{b}{Spectral type estimated from photometric colors.}
\tablenotetext{c}{No parallax measurement included in \textit{Gaia} DR2, however, source has $G<21$. Typically high \textsc{astrometric\_sigma5d\_max} parameters and/or high goodness-of-fit statistics ($>$100).}
\tablenotetext{d}{Potential low-gravity object discussed in \citealt{gagne:2015:33}.}
\tablenotetext{e}{This parallax measurement was published while this manuscript was under review.}
\tablenotetext{f}{Included in \textit{Gaia} DR2, but no 5-parameter astrometric solution due to $G\geqslant21$.}
\end{deluxetable*}

	Using the sources from Tables~\ref{tbl:parallax} and~\ref{tbl:parallax2}, a 2nd order polynomial fit to the parallax uncertainties divided by 0.15 was computed, representing the approximate parallax limit where $\leqslant15\%$ uncertainties can be achieved, as a function of $W2$ magnitude. The polynomial is shown in Figure~\ref{fig:residuals}, converted from parallax to distance (blue dashed line corresponding to the blue y-axis on the right side of Figure~\ref{fig:residuals}). Bright sources ($W2\lesssim8$) can potentially have their parallaxes measured out to distances of $\sim$23~pc, with sources at the \textit{Gaia} 95\% completeness limit ($G\approx19$; $W2\approx11$) requiring distances within $\sim$17~pc. These limits will be validated in the future with a larger control sample and future \textit{Gaia} data releases.

\section{Discussion}
\label{discussion}

	The technique presented here has the potential to find new, nearby, ultracool objects, and measure relatively accurate parallaxes without the need for follow-up observations. This is particularly important as \textit{Spitzer} is expected to be retired in 2018. Its replacement, the \textit{James Webb Space Telescope} \citep[\textit{JWST};][]{gardner:2006:485}, while sensitive to these faint dwarfs, is an unlikely facility for a dedicated parallax program.
	
	There are approximately 300 objects with spectral types between L0 and T8 that currently have published parallaxes\footnote{The majority can be found within the Database of Ultracool Parallaxes maintained by Trent Dupuy: \url{http://www.as.utexas.edu/~tdupuy/plx/Database\_of\_Ultracool\_Parallaxes.html} \citep{dupuy:2012:19,dupuy:2013:1492,liu:2016:96} }. As discussed here and in previous studies \citep{theissen:2017:92, smart:2017:401}, \textit{Gaia} will not provide parallaxes for many of the nearby, lowest-mass, ultracool objects leaving only ground-based programs. The method described here is a useful alternative for the nearest ($\lesssim17$~pc) ultracool objects. 
	
\begin{figure}
\includegraphics[width=\linewidth]{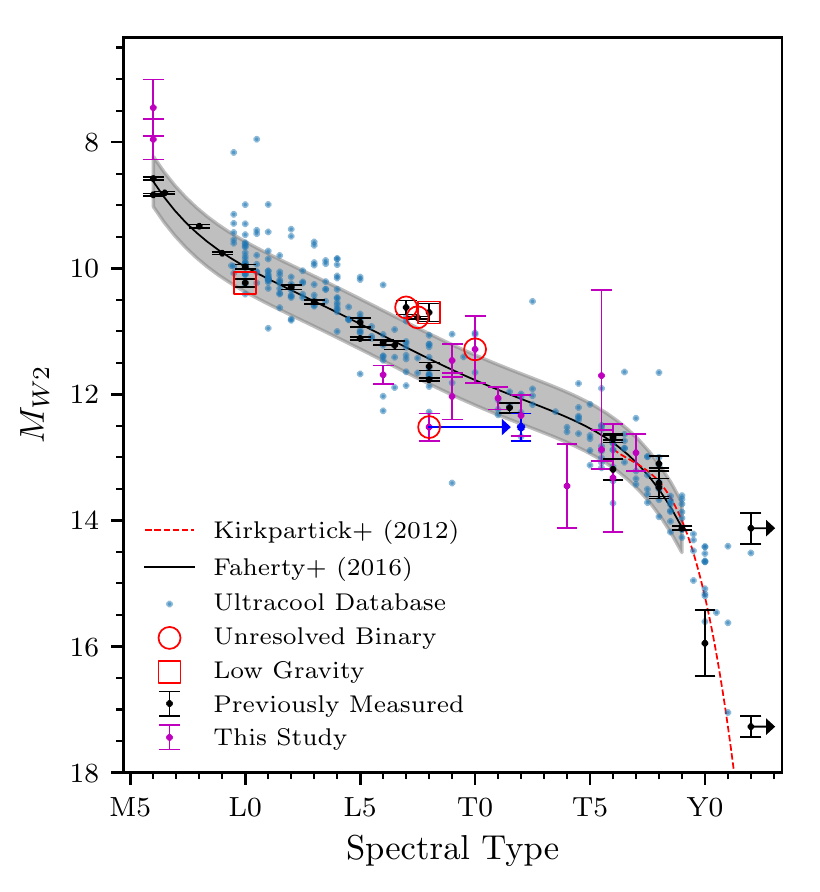}
\caption{Absolute $W2$ magnitude as a function of spectral type for late-type dwarfs. The relationships from \citet{kirkpatrick:2014:122} and \citet{faherty:2016:10} are shown with the dotted red line and solid black line, respectively. Unresolved binaries in the sample are indicated with red circles. Spectral type uncertainties are typically $\pm$1 spectral type. Also plotted are 275 objects from the Database of Ultracool Parallaxes (excluding binaries) with spectral types $\geqslant$ L0 and measured $W2$ magnitudes (cyan points). The blue arrow and corresponding blue marker and errorbar show the best single fit object (T2 dwarf) for the potential spectral binary 2MASS~J1324$+$6358.
\label{fig:absw2}}
\end{figure}

	Figure~\ref{fig:absw2} shows the absolute $W2$ magnitude as a function of spectral type for ultracool dwarfs. The vast majority of sources measured in this study follow the expected empirical relationships from \citet[red dashed line;][]{kirkpatrick:2014:122} and \citet[solid black line with gray uncertainty region;][]{faherty:2016:10}, with a few known exceptions (e.g., the overluminous Luhman 16AB). Additionally, the spectral binary candidate 2MASS~J13243553$+$6358281 \citep[L8$+$T3.5;][hereafter 2MASS~J1324$+$6358]{burgasser:2010:1142} appears underluminous for the spectral type of the primary component. Recent results suggest that 2MASS~J1324$+$6358 is a single T2 dwarf and member of the young ($\sim$150 Myr) AB Doradus moving group \citep{gagne:2018:l27}, consistent with the parallax measurement in this study.
	
	Figure~\ref{fig:absw2} demonstrates the utility of trigonometric parallaxes for identifying overluminous, unresolved binaries---similar to Luhman 16AB---and low surface gravity brown dwarfs, the latter of which are also useful for determining youth and a helpful diagnostic for membership in nearby young moving groups. Future work will focus on measuring parallaxes for all ultracool dwarfs within 17~pc without published trigonometric parallax measurements, and is applicable to any future discoveries of ultracool dwarfs.
 
	Additionally, it may be possible to use unWISE \citep{lang:2014:108,meisner:2017:38,meisner:2017:161} single-epoch coadds \citep{meisner:2017:} for higher precision astrometric measurements of the faintest sources (e.g., Y dwarfs). Currently, not all epochs of NEOWISE-R data have been processed through unWISE, and future work will wait until all epochs become available.

\acknowledgments

	First and foremost, C.A.T would like to thank the anonymous referee for helpful comments that greatly improved the quality of this manuscript. C.A.T would also like to give his sincerest thanks to Adam Burgasser, Julie Skinner, Aurora Kesseli, Andrew West, Jonathan Gagn\'e, and Aaron Meisner for providing useful and insightful conversations, without which this manuscript never would have came to fruition. This material is based upon work supported by the National Aeronautics and Space Administration under Grant No. NNX16AF47G issued through the Astrophysics Data Analysis Program.
 
	This publication makes use of data products from the \textit{Wide-field Infrared Survey Explorer}, which is a joint project of the University of California, Los Angeles, and the Jet Propulsion Laboratory/California Institute of Technology, funded by the National Aeronautics and Space Administration. 
	
	Funding for the Sloan Digital Sky Survey IV has been provided by the Alfred P. Sloan Foundation, the U.S. Department of Energy Office of Science, and the Participating Institutions. SDSS-IV acknowledges
support and resources from the Center for High-Performance Computing at
the University of Utah. The SDSS web site is \url{www.sdss.org}.

SDSS-IV is managed by the Astrophysical Research Consortium for the 
Participating Institutions of the SDSS Collaboration including the 
Brazilian Participation Group, the Carnegie Institution for Science, 
Carnegie Mellon University, the Chilean Participation Group, the French Participation Group, Harvard-Smithsonian Center for Astrophysics, 
Instituto de Astrof\'isica de Canarias, The Johns Hopkins University, 
Kavli Institute for the Physics and Mathematics of the Universe (IPMU) / 
University of Tokyo, Lawrence Berkeley National Laboratory, 
Leibniz Institut f\"ur Astrophysik Potsdam (AIP), 
Max-Planck-Institut f\"ur Astronomie (MPIA Heidelberg), 
Max-Planck-Institut f\"ur Astrophysik (MPA Garching), 
Max-Planck-Institut f\"ur Extraterrestrische Physik (MPE), 
National Astronomical Observatories of China, New Mexico State University, 
New York University, University of Notre Dame, 
Observat\'ario Nacional / MCTI, The Ohio State University, 
Pennsylvania State University, Shanghai Astronomical Observatory, 
United Kingdom Participation Group,
Universidad Nacional Aut\'onoma de M\'exico, University of Arizona, 
University of Colorado Boulder, University of Oxford, University of Portsmouth, 
University of Utah, University of Virginia, University of Washington, University of Wisconsin, 
Vanderbilt University, and Yale University.
	
	This work has made use of data from the European Space Agency (ESA)
mission {\it Gaia} (\url{https://www.cosmos.esa.int/gaia}), processed by
the {\it Gaia} Data Processing and Analysis Consortium (DPAC,
\url{https://www.cosmos.esa.int/web/gaia/dpac/consortium}). Funding
for the DPAC has been provided by national institutions, in particular
the institutions participating in the {\it Gaia} Multilateral Agreement.
		
	This research made use of Astropy, a community-developed core Python package for Astronomy \citep{astropy-collaboration:2013:a33}. Plots in this publication were made using Matplotlib \citep{hunter:2007:90}. This research has made use of the SIMBAD database, operated at CDS, Strasbourg, France. This research has made use of NASA's Astrophysics Data System. This research has also made use of the VizieR catalogue access tool, CDS, Strasbourg, France \citep{wenger:2000:9}.

\facilities{IRSA, \textit{WISE}}.

\software{Astropy \citep{astropy-collaboration:2013:a33}, Matplotlib \citep{hunter:2007:90}, emcee \citep{foreman-mackey:2013:306}, Sublime Text}.

\bibliography{arxiv.bbl}
\bibliographystyle{aasjournal}

\begin{figure*}
\includegraphics{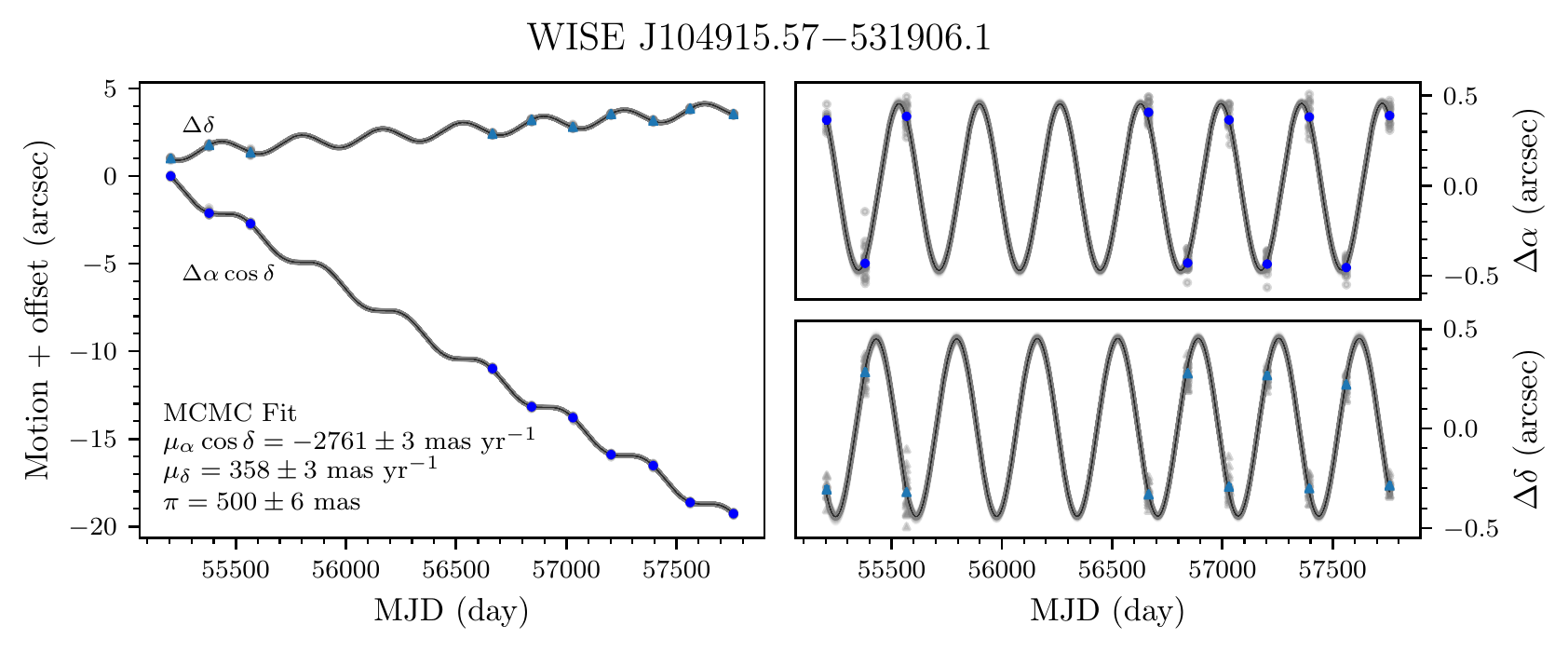}
\includegraphics{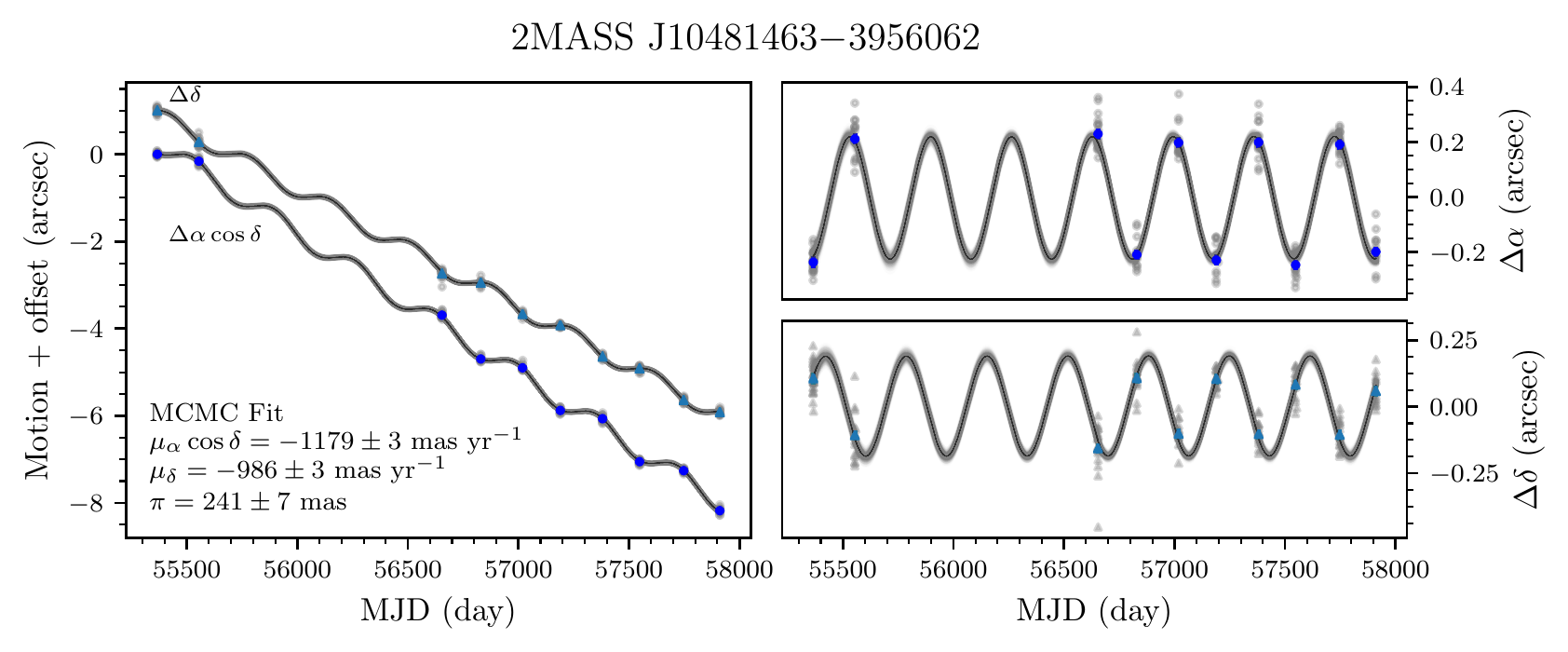}
\includegraphics{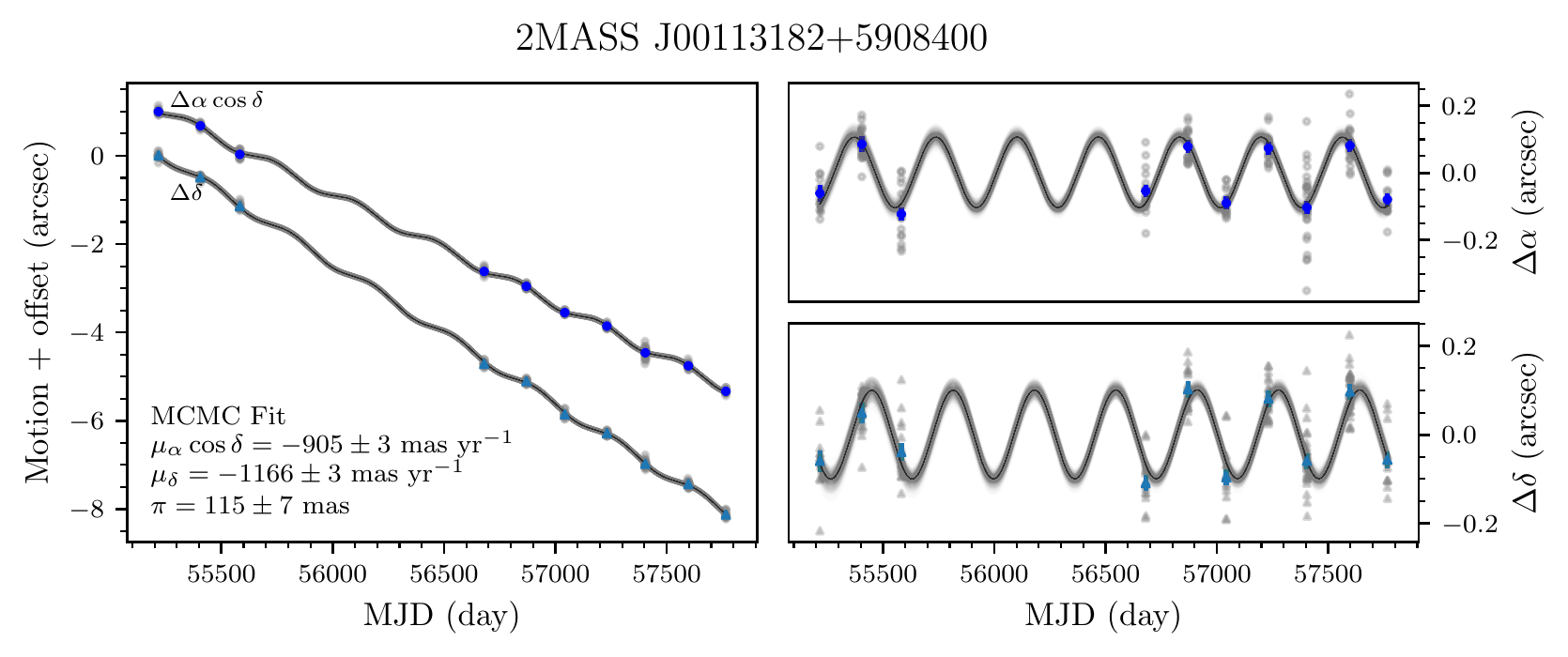}
\end{figure*}
\begin{figure*}
\includegraphics{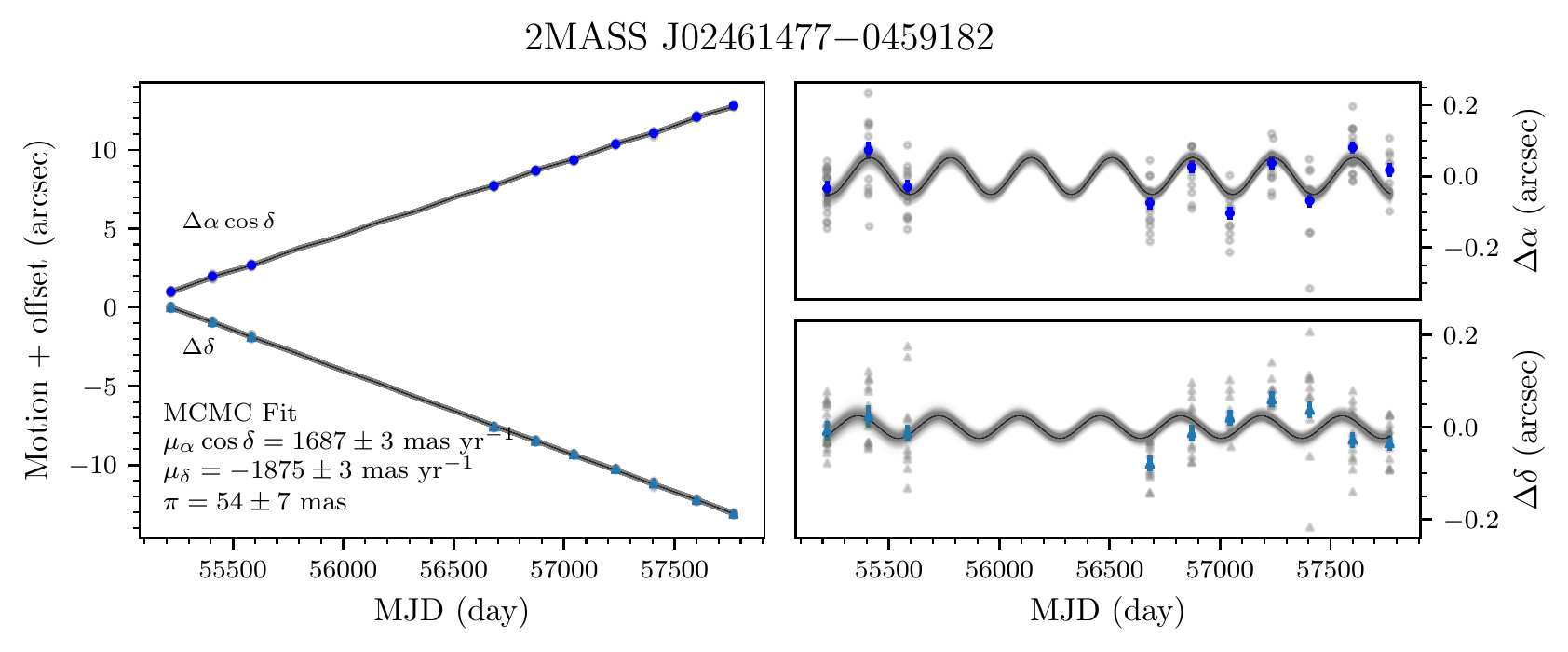}
\includegraphics{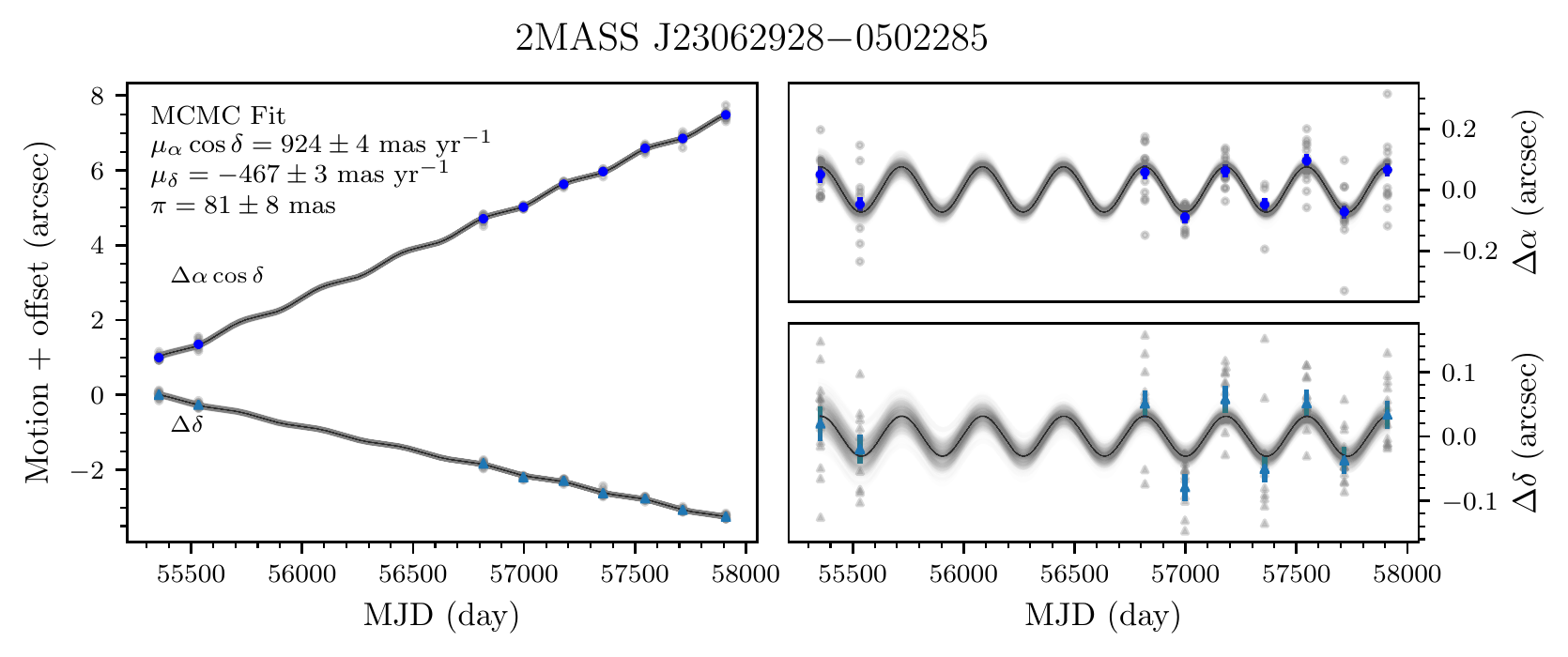}
\includegraphics{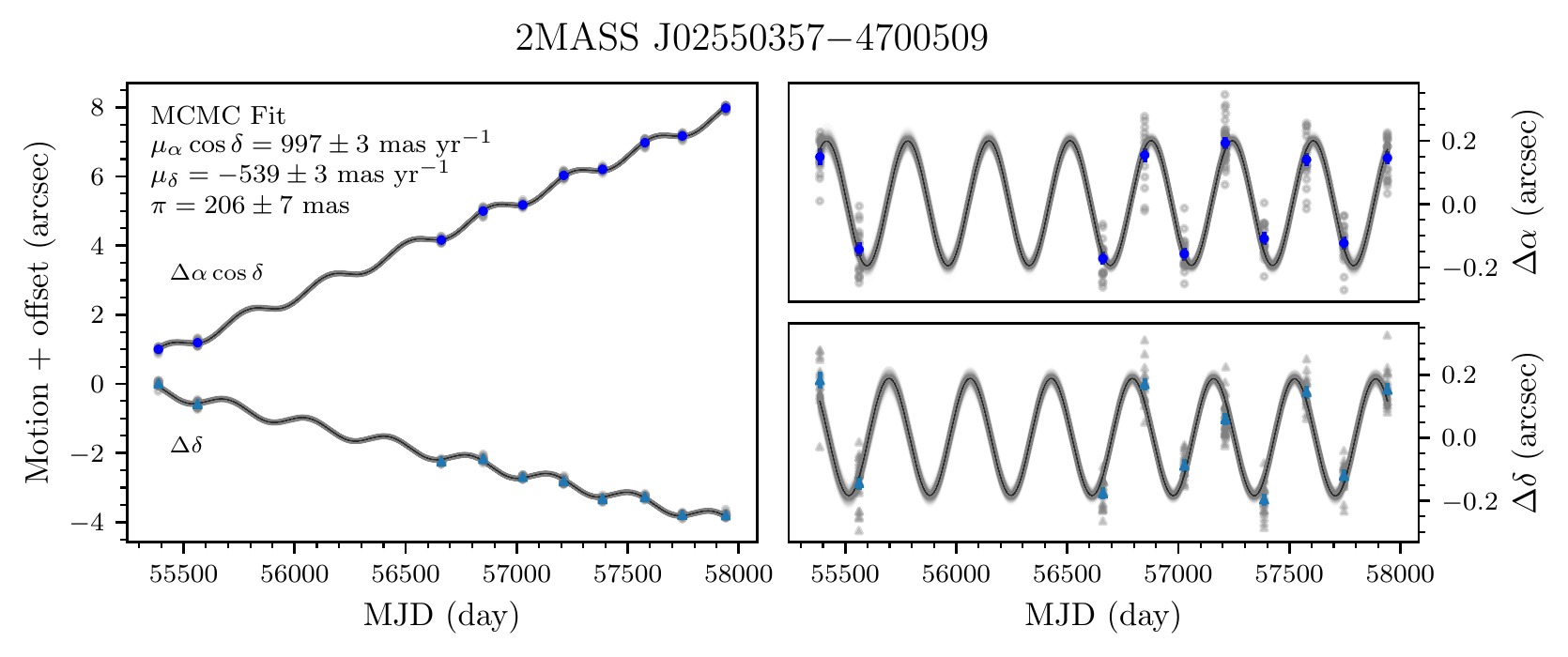}
\end{figure*}
\begin{figure*}
\includegraphics{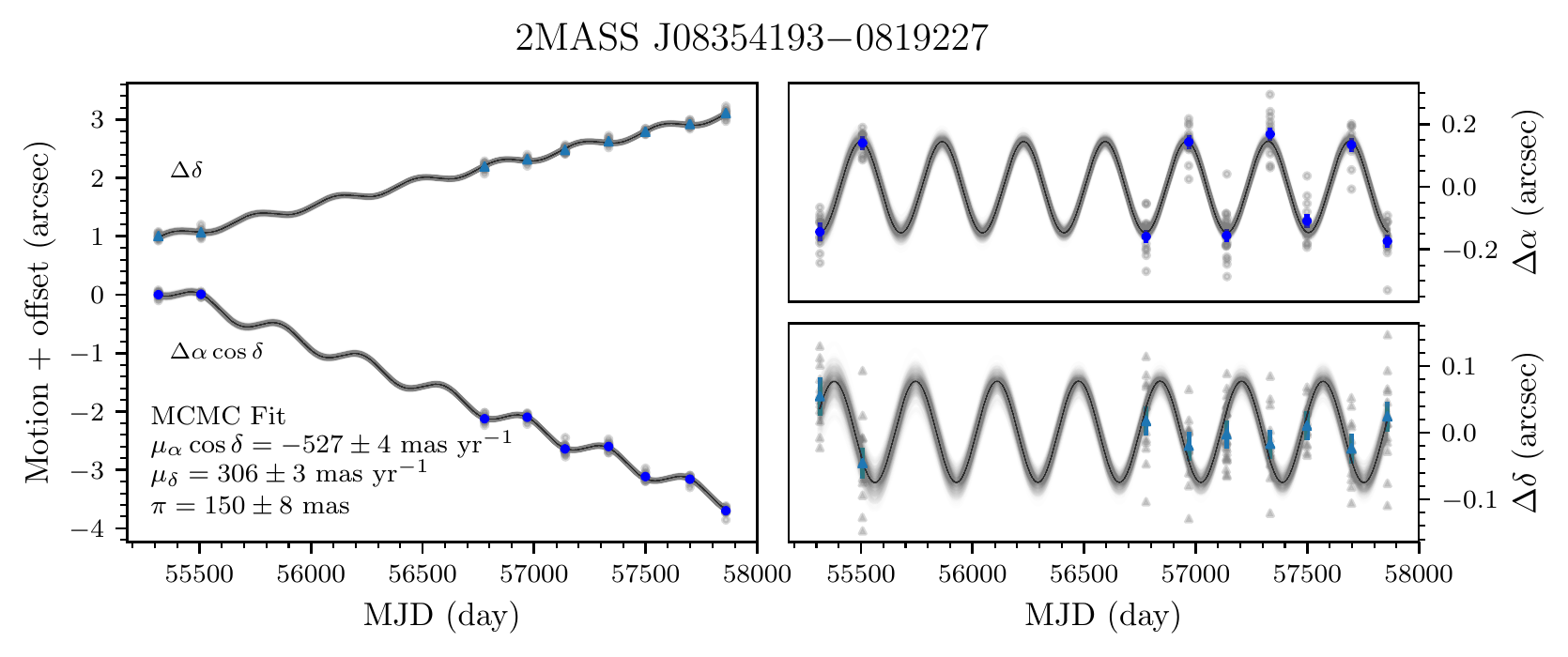}
\includegraphics{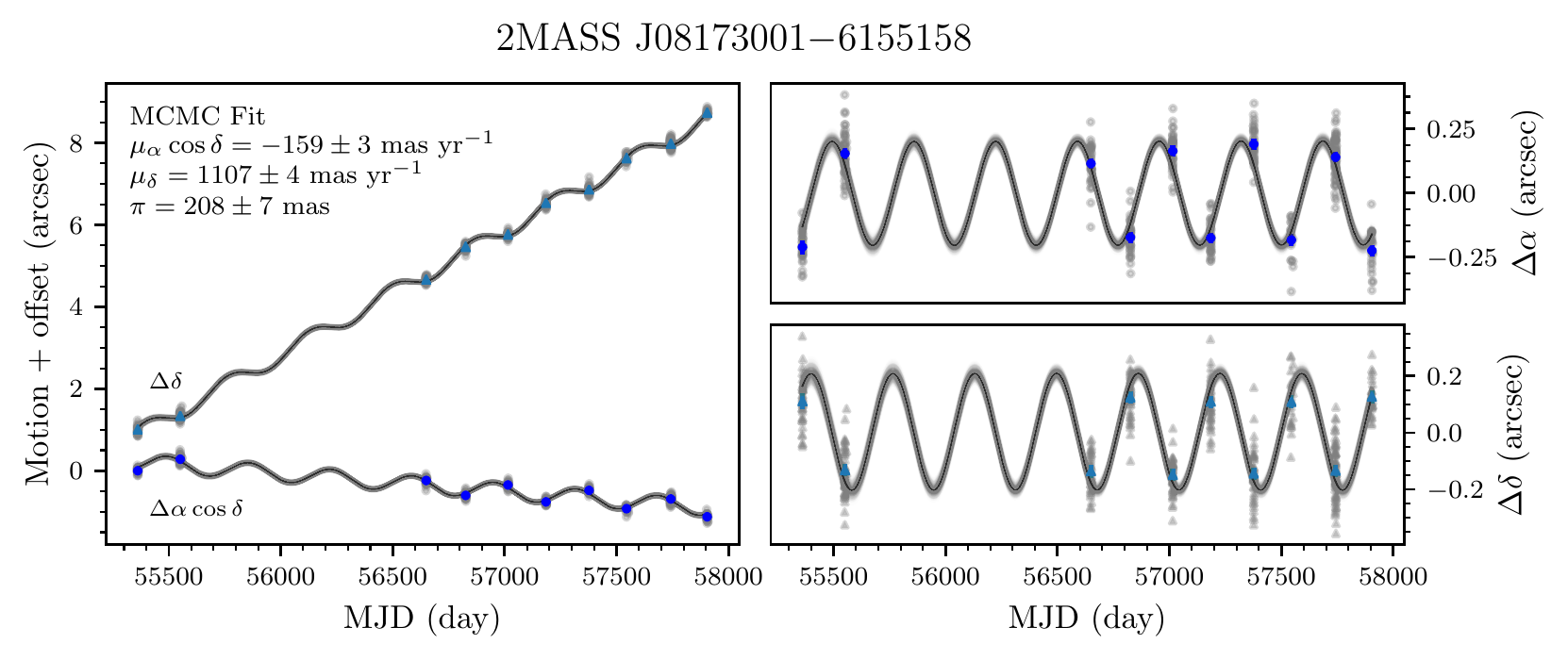}
\includegraphics{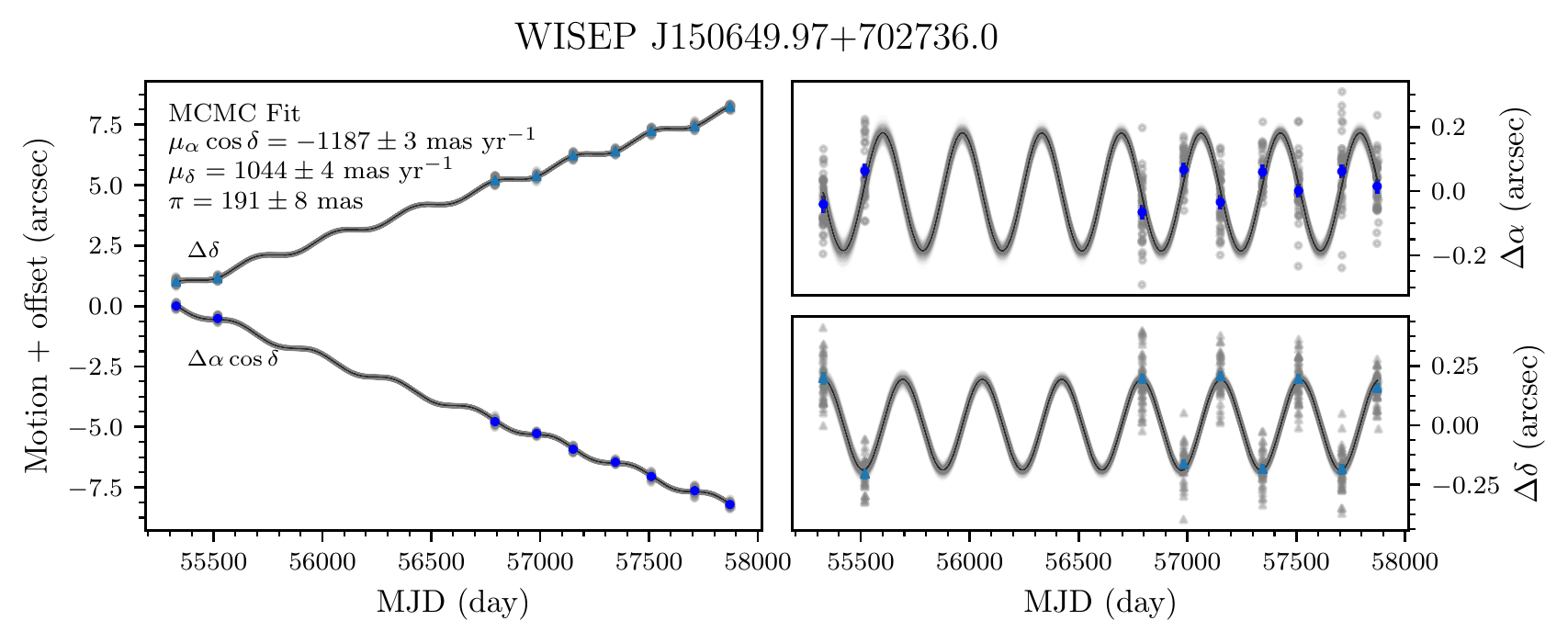}
\end{figure*}
\begin{figure*}
\includegraphics{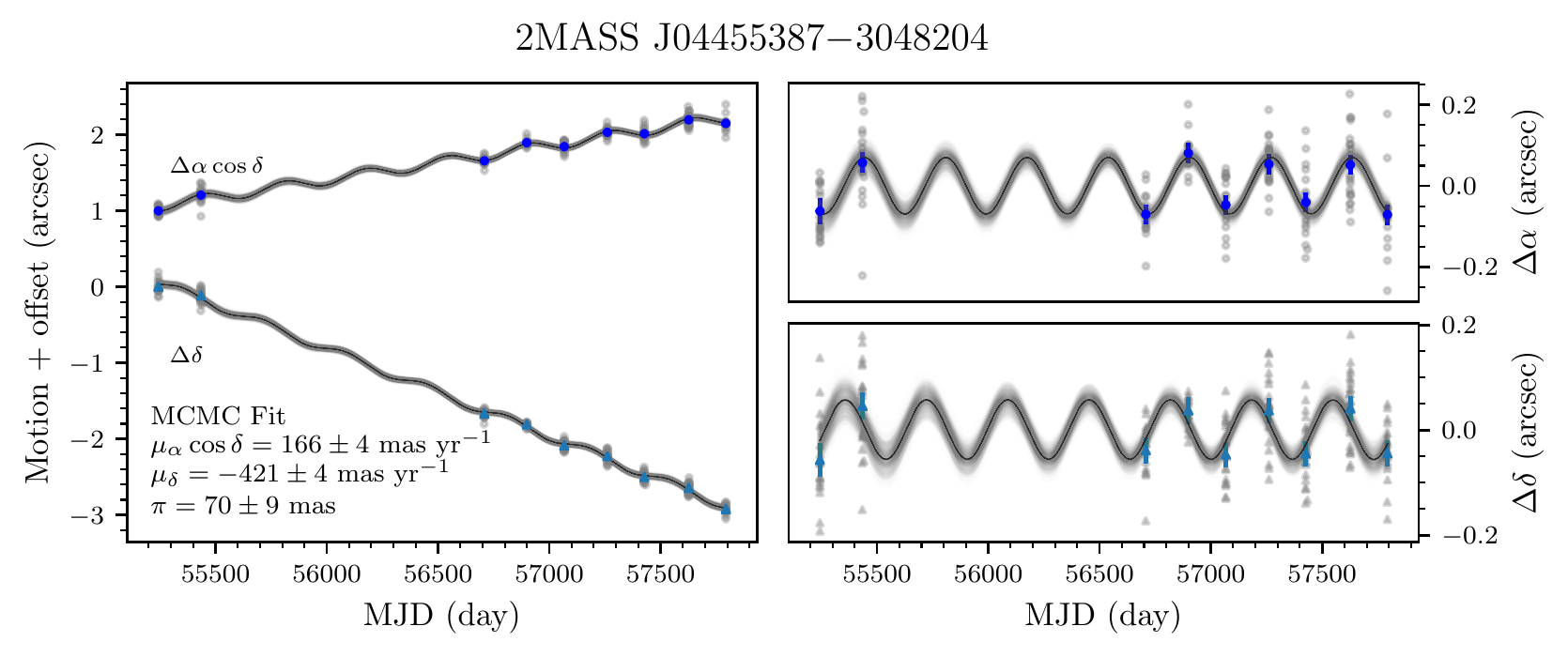}
\includegraphics{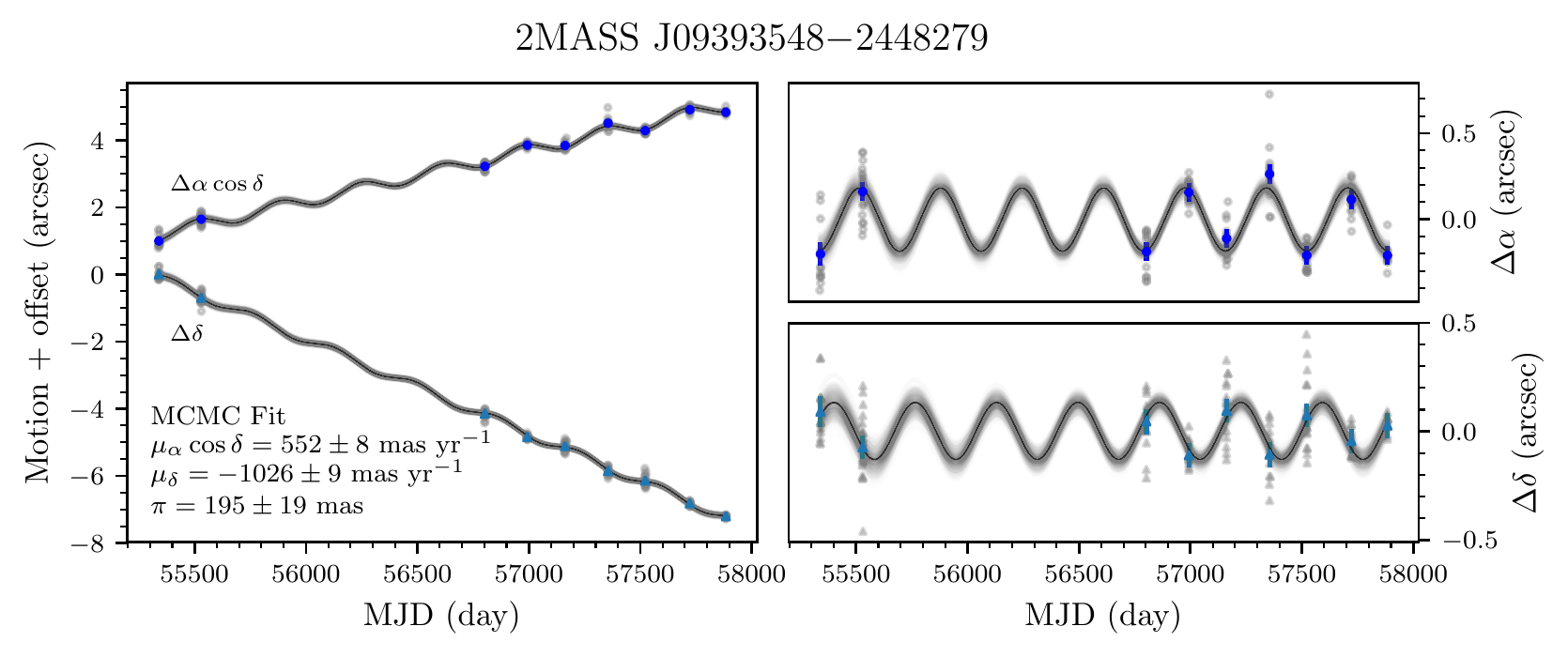}
\includegraphics{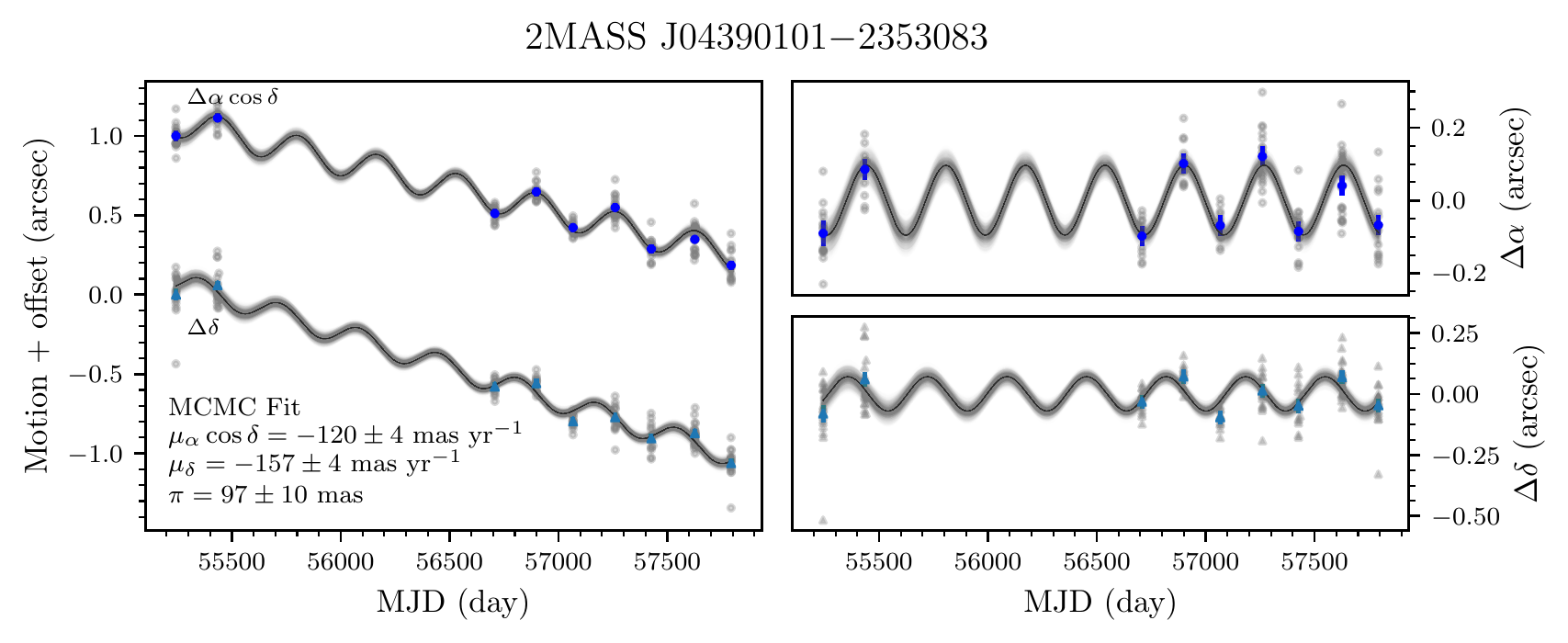}
\end{figure*}
\begin{figure*}
\includegraphics{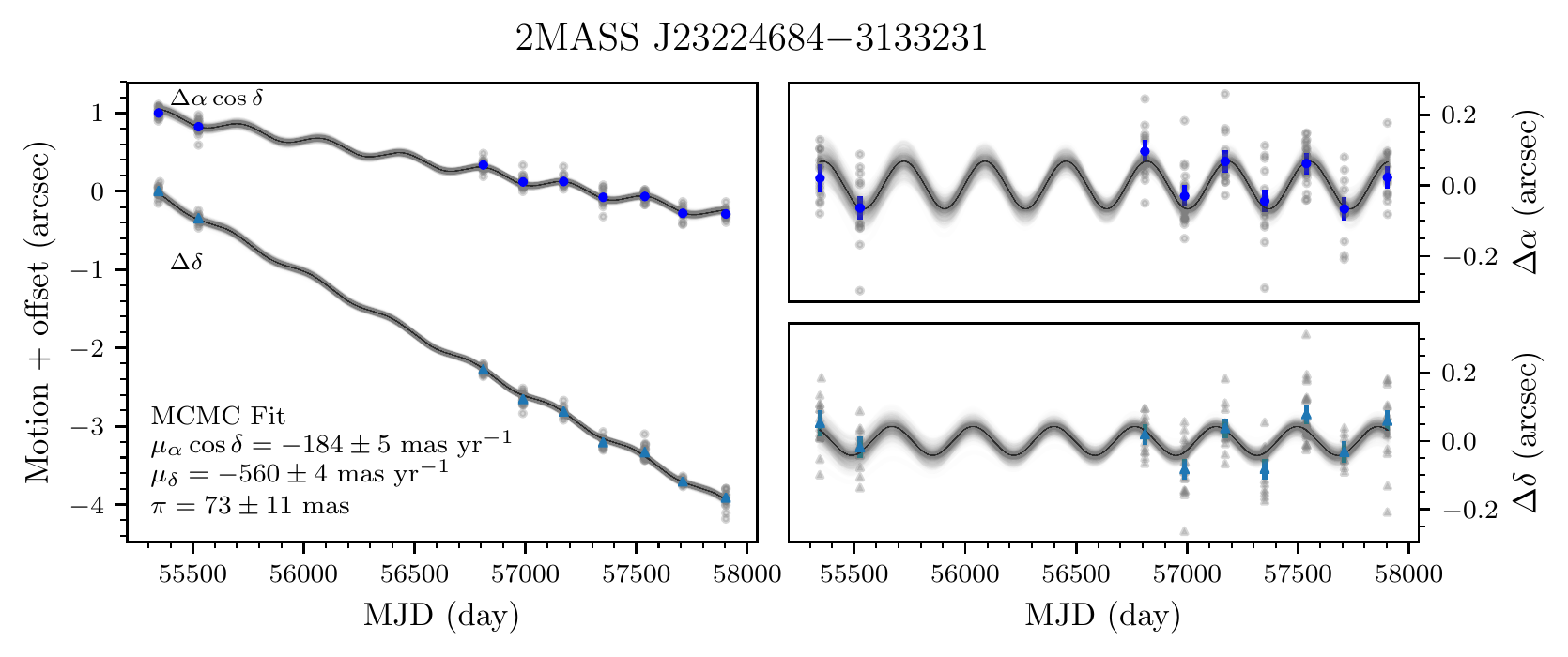}
\includegraphics{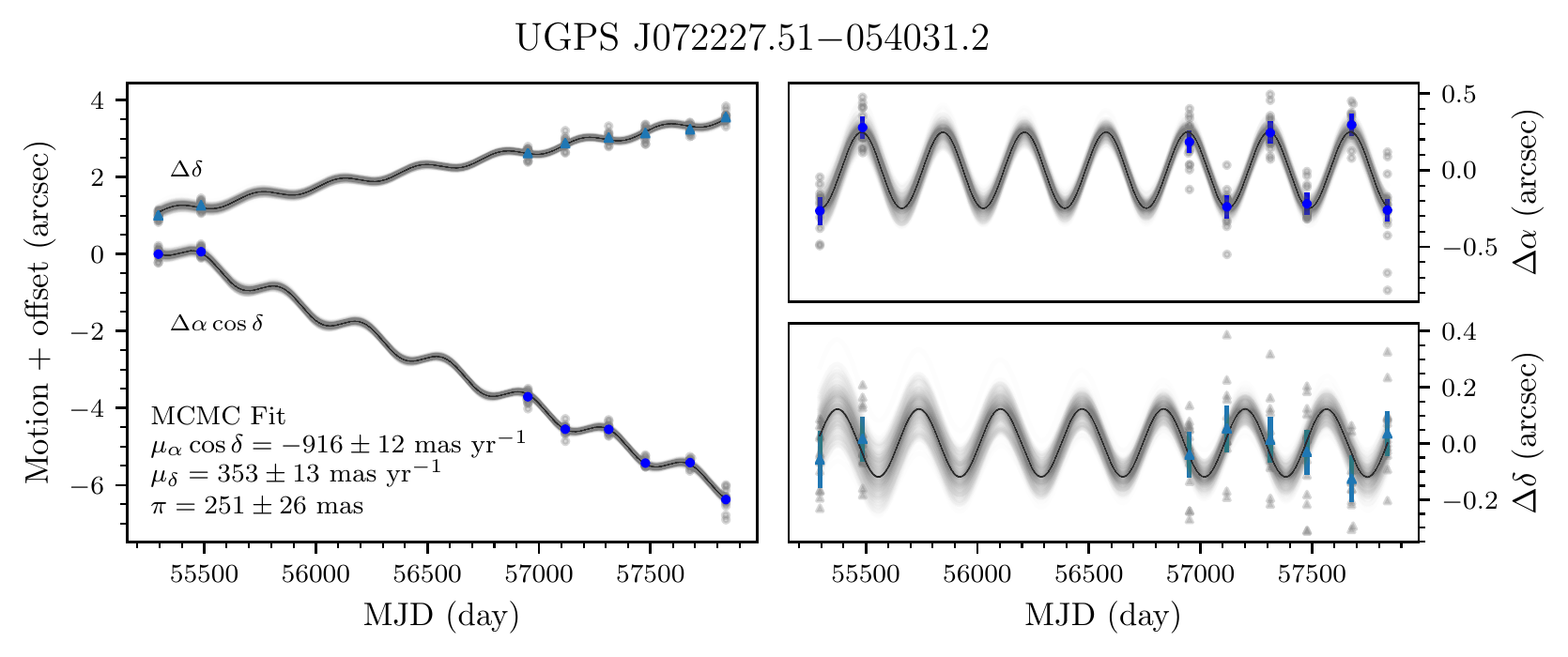}
\includegraphics{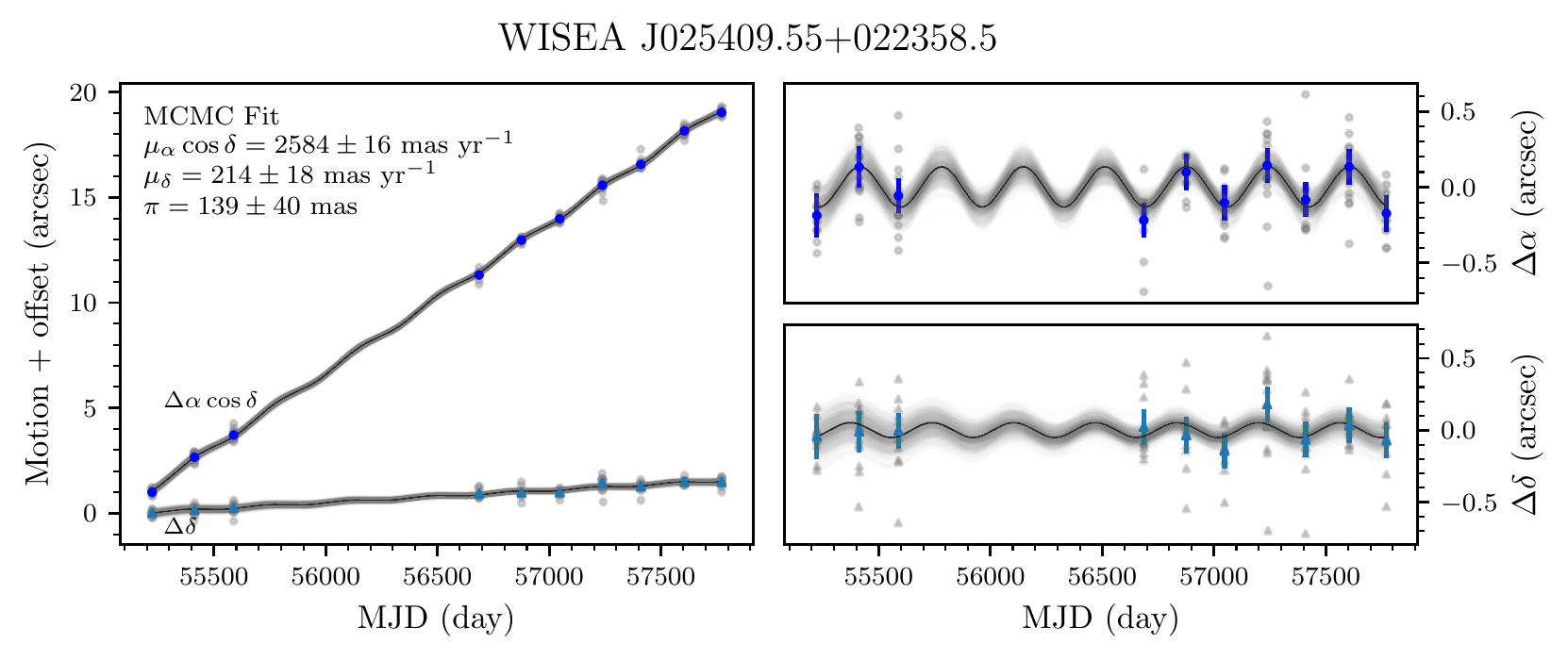}
\end{figure*}
\begin{figure*}
\includegraphics{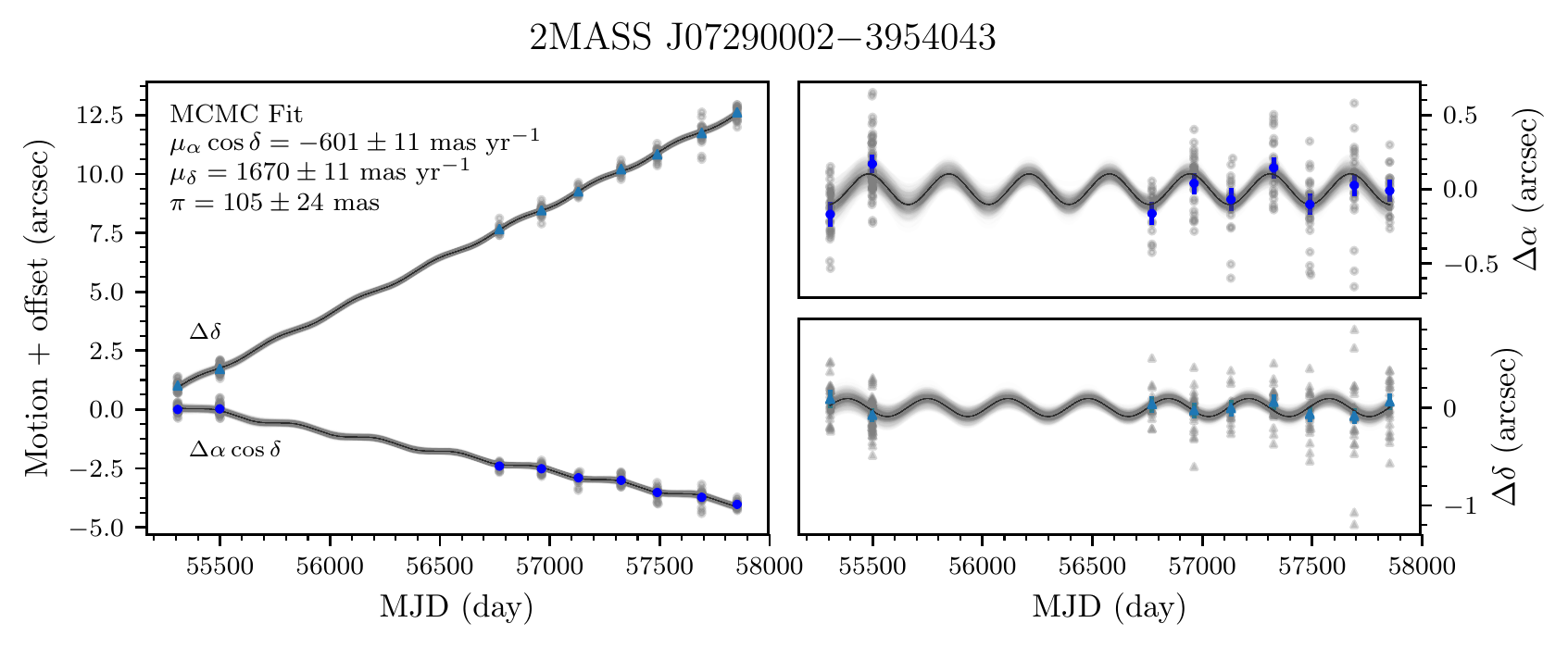}
\includegraphics{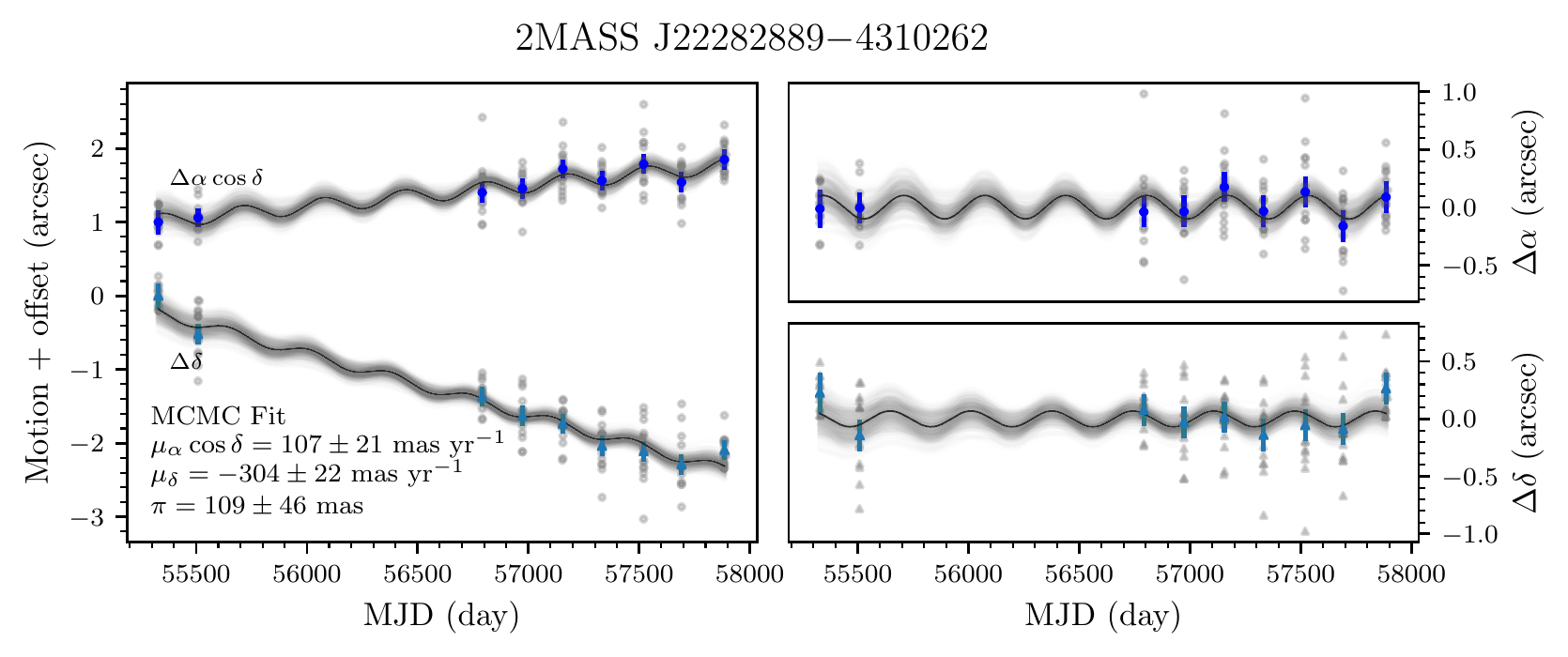}
\includegraphics{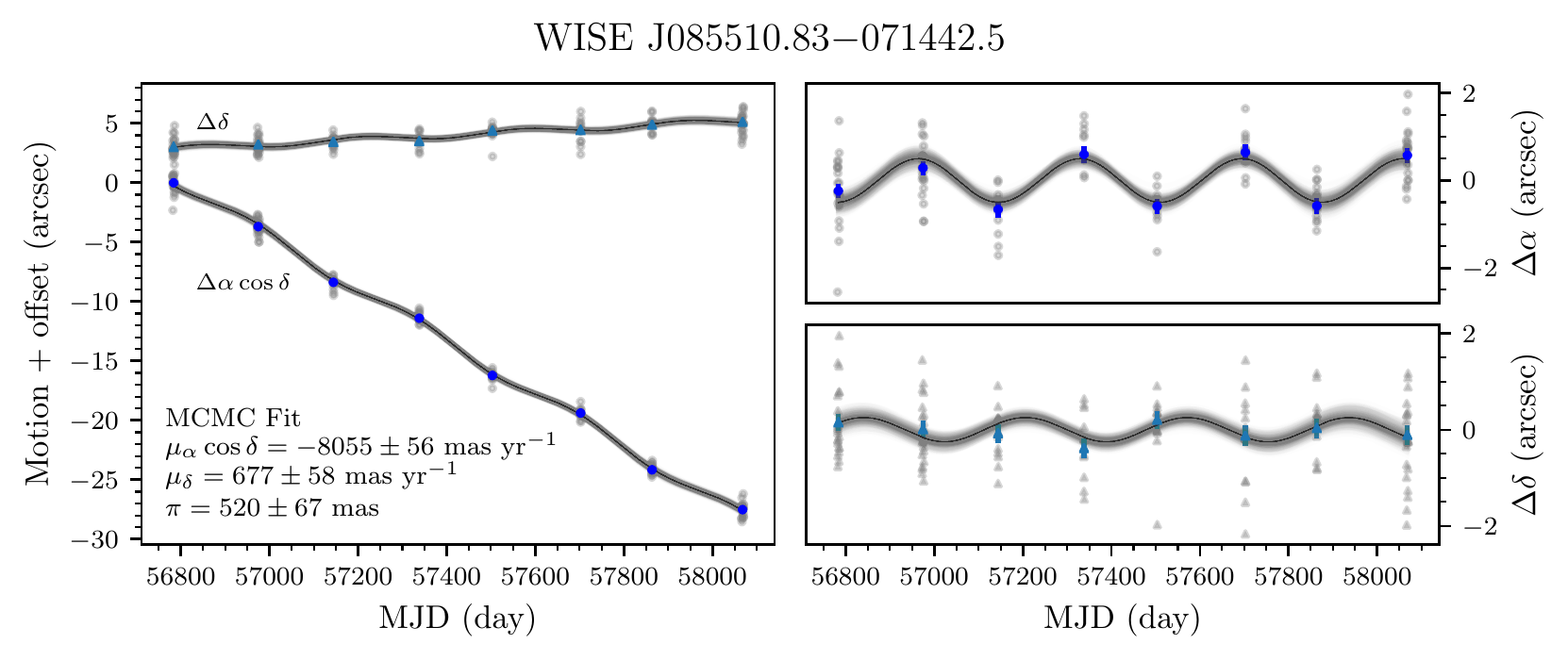}
\end{figure*}
\begin{figure*}
\includegraphics{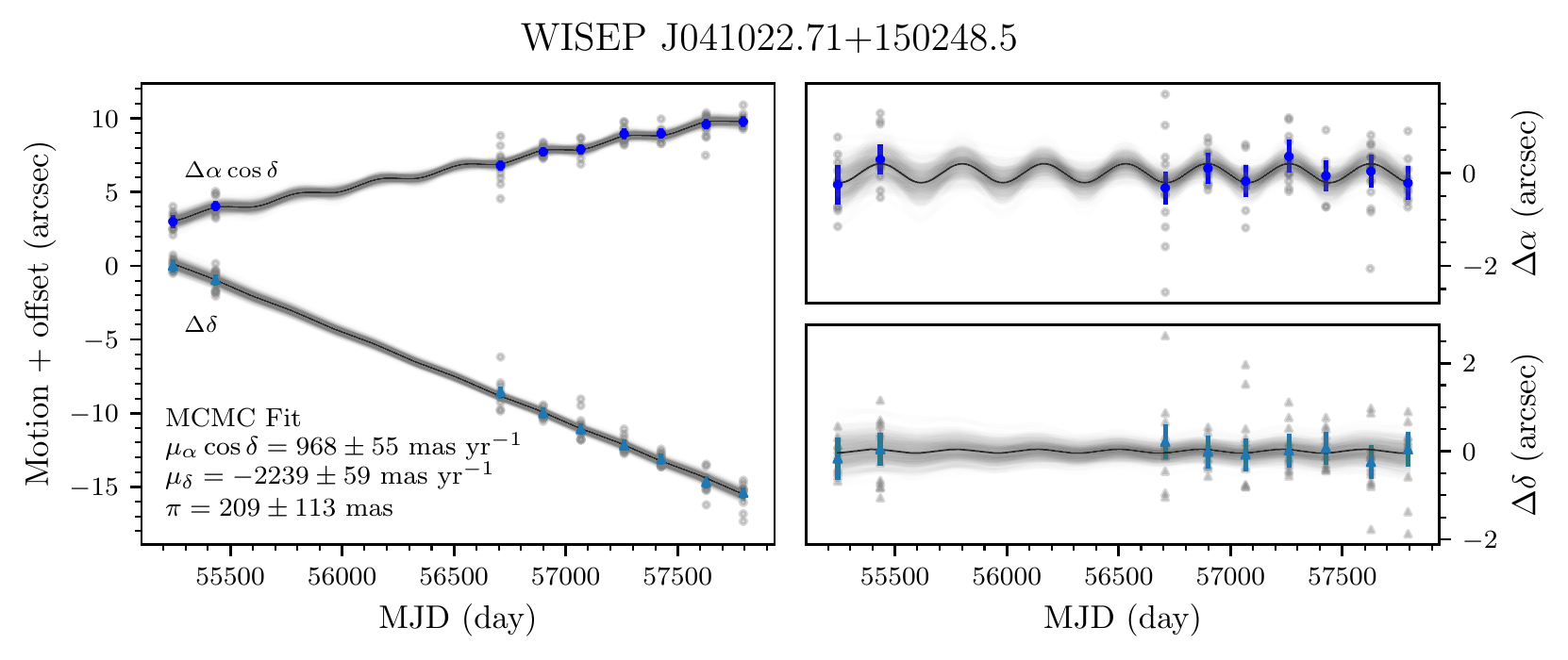}
\includegraphics{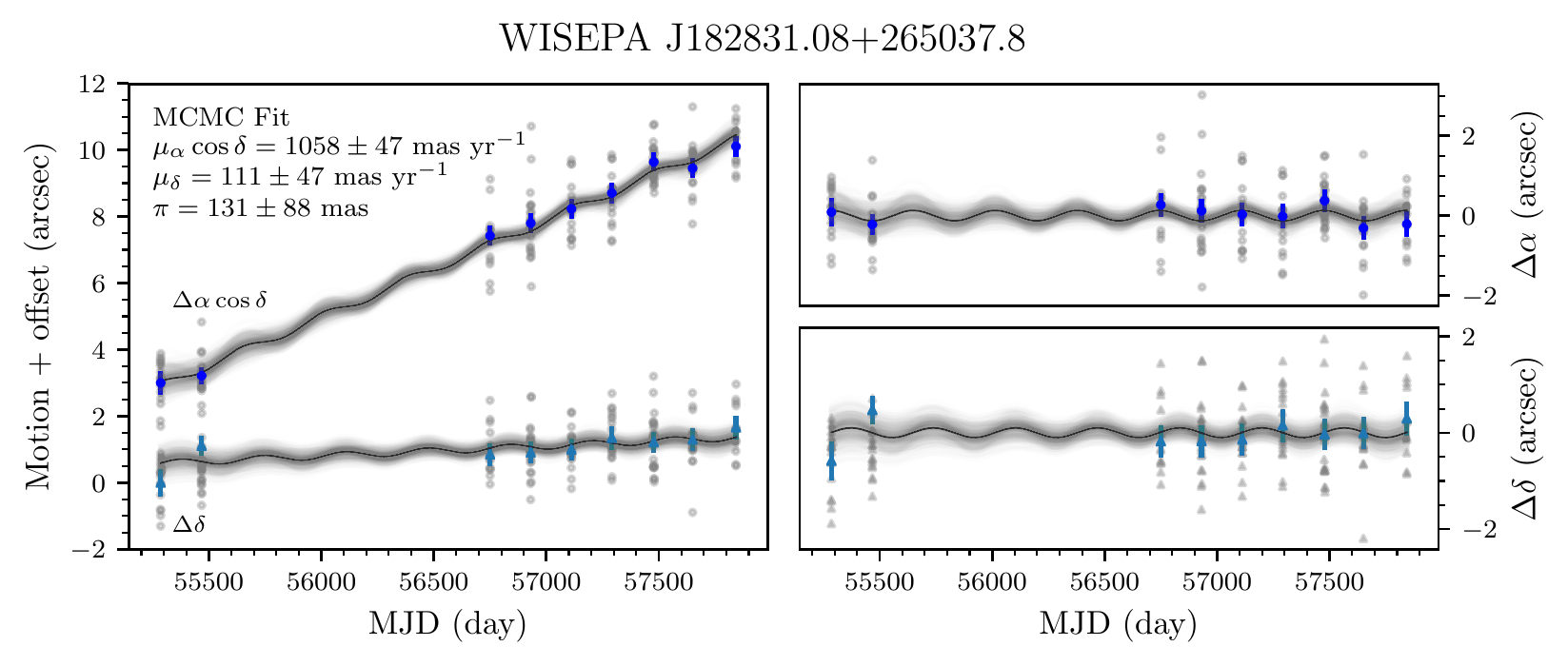}
\includegraphics{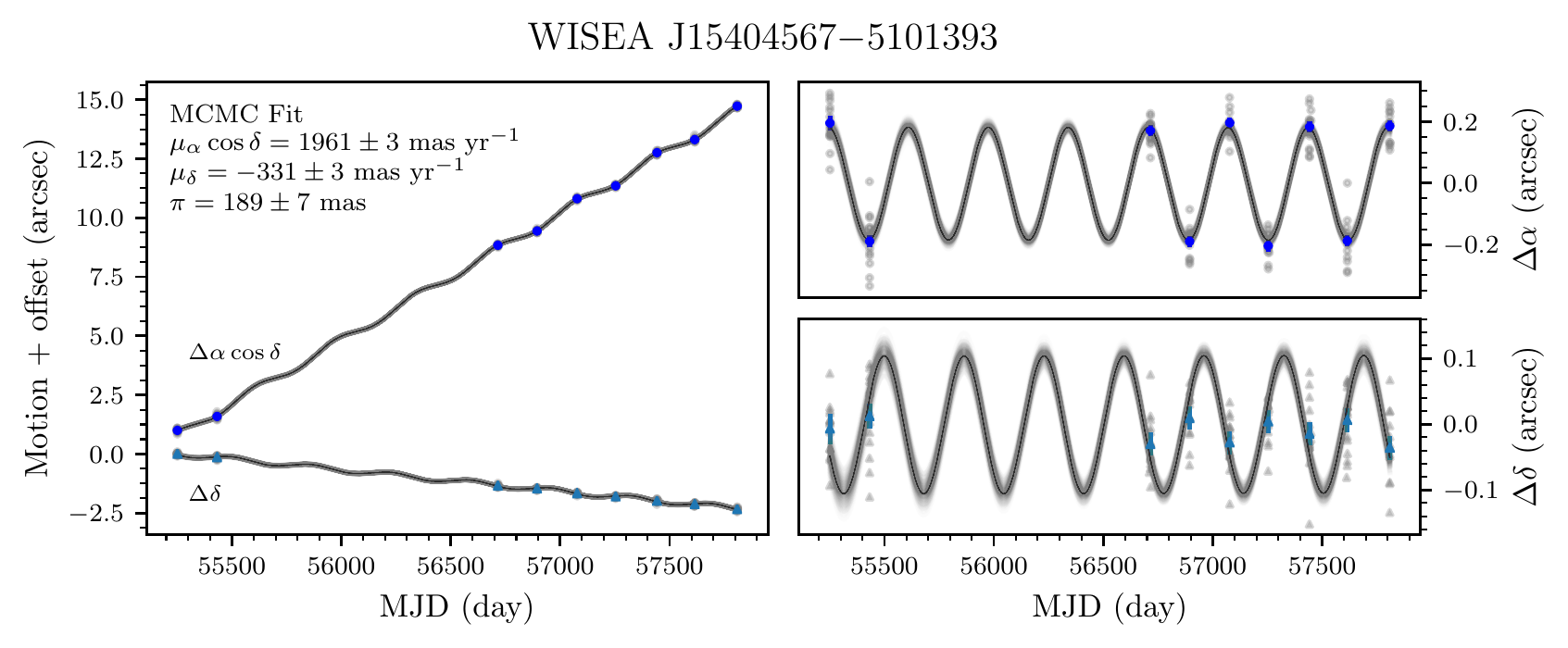}
\end{figure*}
\begin{figure*}
\includegraphics{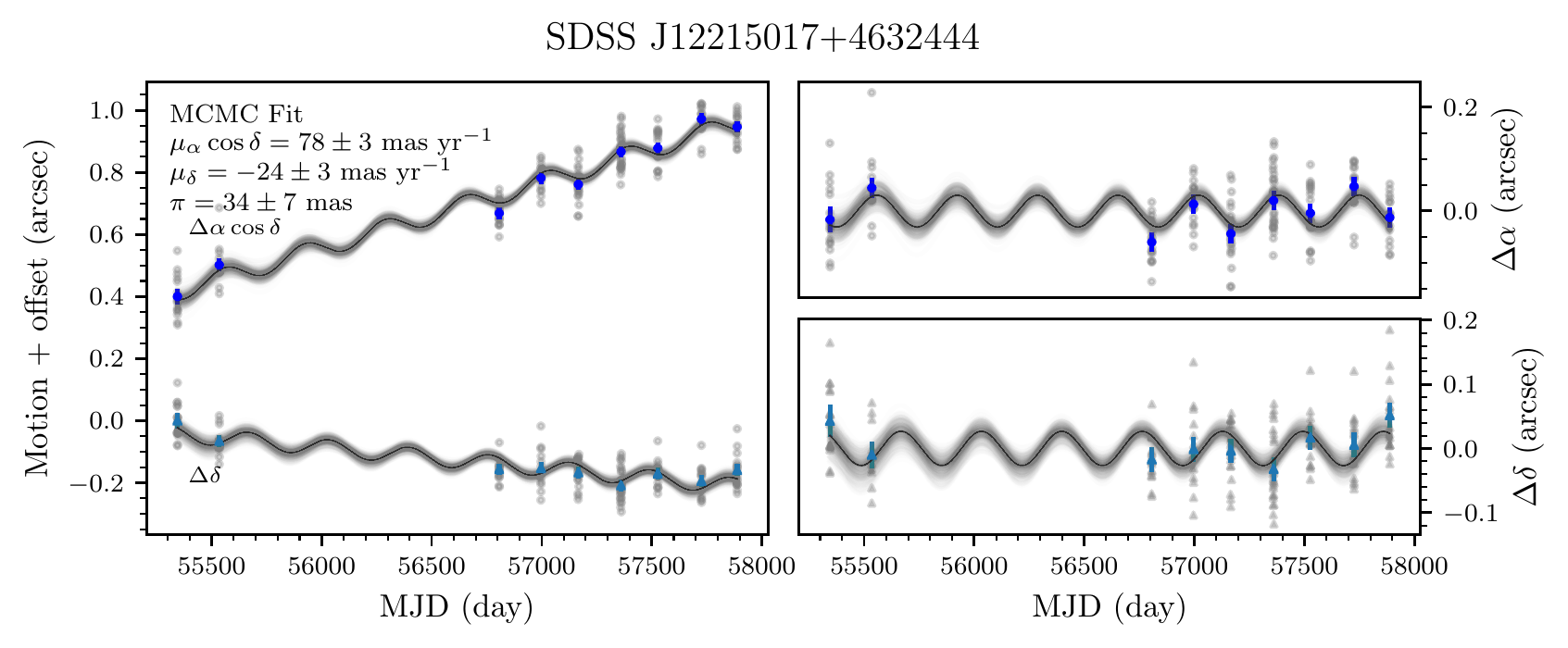}
\includegraphics{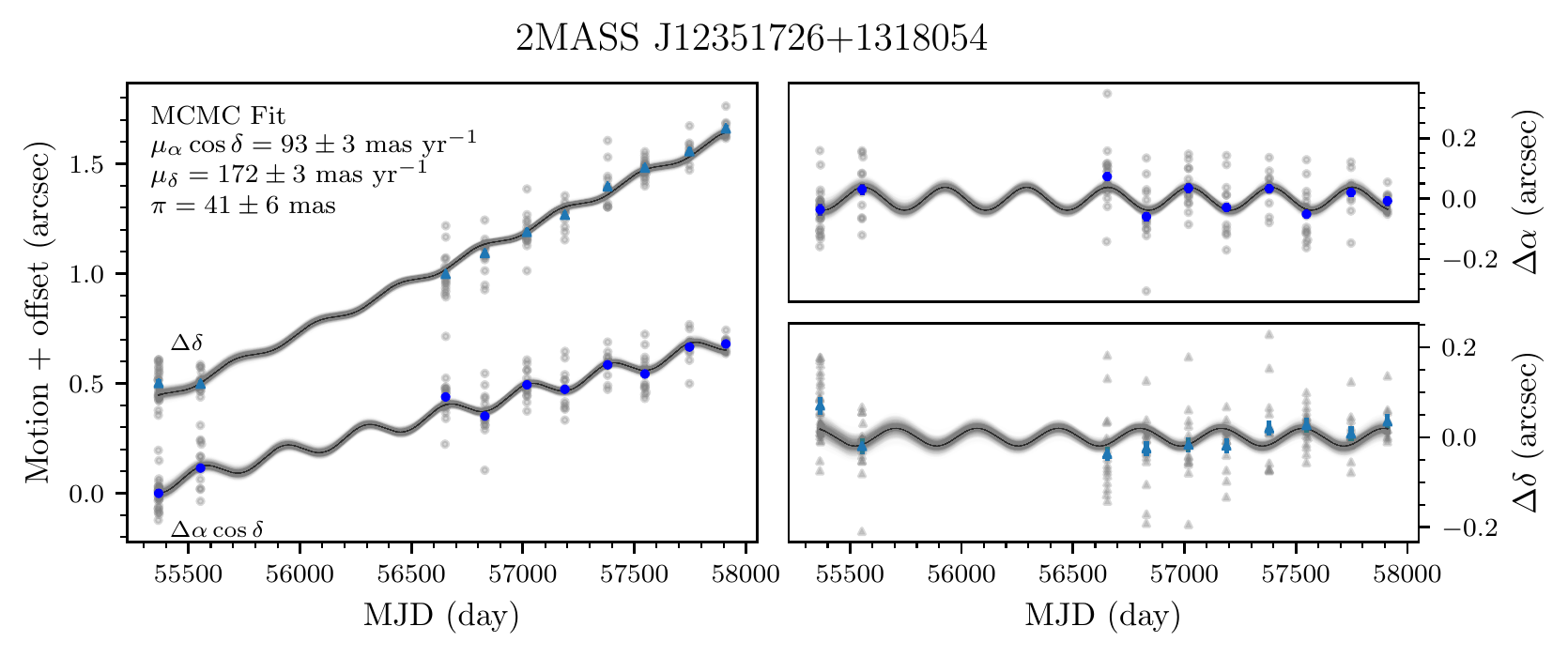}
\includegraphics{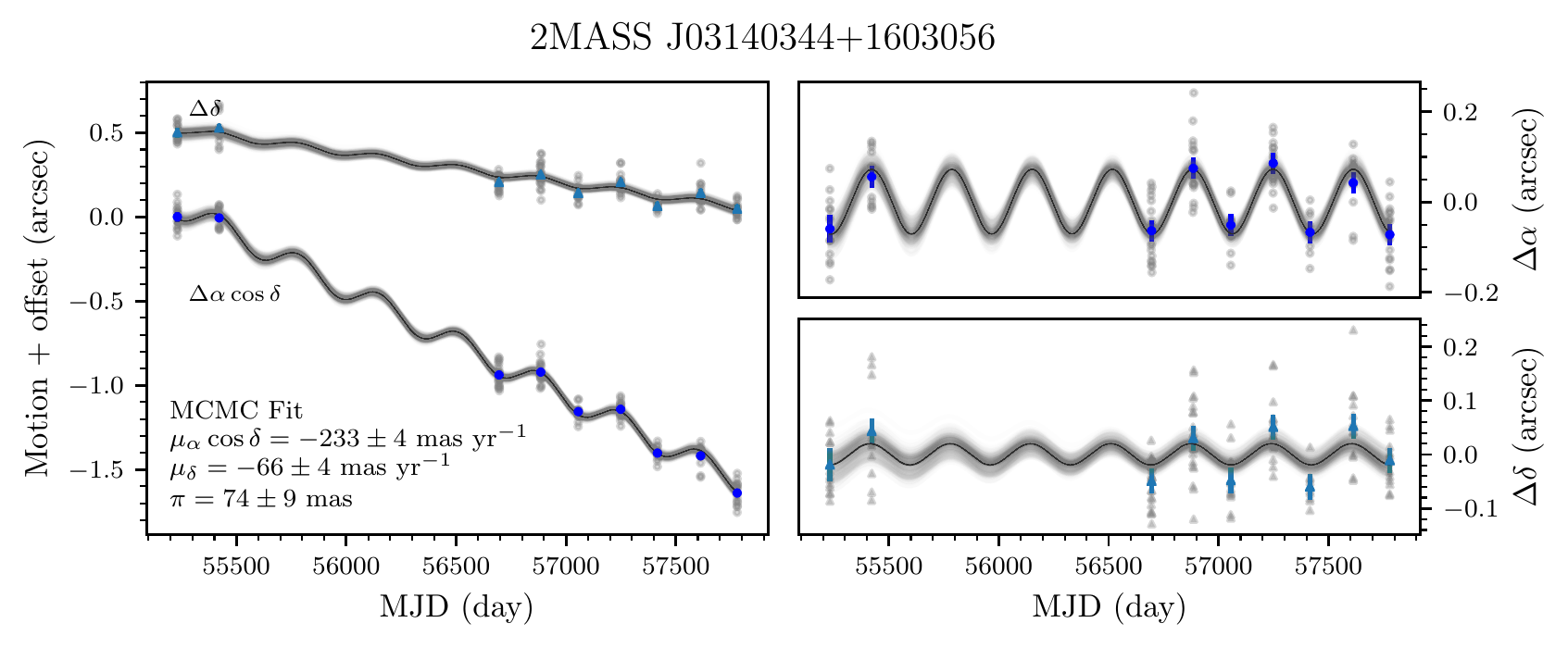}
\end{figure*}
\begin{figure*}
\includegraphics{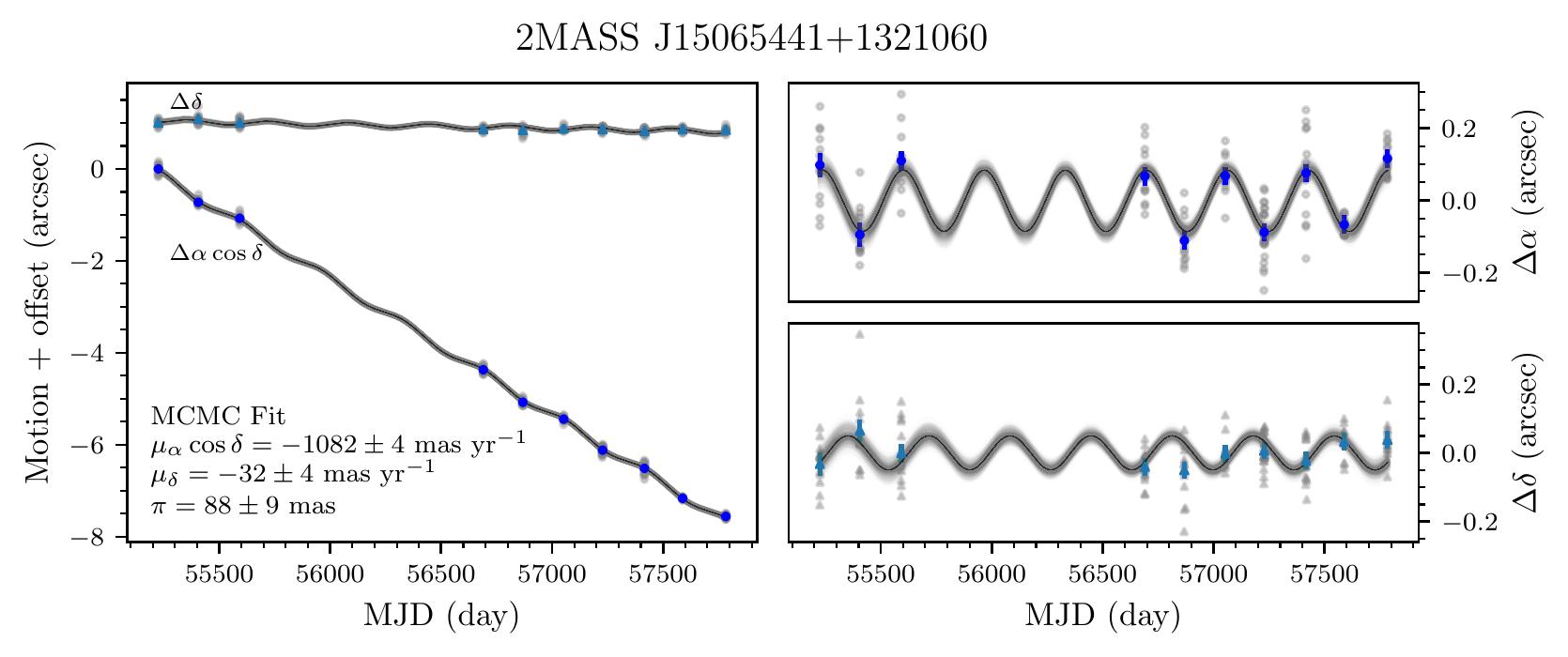}
\includegraphics{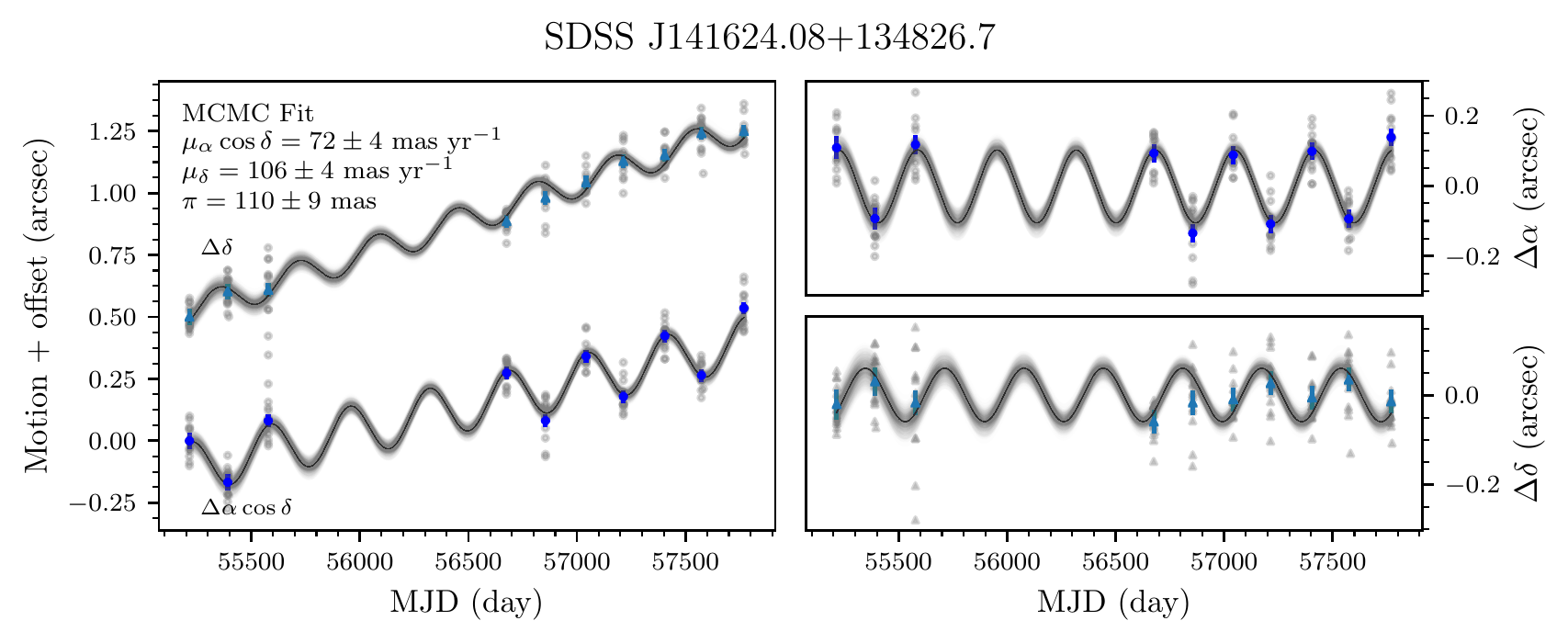}
\includegraphics{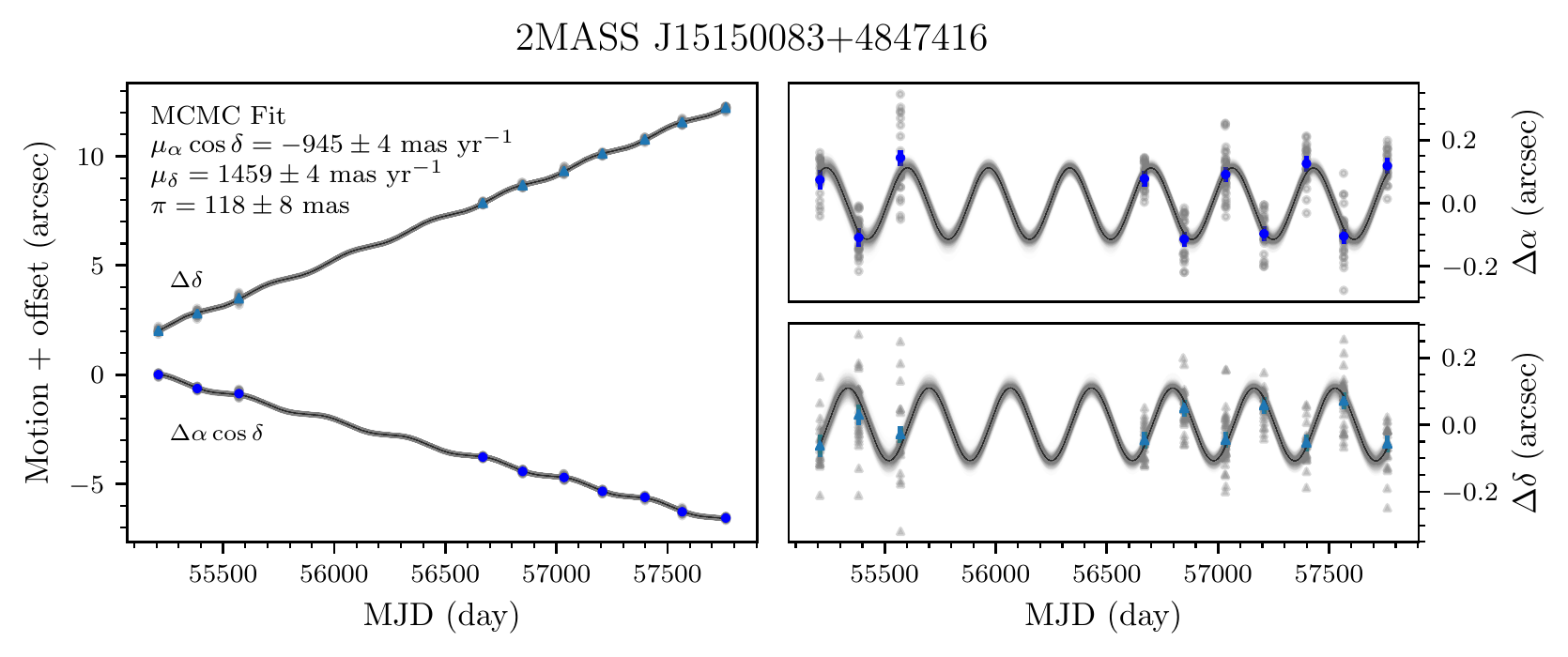}
\end{figure*}
\begin{figure*}
\includegraphics{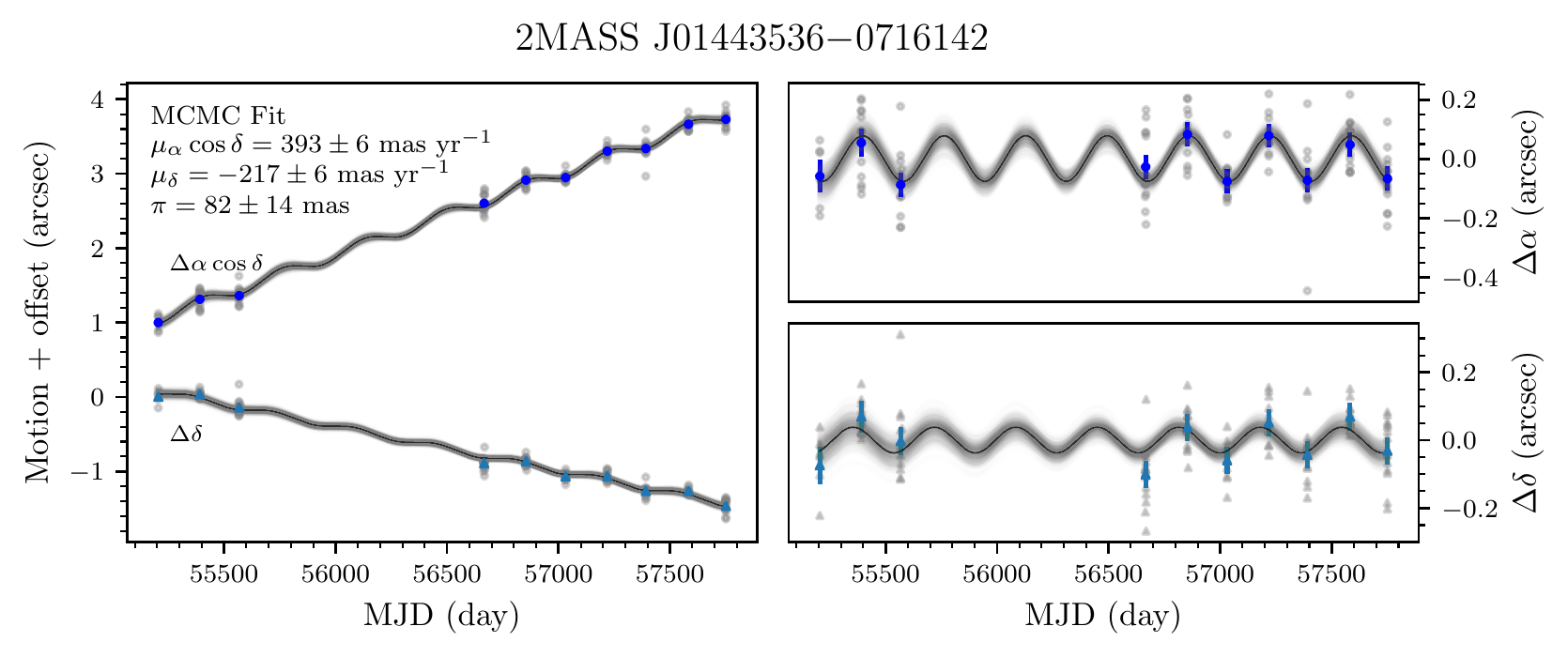}
\includegraphics{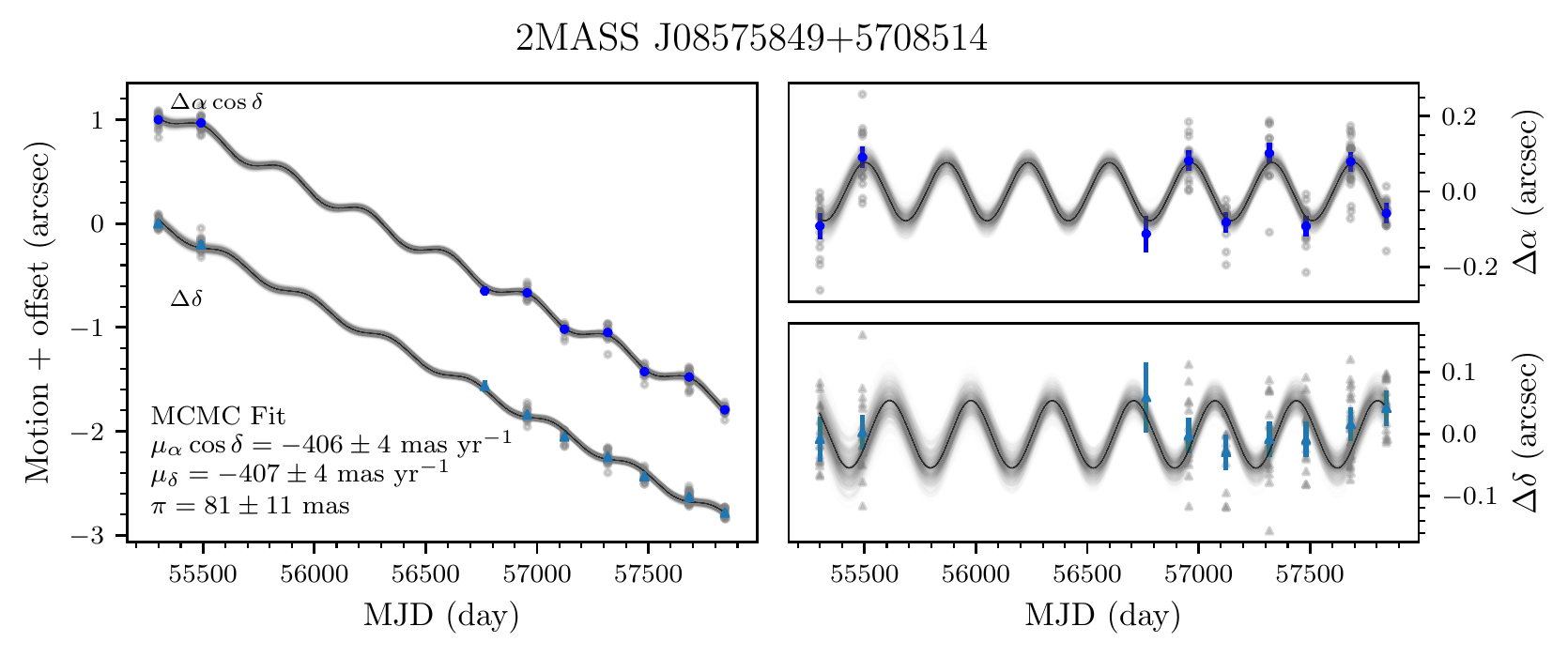}
\includegraphics{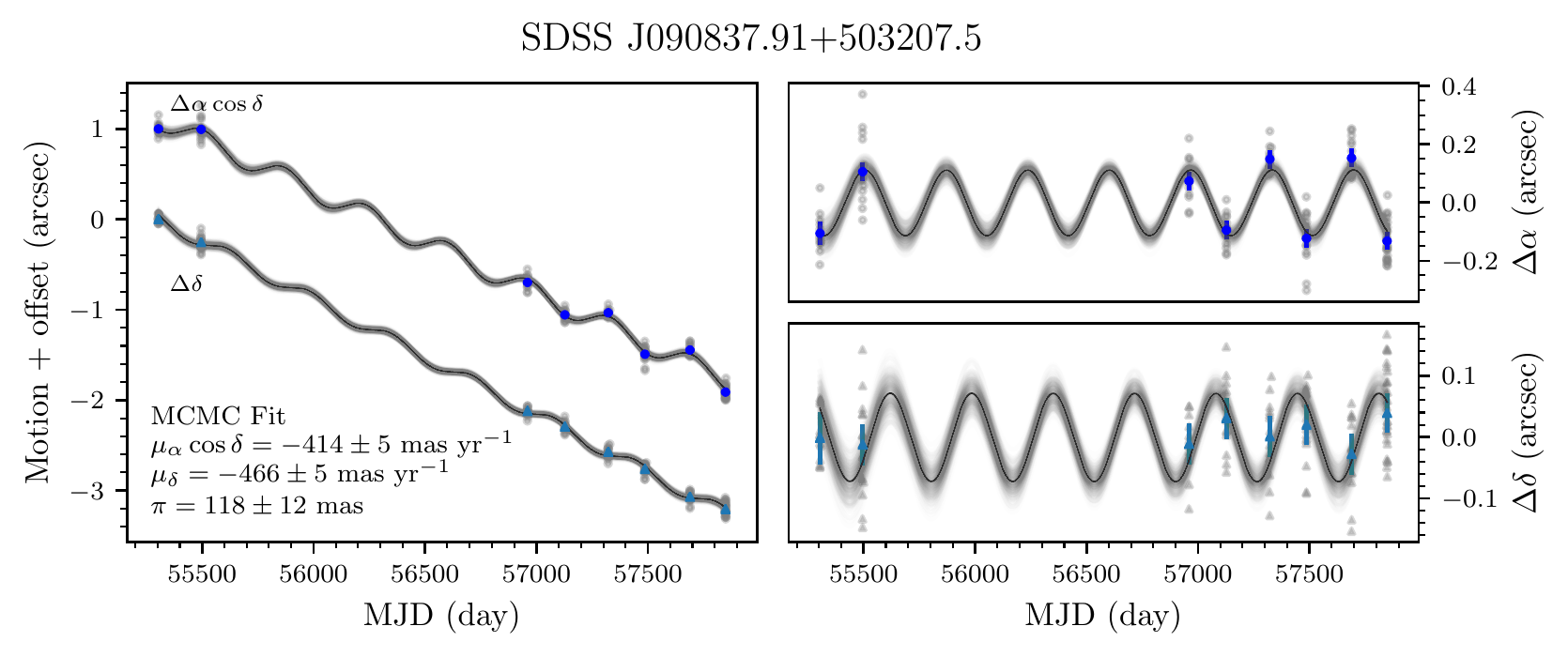}
\end{figure*}
\begin{figure*}
\includegraphics{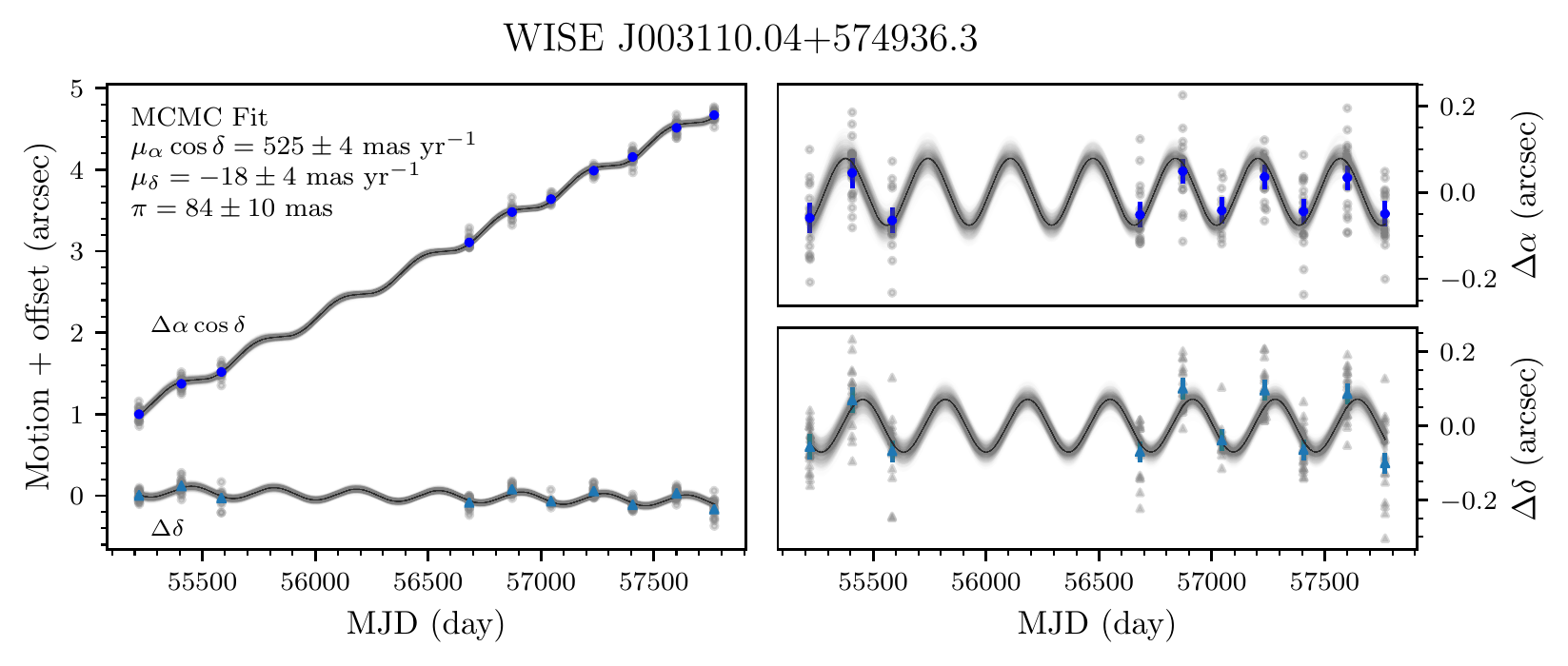}
\includegraphics{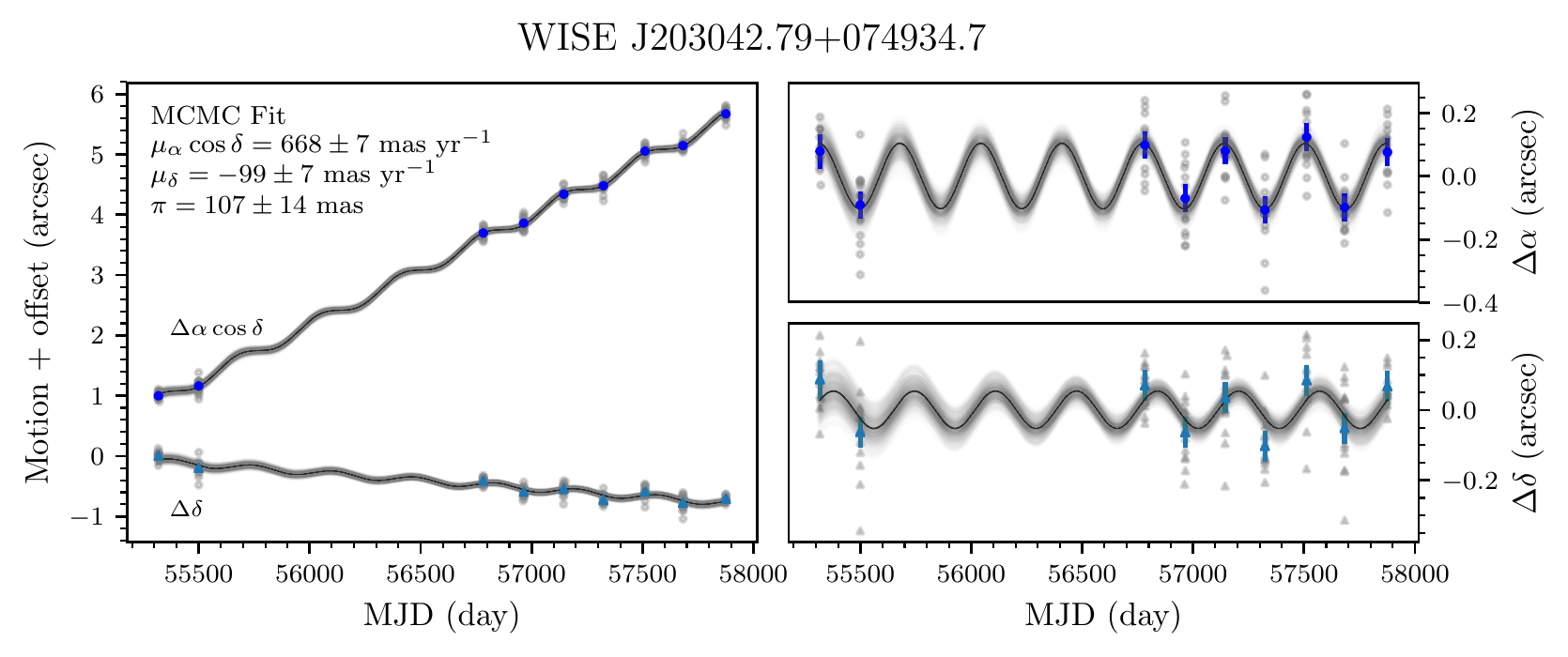}
\includegraphics{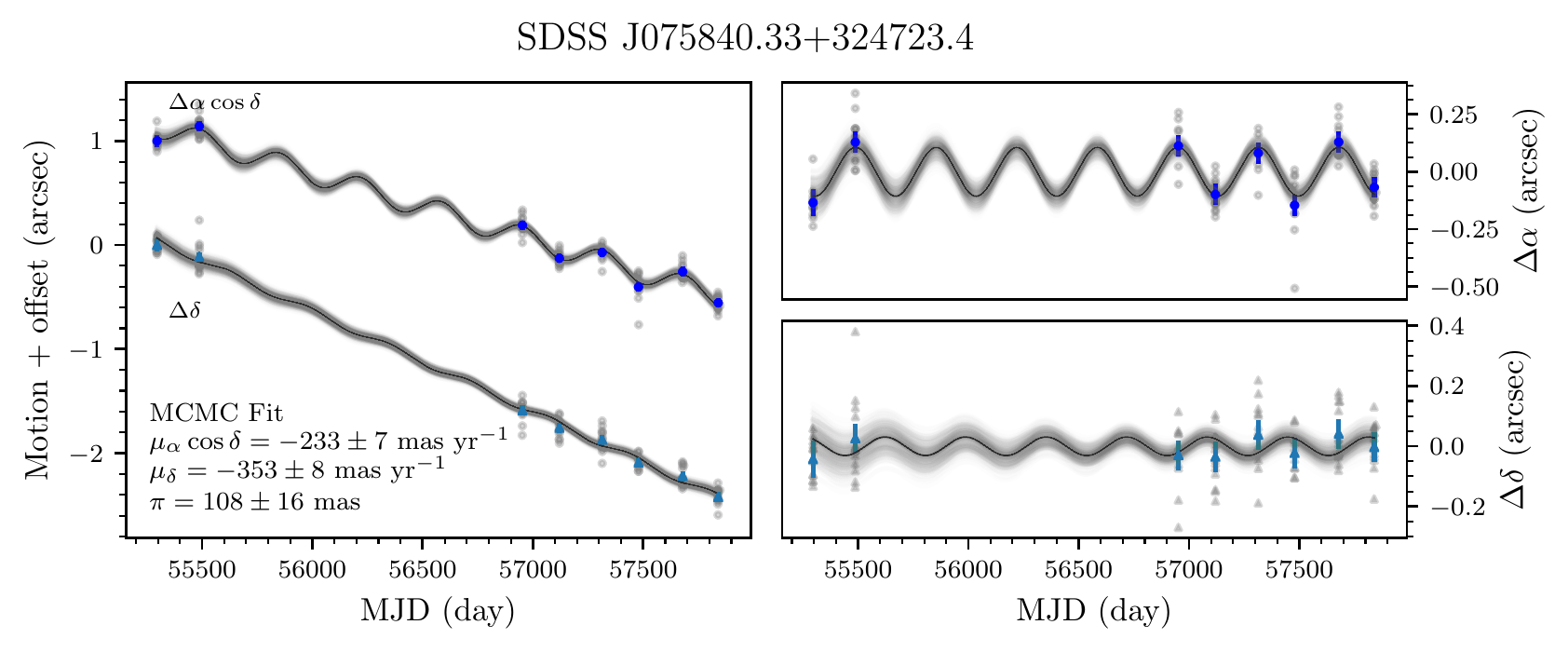}
\end{figure*}
\begin{figure*}
\includegraphics{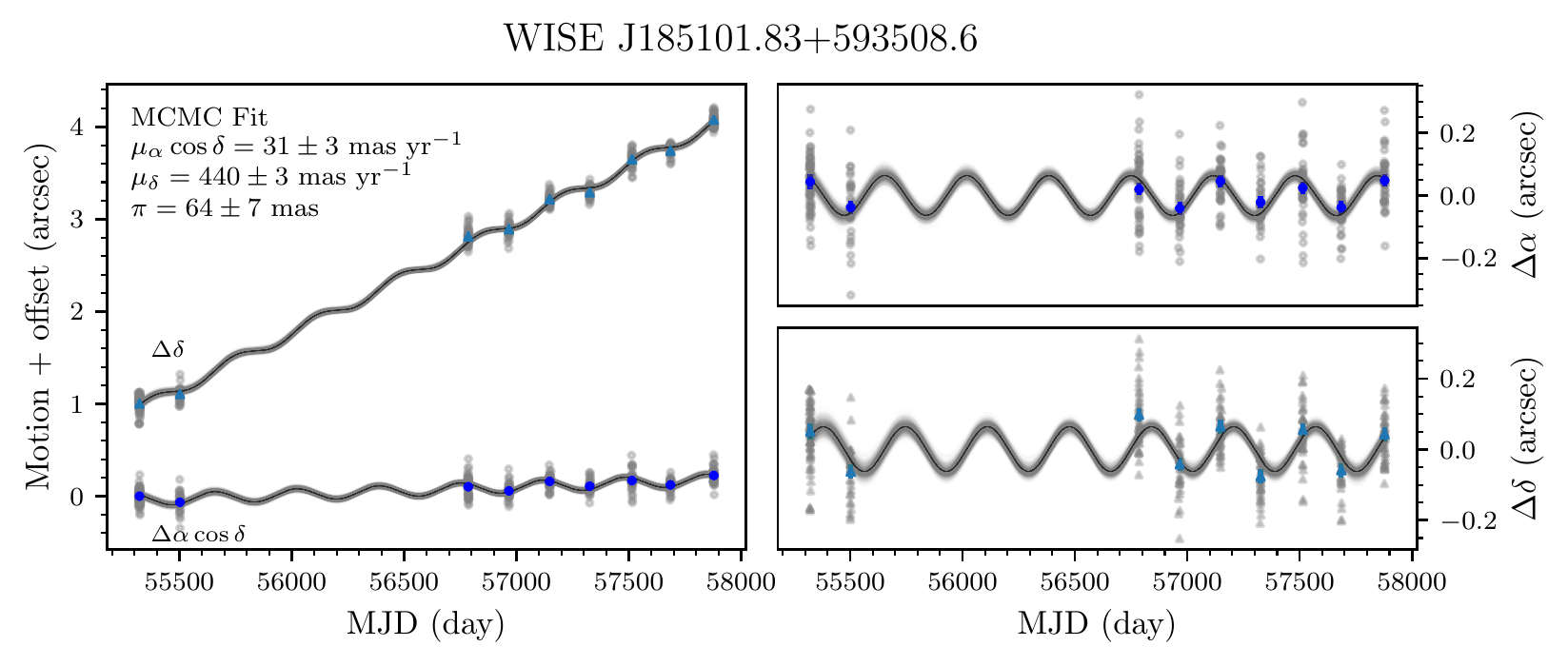}
\includegraphics{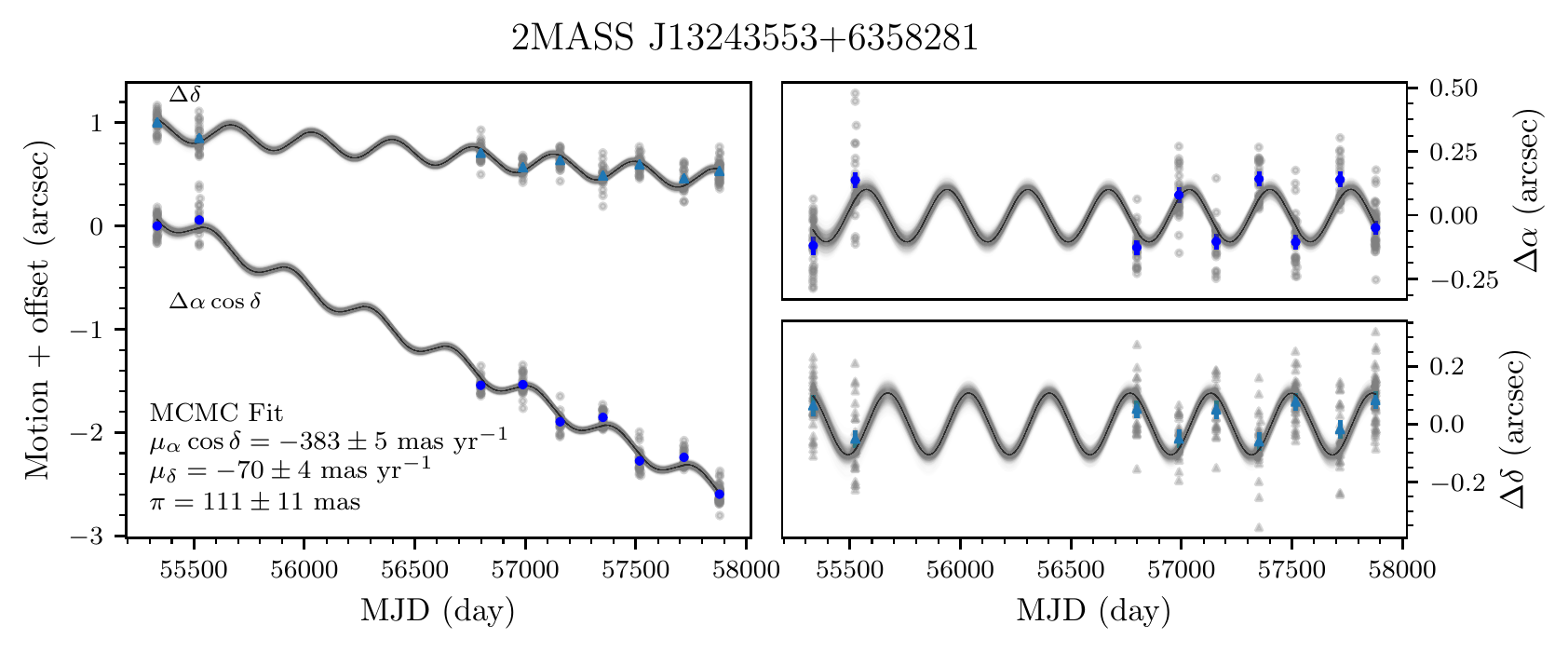}
\includegraphics{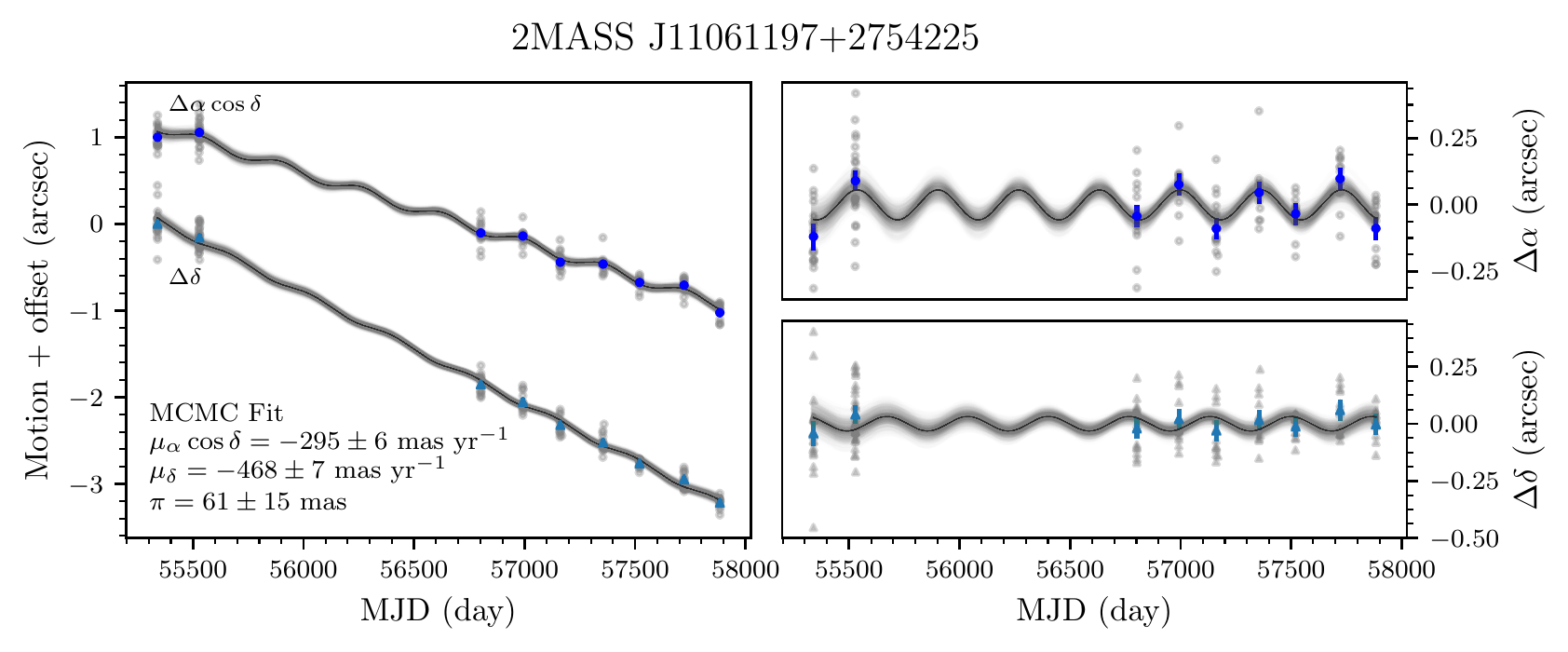}
\end{figure*}
\begin{figure*}
\includegraphics{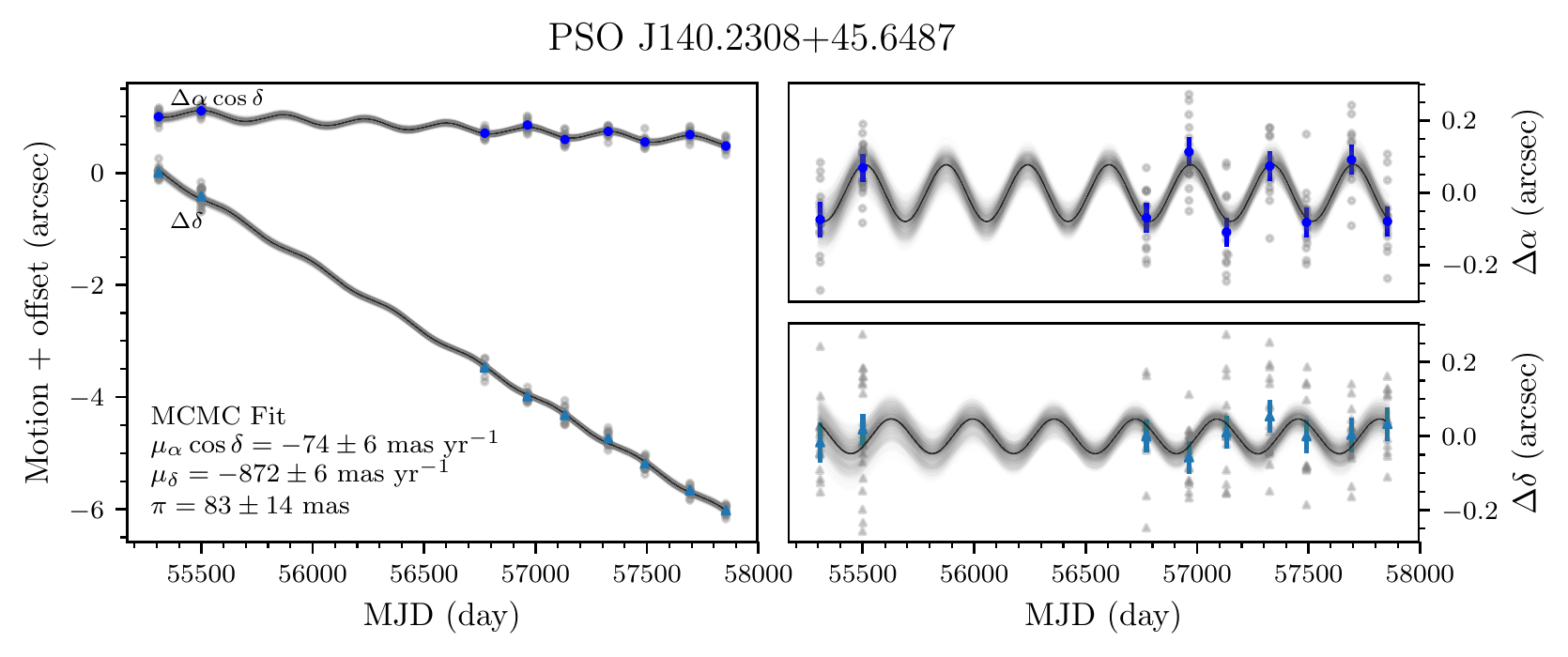}
\includegraphics{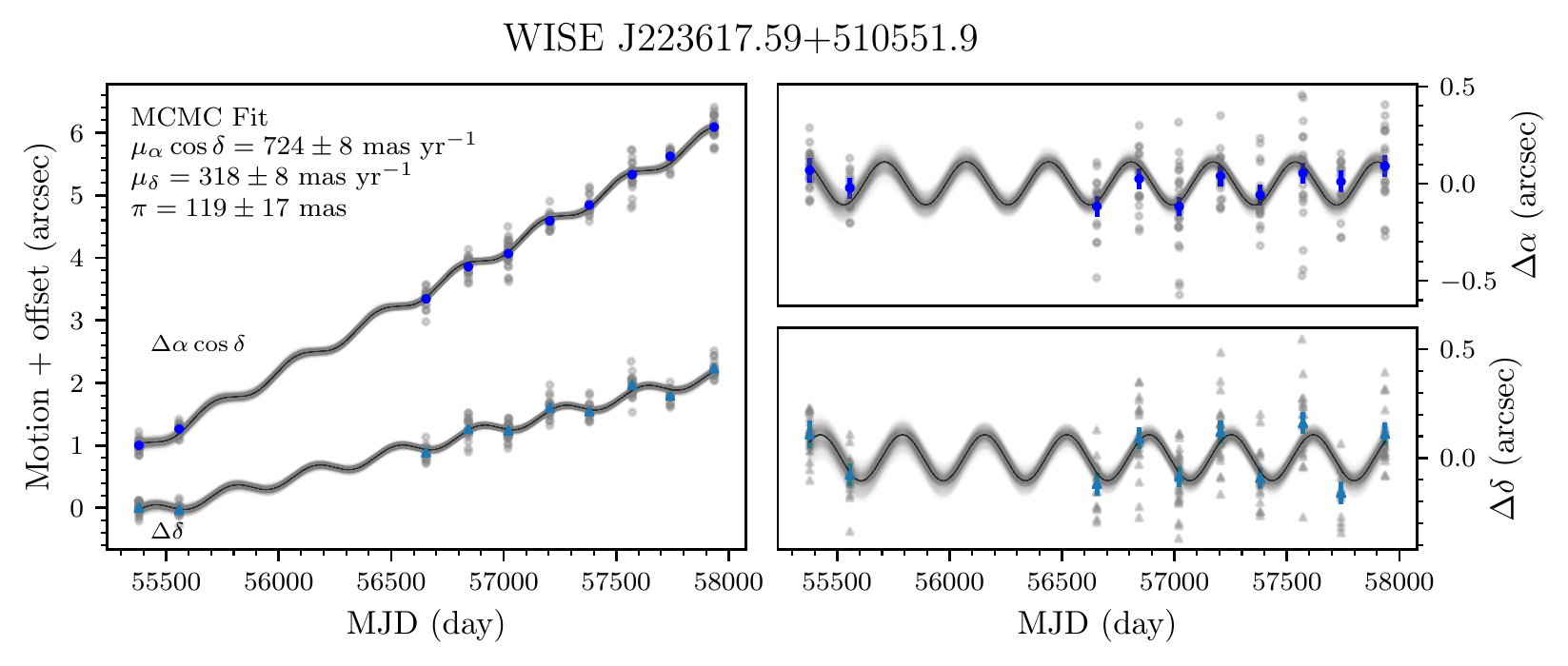}
\includegraphics{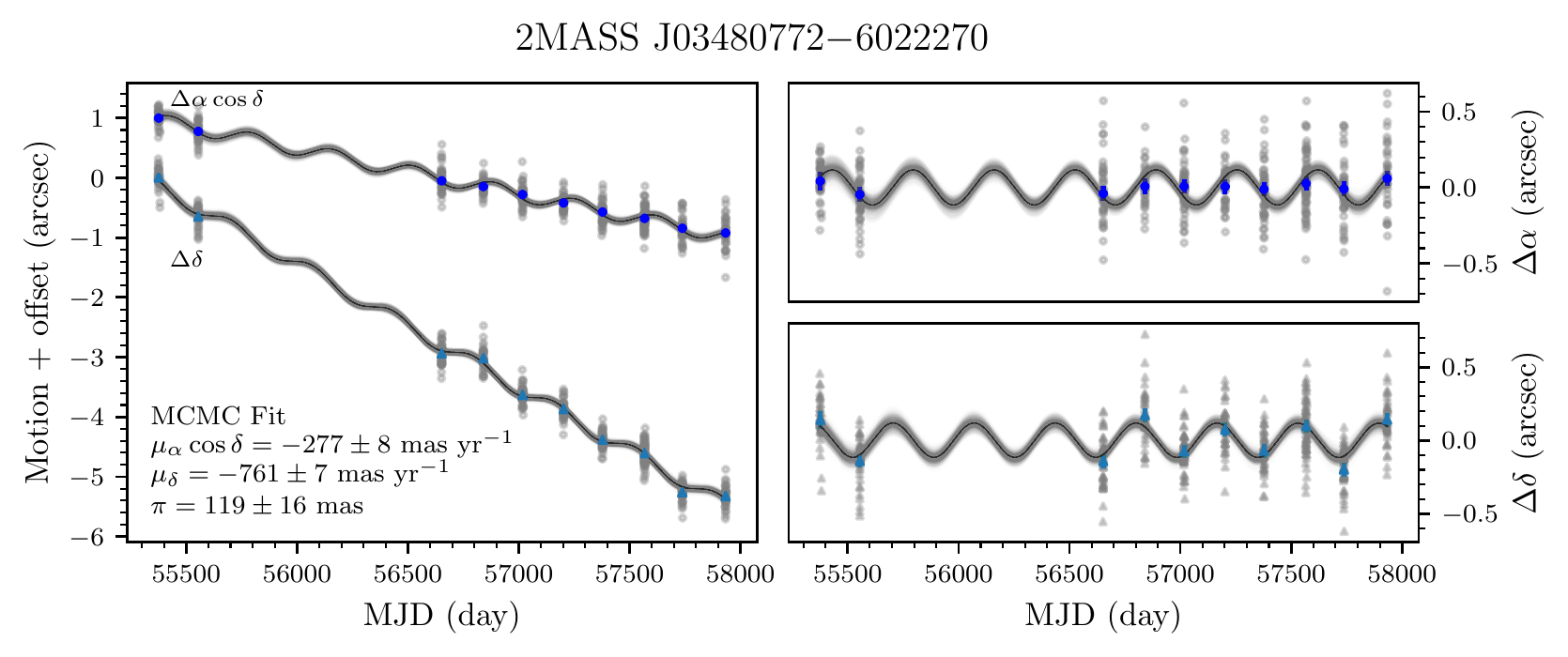}
\end{figure*}
\begin{figure*}
\includegraphics{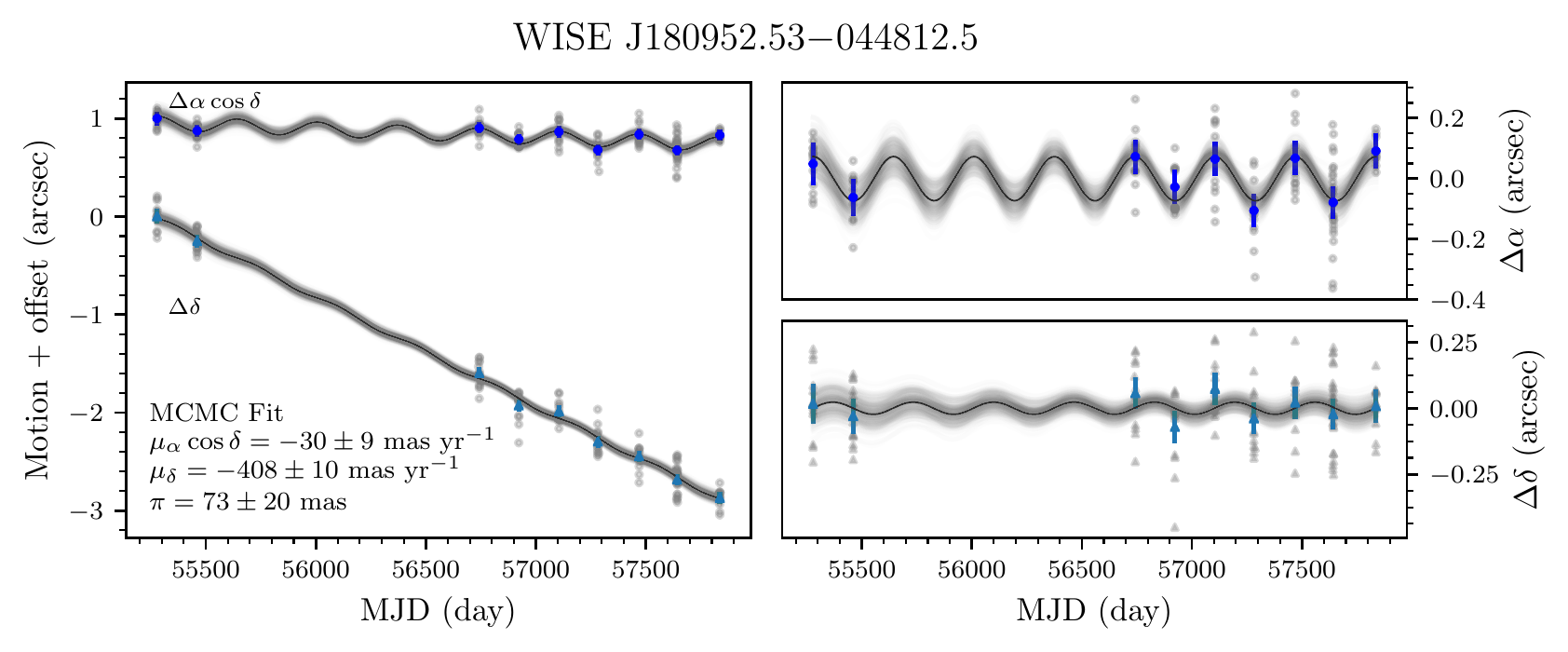}
\includegraphics{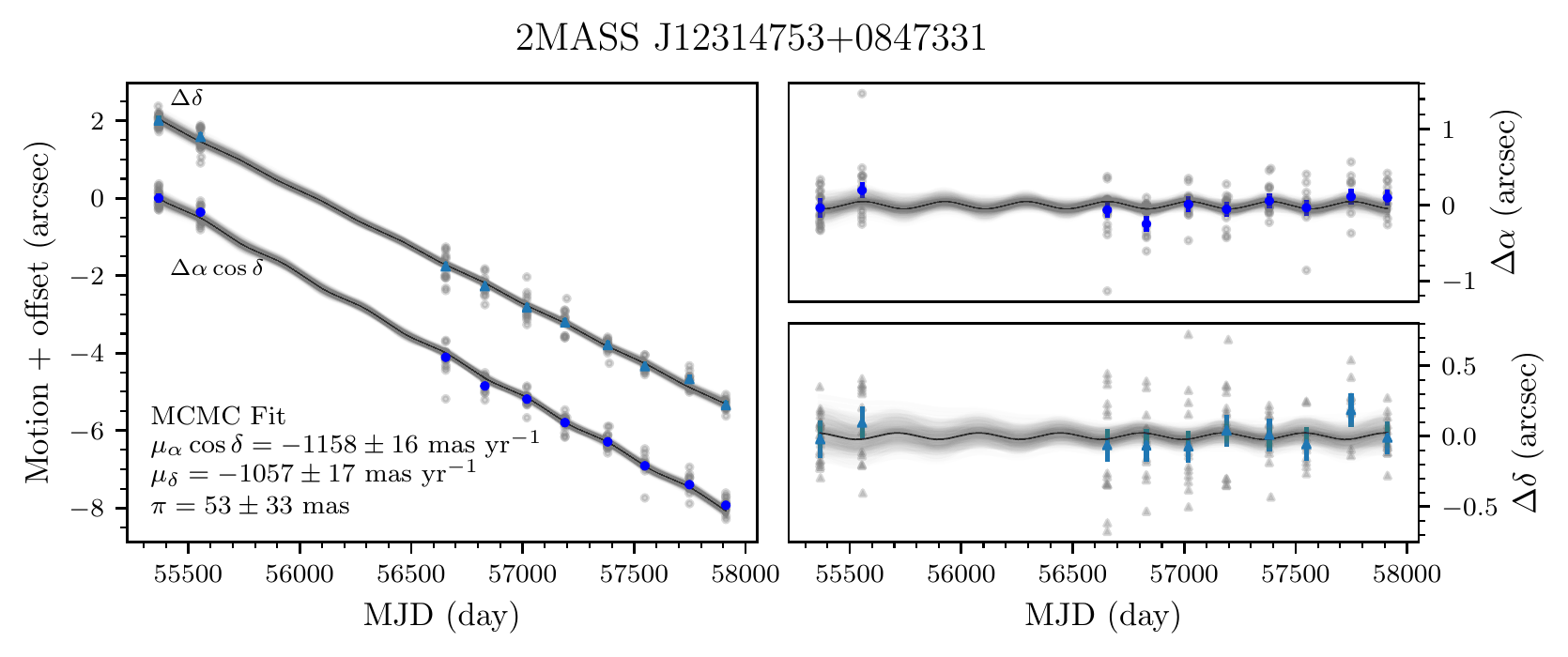}
\includegraphics{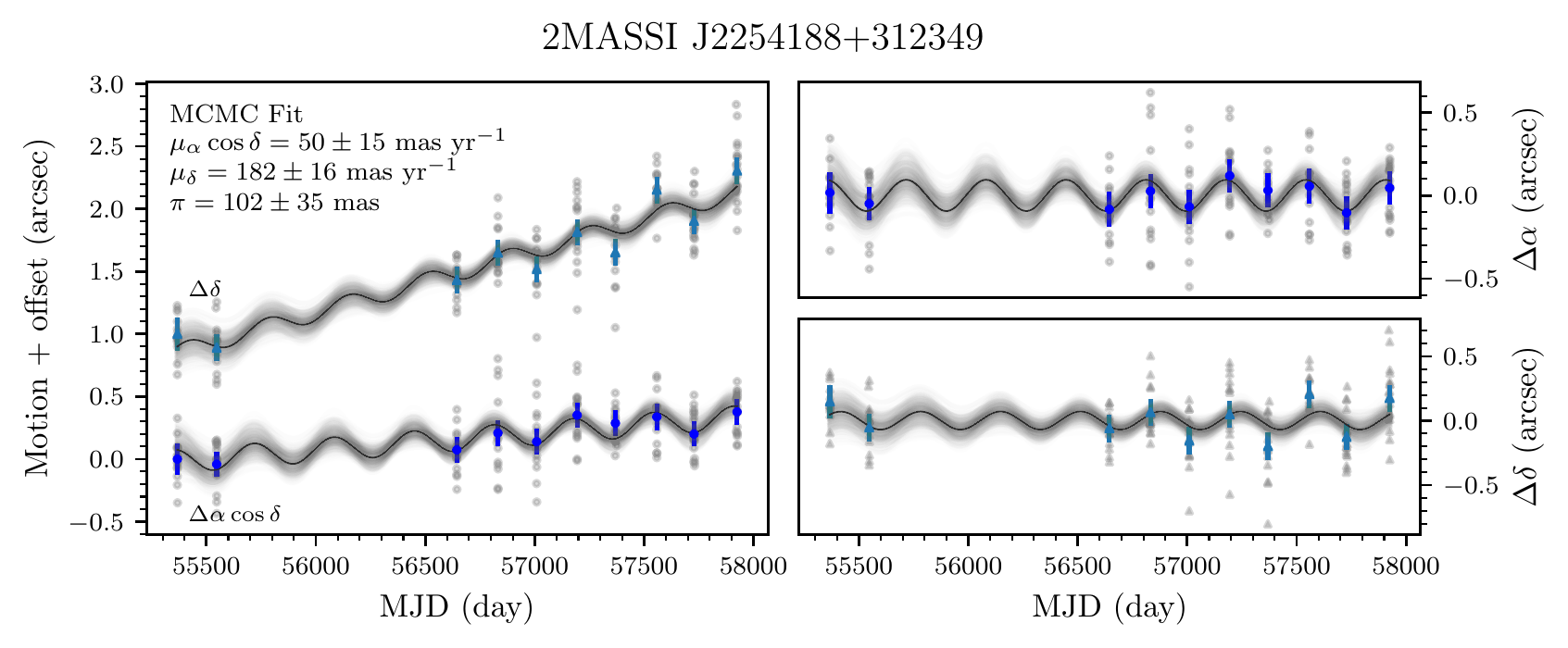}
\end{figure*}
\begin{figure*}
\includegraphics{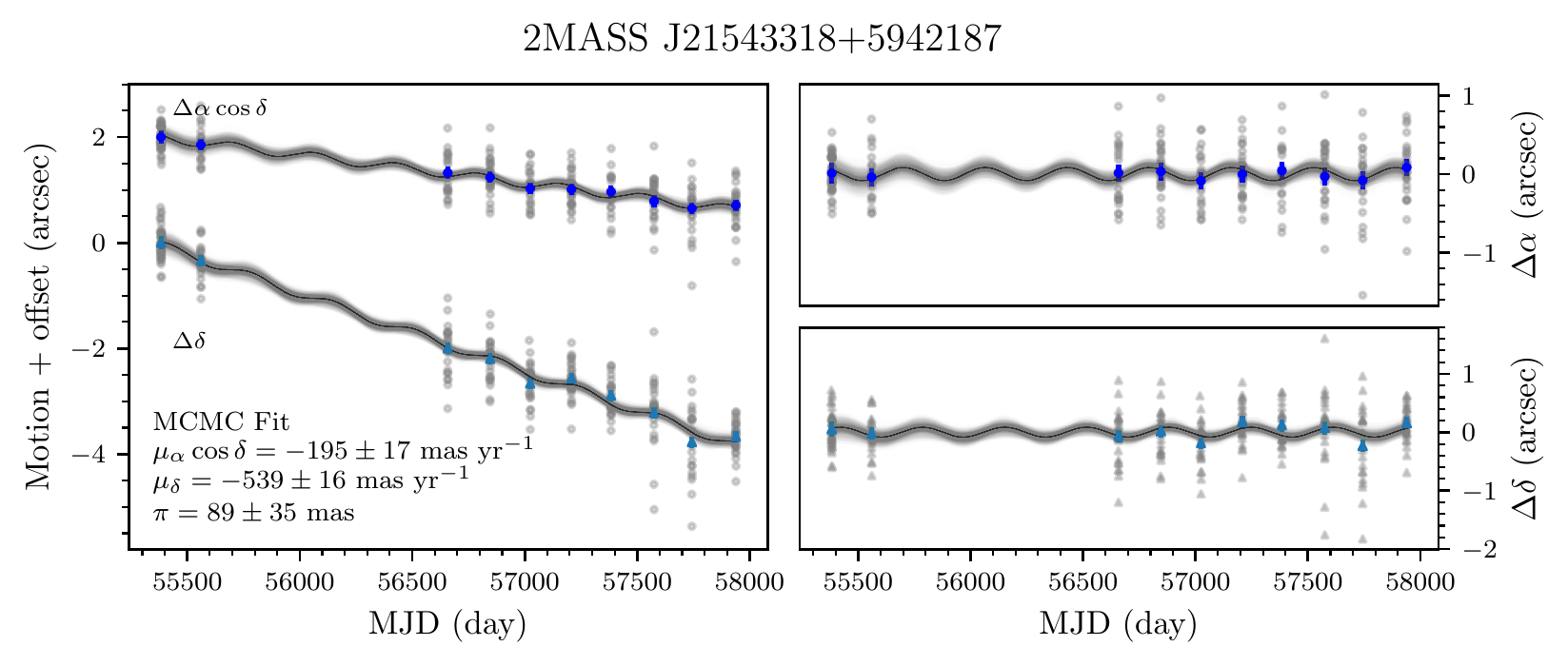}
\end{figure*}

\end{document}